\documentclass[12pt]{article}
\usepackage{amsfonts}
\usepackage{amsmath}
\usepackage{amssymb}
\usepackage{doublespace}
\usepackage{geometry}
\usepackage{tocbibind}
\usepackage[T1]{fontenc}

\input{tcilatex}
\begin{document}

\title{The Stroh formalism for acoustic waves in 1D-inhomogeneous media:\\
unified background and some applications }
\author{A. L.\ Shuvalov \\
Univ. Bordeaux, CNRS, Bordeaux INP, I2M, UMR 5295,\\
F-33400, Talence, France\\
\medskip \\
\ \ \ \ \ \ \ \ \ \ \ \ \ \ \ \ \ \ \ \ \ \ \ \ \ \ \ \ \ \ \ \ \ \ \ \ \ 
\textit{Dedicated with immense gratitude to}\\
\textit{\ \ \ \ \ \ \ \ \ \ \ \ \ \ \ \ \ \ \ \ \ \ \ \ \ \ \ \ \ \ \ \ \ \
\ \ \ Vladimir Alshits, David Barnett,}\\
\textit{\ \ \ \ \ \ \ \ \ \ \ \ \ \ \ \ \ \ \ \ \ \ \ \ \ \ \ \ \ \ \ \ \ \
\ \ \ Peter Chadwick, Arthur Every, }\\
\textit{\ \ \ \ \ \ \ \ \ \ \ \ \ \ \ \ \ \ \ \ \ \ \ \ \ \ \ \ \ \ \ \ \ \
\ \ \ Michael Hayes, Jens Lothe, }\\
\textit{\ \ \ \ \ \ \ \ \ \ \ \ \ \ \ \ \ \ \ \ \ \ \ \ \ \ \ \ \ \ \ \ \ \
\ \ \ my mentors, senior colleagues }\\
\textit{\ \ \ \ \ \ \ \ \ \ \ \ \ \ \ \ \ \ \ \ \ \ \ \ \ \ \ \ \ \
\smallskip\ \ \ \ \ \ \ and dear friends\bigskip }}
\maketitle
\tableofcontents

\pagebreak

\section*{\noindent Introduction\addcontentsline{toc}{section}{Introduction}}

The Stroh formalism is a ubiquitous concept in the elasticity of anisotropic
solids, named after A.N. Stroh and largely derived from his celebrated 1962
publication \cite{Stroh}. Addressing the steady two-dimensional (2D)
acoustic waves in elastic media with 1D material or geometrical
inhomogeneity, it combines the equation of motion and the stress-strain
relation in the convenient framework of a first-order ordinary differential
system (ODS) whose coefficient matrix is built from the density $\rho $ and
the stiffness tensor components $c_{ijkl}$ of the elastic material, and the
frequency $\omega $ and the wavenumber in the direction of inhomogeneity $k$
as the parameters. However, the significance of the Stroh formalism extends
beyond this: one of its key merits is that "Stroh's method is almost a
'Hamiltonian' formulation of elasticity" \cite{B}, i.e. the resulting ODS is
of the Hamiltonian type and hence possesses far-reaching algebraic
properties. They render the Stroh formalism indispensable for tackling
general anisotropic problems, where direct derivations are usually
infeasible due to the overwhelming number of material parameters. Its
combination with the concept of the surface impedance matrix \cite{IT} and
the integral formalism of the dislocation theory \cite{BL73} has generated
remarkable progress in the theory of the surface acoustic waves achieved
through the seminal papers by D. Barnett, J. Lothe, P. Chadwick and V.\
Alshits of the 1970s and early 1980s \cite{BL74}-\cite{AL81}. This
development culminated in the theorems of the existence and uniqueness of
the surface (Rayleigh) and interface (Stoneley) waves in homogeneous
half-spaces of arbitrary anisotropy, which were worked up to the final form
by Barnett and Lothe with coauthors in 1985 \cite{BL85,BLGM}. The momentum
continued in the 1990s by further elaborating the landscape of acoustic
phenomena in anisotropic media. A survey of the then state-of-the-art
advances in the theoretical crystal acoustics based on the Stroh formalism
may be found in \cite{B}, \cite{ADS}-\cite{LA1}\footnote{{\footnotesize %
Another breakthrough was an extension of the Stroh formalism and the
impedance matrix method to piezoelectric materials that Lothe and Barnett
published in two influential articles of 1976. It has considerably boosted
the research in piezoacoustics; however, this field will not be addressed in
the present review.}}. Various aspects of the Stroh formalism application to
static and dynamic elasticity of anisotropic homogeneous continuum are
expounded in a panoramic treatise by T.C.T. Ting \cite{T}, and a recap is
given in \cite{Ta}. Parallel to this, the Stroh-like formulation of the wave
equation has been disseminated in the seismology literature, see \cite{AR}.

The present review is primarily concerned with developments since the 2000s,
which marked the integration of Stroh's approach into the guided wave
problem methodology. A significant milestone was its application to the
theory of Lamb waves in homogeneous plates of unrestricted anisotropy and
then in 1D-functionally graded or multilayered plates (see \cite{ShPD} for
the overview and bibliography). Formally, this is a boundary value problem
of Stroh's ODS with variable coefficients, whose treatment rests on the
matricant solution (also called the propagator or transfer matrix) and the
plate (two-point) impedance matrix. Given a homogeneous or inhomogeneous
medium, the matricant is a matrix exponential or their product, or it is the
so-called product integral computable by standard methods. It determines the
plate impedance matrix, which is then plugged into the two-point boundary
condition to obtain a suitable form of the dispersion equation, linking the
frequency $\omega $ and the tangential wavenumber $k$. A valuable feature of
Stroh's ODS is that its coefficients do not contain derivatives of the
spatial dependence of the material parameters. Hence the matricant solution
is continuous across the (welded) interfaces, and so there is no need to
impose continuity as an additional requirement. Moreover, the Hamiltonian
nature of the Stroh formalism ensures that the analytical properties of the
inhomogeneous-plate impedance, despite its growing explicit complexity,
remain similar in the main to those of the Barnett-Lothe surface impedance
for a homogeneous half-space. These properties significantly facilitate the
computation and analysis of the guided wave dispersion spectrum $\omega
\left( k\right) $.

Since the 2010s, the sweeping trend towards studying wave phenomena in new
types of materials and structures has further unfolded the potential of
Stroh formalism. Its Hamiltonian structure allows taking advantage of the
spectral theory, which is well elaborated within the stability theory of
applied mathematics, particularly for cases with periodically varying
coefficients. The impedance based on the monodromy matrix (the matricant
over a single period) of the Hamiltonian-type ODS has proven highly
effective for treating boundary-value problems. This background makes the
Stroh formalism a powerful tool for dealing with acoustic waves in phononic
crystals with the 1D-periodic arrangement (superlattices), while using the
plane-wave expansion (PWE) method allows for straightforward extension to
the case of 2D- and 3D-periodic structures.

The aim of the review is to consolidate the core aspects of the Stroh
formalism, trace the development of the impedance matrix concept, and
outline the range of acoustic problems that have been treated by these
methods over the last two and a half decades. It is hoped to be of interest
to both the physical acoustics community by putting together the
mathematical basis of the above problems and to the applied mathematics
community by showcasing a physical context promising for the application of
well-established mathematical tools. It also seeks to be pedagogical enough
by providing a consistent and reasonably comprehensive account of the topic
using only primary analytical means while supplying references on technical
details.

The text consists of two parts and two appendices. Part I recaps the setup
of Stroh's ODS for plane waves in 1D-inhomogeneous anisotropic media and
summarizes the algebraic properties of its solutions based on the
Hamiltonian type of the system's coefficient matrix. This material largely
rests on a standard mathematical background available in many textbooks (we
refer to \cite{P,YS}). Part II, which is more specialized, begins by
revisiting the Barnett-Lothe surface impedance for a homogeneous half-space.
A brief update on the derivation of its properties is provided with the aim
of pinpointing general similarities with and specific differences from the
impedance matrices in transversely and laterally periodic half-spaces, which
are considered next.\ These impedance matrices are then used to analyze the
existence and number of surface waves. Another issue in Part II is the
two-point impedance for transversely inhomogeneous plates and its
application to studying the guided wave dispersion spectra. Appendix 1
mentions Stroh's original formulation as a partial differential system, and
the ODS for the case of time-space modulated material coefficients. Appendix
2 provides an overview of energy-based identities.\ 

We note that the Stroh formalism of elastodynamics allows for the
straightforward incorporation of the coupled-field phenomena, such as
piezoelectricity, magneto-piezoelectricity and thermoelasticity \cite{LB1}-%
\cite{LLWZ}. It can also accommodate the transient viscoelasticity \cite%
{CM,R}, adapts to the Lamb problem \cite{EKM, FKP}, and is applicable to the
constrained and prestressed materials \cite{Chad,EF}, nonlinear \cite{Fu0}
and nonlocal elasticity \cite{LK,NR}, the Willis constitutive model \cite%
{MW,ShKNP} and cloaking phenomena \cite{NSh1}. It serves a starting point
for the Wentzel--Kramers--Brillouin (WKB) and ray methods \cite{K,BK}, and
has a similar counterpart in optics \cite{Be}-\cite{D21}. However, these
topics are not addressed in this review. Unless specified otherwise, we will
confine to media with 1D inhomogeneity and omit discussion of the resolvent
and projector-based techniques, which extend the formalism to phononic
crystals with multi-dimensional periodicity \cite{KSh,KKShP}. We will not
cover other types of Stroh-like ODS describing cylindrical and spherical
waves in the axially inhomogeneous rectangularly anisotropic media \cite{K}
and in the radially inhomogeneous cylindrically \cite{Sh}{\footnotesize \ }%
and spherically \cite{NSh} anisotropic media. Finally, we will not delve
into a discussion of numerical implementation.

Let us introduce a few notational conventions for future use. The
superscripts $^{T},~^{\ast }$\ and $^{+}$ imply transpose, complex conjugate
and Hermitian conjugate, respectively. Zero and identity matrices of any
size are denoted by $\widehat{\mathbf{0}}$ and $\mathbf{I.}$ The notation $%
\mathrm{diag}\left( \cdot \right) $ indicates a diagonal matrix whose
non-zero entries are given in parentheses. A $2n\times 2n$ matrix $\mathbf{C}
$ will be written in the block or column form as 
\begin{equation}
\mathbf{C}=\left( 
\begin{array}{cc}
\mathbf{C}_{1} & \mathbf{C}_{2} \\ 
\mathbf{C}_{3} & \mathbf{C}_{4}%
\end{array}%
\right) =\left\Vert \mathbf{c}_{1}...\mathbf{c}_{2n}\right\Vert ,  \label{0}
\end{equation}%
where $\mathbf{C}_{1},...,\mathbf{C}_{4}$ (sometimes referred to as $\mathbf{%
C}_{1...4}$) are $n\times n$ blocks and $\mathbf{c}_{1},...,\mathbf{c}_{2n}$
are vector columns. Round and square brackets will signify dependence on a
variable and a parameter, respectively.\pagebreak

\part{Ordinary differential system of the Stroh formalism\label{Part1}}

\section{ The wave equation in the Stroh form\label{Sec1}}

Consider a purely elastic solid continuum of arbitrary anisotropy,
characterized by the mass density $\rho $ and the stiffness tensor $\mathbf{c%
}$ with components $c_{ijkl}$ defined in rectangular coordinates. Within the
framework of linear local elasticity without body forces, the equations of
acoustic wave motion and Hooke's law read%
\begin{equation}
\sigma _{ij,j}=\rho \ddot{u}_{i}\mathbf{,\ }\sigma _{ij}=c_{ijkl}u_{k,l},\ \
i,j,k,l=1,2,3,  \label{1}
\end{equation}%
where $u_{i}$ and $\sigma _{ij}$ are the components of the displacement
vector $\mathbf{u}\left( \mathbf{r}\right) $ and the stress tensor $\mathbf{%
\sigma }\left( \mathbf{r}\right) $ depending on the radius vector $\mathbf{r,%
}$ the commas and dots denote partial derivatives in coordinate $r_{i}$ and
time $t,$ respectively. Let us further assume that $\rho $ and $c_{ijkl}$
may vary continuously or stepwise along one direction due to intrinsic
material inhomogeneity (functional grading) or at an interface between
different materials (or both). Denote the unit vector parallel to this
distinguished direction by $\mathbf{e}_{2}$ and the unit vector in an
arbitrary direction orthogonal to $\mathbf{e}_{2}$ by $\mathbf{e}_{1}$; also
denote $x=\mathbf{r\cdot e}_{1}$ and $y=\mathbf{r\cdot e}_{2}$, whence $\rho
=\rho \left( y\right) $ and $c_{ijkl}=c_{ijkl}\left( y\right) $. The plane $%
XY$ spanned by the vectors $\mathbf{e}_{1}$ and $\mathbf{e}_{2}$ is called
the sagittal plane. Unless otherwise specified, we seek 2D wave modes in the
form of Fourier harmonics in $x$ and $t.$ Plugging the displacement and
traction vectors%
\begin{equation}
\mathbf{u}\left( x,y,t\right) =\mathbf{u}\left( y\right) e^{i\left(
k_{x}x-\omega t\right) },\ \ \mathbf{e}_{j}\cdot \mathbf{\sigma }\left(
x,y\right) \equiv \mathbf{t}_{j}\left( x,y,t\right) =\mathbf{t}_{j}\left(
y\right) e^{i\left( k_{x}x-\omega t\right) },\ \ j=1,2,3,  \label{2}
\end{equation}%
in Eqs. (\ref{1})$_{1}$ and (\ref{1})$_{2}$ yields differential equations on
the vector functions $\mathbf{u}\left( y\right) $ and $\mathbf{t}_{j}\left(
y\right) $, namely, 
\begin{equation}
\mathbf{t}_{2}^{\prime }+ik_{x}\mathbf{t}_{1}+\rho \omega ^{2}\mathbf{u=0}
\label{3}
\end{equation}%
and 
\begin{equation}
\mathbf{t}_{1}=\left( e_{1}e_{2}\right) \mathbf{u}^{\prime }+ik_{x}\left(
e_{1}e_{1}\right) \mathbf{u,\ \ t}_{2}=\left( e_{2}e_{2}\right) \mathbf{u}%
^{\prime }+ik_{x}\left( e_{2}e_{1}\right) \mathbf{u,}  \label{4}
\end{equation}%
where prime indicates the derivative in $y,$ and Lothe \& Barnett's notation 
$\left( e_{\alpha }e_{\beta }\right) ,$ $\alpha ,\beta =1,2$, implies the $%
3\times 3$ matrices with components 
\begin{equation}
\left( e_{\alpha }e_{\beta }\right) _{ik}\equiv \left( \mathbf{e}_{\alpha
}\right) _{j}c_{ijkl}\left( \mathbf{e}_{\beta }\right) _{l}.  \label{4.1}
\end{equation}%
They constitute the $3\times 3$ blocks $\mathbf{N}_{1},...,\mathbf{N}_{4}$
of the $6\times 6$ matrix $\mathbf{N}\left( y\right) $ (see (\ref{0})): 
\begin{equation}
\begin{array}{c}
\mathbf{N}_{1}=-\mathbf{T}^{-1}\mathbf{R}^{T},\ \mathbf{N}_{2}=-\mathbf{T}%
^{-1},\ \mathbf{N}_{3}=\mathbf{P-RT}^{-1}\mathbf{R}^{T},\ \mathbf{N}_{4}=-%
\mathbf{RT}^{-1}, \\ 
\mathbf{P}=\left( e_{1}e_{1}\right) \mathbf{,\ R}=\left( e_{1}e_{2}\right)
,\ \mathbf{T}=\left( e_{2}e_{2}\right) ,%
\end{array}
\label{7}
\end{equation}%
which is called the fundamental elasticity matrix in \cite{T} and is also
often referred to as the Stroh matrix. The symmetry of $c_{ijkl}$ in indices
and its strong ellipticity is assumed thereafter (unless where
viscoelasticity is mentioned); hence $\left( e_{\alpha }e_{\alpha }\right) $
is a symmetric and positive definite matrix and $\left( e_{\alpha }e_{\beta
}\right) =\left( e_{\beta }e_{\alpha }\right) ^{T}$. In consequence, the
blocks of the Stroh matrix satisfy the identities $\mathbf{N}_{1}\mathbf{e}%
_{2}=-\mathbf{e}_{1},$ $\mathbf{N}_{3}\mathbf{e}_{2}=\mathbf{0}$ and 
\begin{equation}
\mathbf{N}_{2}=\mathbf{N}_{2}^{T},\ \mathbf{N}_{3}=\mathbf{N}_{3}^{T},\ 
\mathbf{N}_{4}=\mathbf{N}_{1}^{T};  \label{7.1}
\end{equation}%
also, $\mathbf{N}_{2}$ is positive definite while $\mathbf{N}_{3}$ is
positive semi-definite (see \cite{T}). Note aside that the choice of letter
designations $\mathbf{P,~R,\ T}$ used in (\ref{5}) is not universally
adopted in the literature.

Substitution of (\ref{4}) to (\ref{3}) leads to a second-order ordinary
differential system (ODS) of three equations on the components of the
displacement $\mathbf{u}\left( y\right) $. An alternative idea put forward
by Stroh \cite{Stroh} and later and independently by Ingebrigtsen and
Tonning \cite{IT} suggests incorporating the vectors $\mathbf{u}\left(
y\right) $ and $\mathbf{t}_{2}\left( y\right) $ (\ref{2}) into a $6$%
-component state vector $\mathbf{\eta }\left( y\right) $ and thereby
transforming (\ref{3}) and (\ref{4}) into the 1st-order ODS with a $y$%
-dependent $6\times 6$ matrix of coefficients (the system matrix) $\mathbf{Q}%
\left( y\right) $ such that 
\begin{equation}
\mathbf{\eta }^{\prime }\left( y\right) =\mathbf{Q}\left( y\right) \mathbf{%
\eta }\left( y\right) ,  \label{5}
\end{equation}%
where $\mathbf{Q}$ takes over the block structure (\ref{7.1}) of $\mathbf{N,}
$ i.e. satisfies%
\begin{equation}
\left( \mathbf{\mathbb{T}Q}\right) ^{T}=\mathbf{\mathbb{T}Q}\Leftrightarrow 
\mathbf{Q}=\mathbf{\mathbb{T}Q}^{T}\mathbf{\mathbb{T}}\ \ \mathrm{with}\ 
\mathbf{\mathbb{T}}=\left( 
\begin{array}{cc}
\widehat{\mathbf{0}} & \mathbf{I} \\ 
\mathbf{I} & \widehat{\mathbf{0}}%
\end{array}%
\right) =\mathbf{\mathbb{T}}^{T}=\mathbf{\mathbb{T}}^{-1}.  \label{7.2}
\end{equation}%
Once $\mathbf{u}$ and $\mathbf{t}_{2}$ have been established, the traction
vector $\mathbf{t}_{1}$ can be found from (\ref{3}) or (\ref{4})$_{1}$,
while the components of $\mathbf{t}_{3}=\mathbf{e}_{3}\cdot \mathbf{\sigma }$
follow from the identities $\left( \mathbf{t}_{3}\right) _{1}=\left( \mathbf{%
t}_{1}\right) _{3},$ $\left( \mathbf{t}_{3}\right) _{2}=\left( \mathbf{t}%
_{2}\right) _{3},$ and the equality $u_{3,3}\left( =s_{33ij}\sigma
_{ij}\right) =0,$ where $s_{ijkl}$ is the compliance tensor \cite{T}.

The explicit form of $\mathbf{\eta }\left( y\right) $ and $\mathbf{Q}\left(
y\right) $ appearing in Eq. (\ref{5}) is optional up to scalar multipliers
involving the parameters $\omega $ and$~k_{x}$. Ingebrigtsen and Tonning 
\cite{IT} specified (\ref{5}) with 
\begin{equation}
\mathbf{\eta }=\left( 
\begin{array}{c}
\mathbf{u} \\ 
\mathbf{t}_{2}%
\end{array}%
\right) ,\ \ \mathbf{Q}\left[ k_{x},\omega ^{2}\right] =\left( 
\begin{array}{cc}
ik_{x}\mathbf{N}_{1} & -\mathbf{N}_{2} \\ 
k_{x}^{2}\mathbf{N}_{3}-\rho \omega ^{2}\mathbf{I} & ik_{x}\mathbf{N}_{1}^{T}%
\end{array}%
\right) ,  \label{8.1}
\end{equation}%
where $\left[ \cdot \right] ${\small \ }indicates the parametric{\small \ }%
dependence, while the dependence on $y$ is understood and omitted. This
formulation is common in structural elasticity, see \cite{YZL}. Other
broadly used formulations are 
\begin{equation}
\begin{array}{l}
\mathbf{\eta }=\left( 
\begin{array}{c}
\mathbf{u} \\ 
ik_{x}^{-1}\mathbf{t}_{2}%
\end{array}%
\right) ,\ \ \mathbf{Q}=ik_{x}\left( 
\begin{array}{cc}
\mathbf{N}_{1} & \mathbf{N}_{2} \\ 
\mathbf{N}_{3}-\rho v^{2}\mathbf{I} & \mathbf{N}_{1}^{T}%
\end{array}%
\right) \equiv ik_{x}\mathbf{N}\left[ v^{2}\right] , \\ 
\mathbf{\eta }=\left( 
\begin{array}{c}
i\omega \mathbf{u} \\ 
\mathbf{t}_{2}%
\end{array}%
\right) ,\ \ \mathbf{Q}=i\omega \left( 
\begin{array}{cc}
s\mathbf{N}_{1} & -\mathbf{N}_{2} \\ 
-s^{2}\mathbf{N}_{3}+\rho \mathbf{I} & s\mathbf{N}_{1}^{T}%
\end{array}%
\right) \equiv i\omega \mathbf{N}\left[ s\right] , \\ 
\mathbf{\eta }=\left( 
\begin{array}{c}
\mathbf{u} \\ 
i\mathbf{t}_{2}%
\end{array}%
\right) ,\ \ \mathbf{Q}=i\left( 
\begin{array}{cc}
k_{x}\mathbf{N}_{1} & \mathbf{N}_{2} \\ 
k_{x}^{2}\mathbf{N}_{3}-\rho \omega ^{2}\mathbf{I} & k_{x}\mathbf{N}_{4}%
\end{array}%
\right) \equiv i\mathbf{N}\left[ k_{x},\omega ^{2}\right] ,%
\end{array}
\label{8}
\end{equation}%
where $v=\omega /k_{x}=s^{-1}$. Definition (\ref{8})$_{1}$ stemming from
Stroh's original development was introduced in this form by Lothe \& Barnett 
\cite{LB} and adopted in numerous successive publications. Definition (\ref%
{8})$_{2},$ which is traced from the papers by Thompson and Haskell \cite%
{Th,Ha}, is often utilized in the seismic wave calculations \cite{K}. Note
that having the matrix $\mathbf{Q}$ with factored-out parameters $k_{x}$ or $%
\omega ,$ as in (\ref{8})$_{1}$ or (\ref{8})$_{2},$ is convenient for
casting the finite-difference scheme of Eq. (\ref{5}) as the eigenvalue
problem and for the power-series expansion at simultaneously small $k_{x}$
and $\omega $ (hence finite $v$ or $s$). In turn, definition (\ref{8})$_{3}$
offers a versatile pattern, which is suitable for either small $k_{x}$ and
finite $\omega $ or vice versa. In the following, when referring to the
displacement-traction state vector without specifying an explicit choice
between the above options, we will write it as $\mathbf{\eta }\left(
y\right) =\left( \mathbf{a}\left( y\right) \ \mathbf{b}\left( y\right)
\right) ^{T}$.

Thus, the Stroh formalism is based on the standard way of reducing the order
of ODS via engaging the derivative of the sought function as a complementary
unknown; however, Stroh's actual procedure is more powerful. The key point
is that the unknown displacement $\mathbf{u}\left( y\right) $ is
complemented not by $\mathbf{u}^{\prime }\left( y\right) $ but by its matrix
multiple, namely, the traction $\mathbf{t}_{2}\left( y\right) .$ This leads
to two advantages: first, the corresponding system matrix $\mathbf{Q}\left(
y\right) $ acquires a Hamiltonian structure (see \S \ref{Sec3}) and second,
it involves the density $\rho \left( y\right) $ and the stiffness
coefficients $\mathbf{c}\left( y\right) $ but not their derivatives, as it
would be if $\mathbf{u}^{\prime }\left( y\right) $ were used. As a result,
finite jumps of $\rho \left( y\right) $ and/or $\mathbf{c}\left( y\right) $
keep the matrix function $\mathbf{Q}\left( y\right) $ piecewise continuous
integrable and hence the solution of (\ref{5}) continuous. In other words,
when seeking the wave solutions in layered media, there is no need to impose
continuity of displacements and tractions as additional boundary conditions
at (rigid) interfaces\footnote{%
\noindent {\footnotesize The reader is cautioned against viral publications
touting a mutilated version of the Stroh formalism, enterprisingly branded
with the reverent Cauchy name. Without any rational reason, it ignores
Stroh's idea and advocates the choice of the state vector formed of }$%
\mathbf{u}${\footnotesize \ and }$\mathbf{u}^{\prime }${\footnotesize . As
implied above, this framework obscures the Hamiltonian nature of the
problem, involves derivatives of }$c_{ijkl}${\footnotesize \ that diverge at
the interfaces, and on top of that, is disproportionately cumbersome. }}. In
this regard, the solid-fluid{\small \ }interface stands out since the
vanishing of the shear modulus precludes defining the inverse matrix $%
\mathbf{T}^{-1},$ which appears in the system matrix, see (\ref{7}) and (\ref%
{8.1}), (\ref{8}). The same exception is the solid-solid sliding contact,
which can be viewed as a fluid interlayer with a thickness tending to zero%
{\small .\ }In both latter cases, the solutions must be independently
obtained in each of the two neighboring media and then stitched together
according to the appropriate boundary condition at their interface (see
references in \S \S \ref{SSSec4.5.2} and \ref{SSec6.5}). Another reservation
concerns the WKB asymptotic solution, which assumes finite derivatives of
material coefficients and hence remains continuous in a functionally graded
medium, but diverges at the rigid-contact interfaces or even at the
so-called weak interfaces, see e.g. \cite{BG,BSP}.

\section{The matricant\label{Sec2}}

\subsection{Definition\label{SSec2.1}}

Denote by $\mathcal{N}\left( y\right) =\left\Vert \mathbf{\eta }_{1}\left(
y\right) ,...,\mathbf{\eta }_{6}\left( y\right) \right\Vert $ the $6\times 6$
fundamental matrix solution of ODS (\ref{5}), whose columns $\mathbf{\eta }%
_{\alpha }\left( y\right) ,$ $\alpha =1,...,6,$ are linearly independent
partial solutions. Provided that the system matrix $\mathbf{Q}\left(
y\right) $ components are piecewise continuous integrable functions of $y,$
the matrix $\mathcal{N}\left( y\right) $ components are continuous. Given
the $6\times 6$ matrix $\mathcal{N}\left( y_{0}\right) $ of initial data at
some $y_{0}$, the matricant is defined as 
\begin{equation}
\mathbf{M}\left( y,y_{0}\right) =\mathbf{\mathcal{N}}\left( y\right) \mathbf{%
\mathcal{N}}^{-1}\left( y_{0}\right) \ \Leftrightarrow \ \mathbf{\eta }%
_{\alpha }\left( y\right) =\mathbf{M}\left( y,y_{0}\right) \mathbf{\eta }%
_{\alpha }\left( y_{0}\right) ,\ \alpha =1,...,6.  \label{9}
\end{equation}%
In other words, it is the solution of the matrix differential equation with
the identity initial condition, 
\begin{equation}
\mathbf{M}^{\prime }\left( y,y_{0}\right) =\mathbf{Q}\left( y\right) \mathbf{%
M}\left( y,y_{0}\right) ,\ \ \mathbf{M}\left( y_{0},y_{0}\right) =\mathbf{I.}
\label{5.1}
\end{equation}%
The continuity of the solution implies that 
\begin{equation}
\mathbf{M}\left( y,y_{0}\right) =\mathbf{M}\left( y,y_{1}\right) \mathbf{M}%
\left( y_{1},y_{0}\right) .  \label{9.1}
\end{equation}%
If $\omega ,~k_{x}$ and $\rho ,~c_{ijkl}$ are all real, then any of the
matrices $\mathbf{Q}$ given by (\ref{8.1}) or (\ref{8}) has a purely
imaginary trace and hence, by Liouville's formula, $\left\vert \det \mathbf{M%
}\right\vert =\left\vert \exp \left( \mathrm{tr}\mathbf{Q}\right)
\right\vert =1$ (see also \S \ref{SSSec3.3.1}). Note that, as evident from (%
\ref{9})$_{2},$ a scalar factor multiplying $\mathbf{u}$ or $\mathbf{t}_{2}$
when switching between the equivalent definitions (\ref{8.1}) and (\ref{8})
of $\mathbf{\eta }$ results in the same factor multiplying off-diagonal
blocks of the matricant $\mathbf{M}$ corresponding to this definition.

In the simple case of constant material properties $\rho $ and$~c_{ijkl}$
within some given interval $\left[ y_{0},y_{1}\right] $ and hence a constant
system matrix $\mathbf{Q=Q}_{0}$ in there, the matricant of (\ref{5}) is the
matrix exponential 
\begin{equation}
\ \mathbf{M}\left( y,y_{0}\right) =e^{\mathbf{Q}_{0}\left( y-y_{0}\right) }\
\forall y\in \left[ y_{0},y_{1}\right] ,  \label{11.0}
\end{equation}%
see \cite{MV-L}. The corresponding partial solutions of the form 
\begin{equation}
\mathbf{\eta }_{\alpha }\left( y\right) =\mathbf{\xi }_{\alpha
}e^{ik_{y\alpha }y},\   \label{11.01}
\end{equation}%
with $ik_{y\alpha }$ and $\mathbf{\xi }_{\alpha }$ being the eigenvalues and
eigenvectors of $\mathbf{Q}_{0},$ are often referred to as the $\alpha $th
(eigen)modes ($\alpha =1,...,6$ if $\mathbf{Q}_{0}$ is diagonalizable),
while the columns of $\mathbf{\mathcal{N}}\left( y\right) $ are their
arbitrary linear independent superpositions. It is seen from (\ref{11.0})
that $\mathbf{M}=\mathbf{\mathbb{T}M}^{T}\mathbf{\mathbb{T}}$ due to (\ref%
{7.2}) and that $\mathbf{M}^{-1}=\mathbf{M}^{\ast }$ in the case of purely
imaginary form of $\mathbf{Q}_{0}$. If the matrix function $\mathbf{Q}\left(
y\right) $ is piecewise constant on $\left[ y_{0},y_{1}\right] $, i.e. takes
constant values $\mathbf{Q}_{0j}$ on $n$ subintervals of the width $d_{j},$ $%
j=1,...,n$ (a model for a stack of homogeneous layers), then, by (\ref{9.1})
and (\ref{11.0}), the matricant through $\left[ y_{0},y_{1}\right] $ is 
\begin{equation}
\mathbf{M}\left( y_{1},y_{0}\right) =e^{\mathbf{Q}_{0n}d_{n}}...e^{\mathbf{Q}%
_{01}d_{1}}.  \label{11}
\end{equation}%
If $\mathbf{Q}\left( y\right) $ is continuous in $\left[ y_{0},y_{1}\right] $
(the case of functionally graded materials), the matricant of (\ref{5}) can
be expressed as Picard's iterative solution leading to the Peano-Baker
series of multiple integrals, namely, 
\begin{equation}
\begin{array}{c}
\mathbf{M}\left( y,y_{0}\right) =\mathbf{I}+\int\nolimits_{y_{0}}^{y}\mathbf{%
Q}\left( \varsigma _{1}\right) \mathrm{d}\varsigma
_{1}+\int\nolimits_{\varsigma _{1}}^{y}\mathbf{Q}\left( \varsigma
_{1}\right) \mathrm{d}\varsigma _{1}\int\nolimits_{y_{0}}^{\varsigma _{1}}%
\mathbf{Q}\left( \varsigma _{2}\right) \mathrm{d}\varsigma _{2}+\ldots \\ 
\equiv \widehat{\int }_{y_{0}}^{y}\left[ \mathbf{I}+\mathbf{Q}\left(
\varsigma \right) \mathrm{d}\varsigma \right] \ \ \forall y\in \left[
y_{0},y_{1}\right] ,%
\end{array}%
\   \label{10}
\end{equation}%
where the symbol $\widehat{\int }_{y_{0}}^{y}$ implies the so-called product
integral or multiplicative integral of Volterra \cite{P} (it bears some
other names in various physical applications). The series (\ref{10}) may be
viewed as the limiting form of the product (\ref{11}) with $y_{1}\equiv y,$
where the limit $n\rightarrow \ \infty $ is taken at a fixed $\Delta
y=y-y_{0}$. It reduces to (\ref{11.01}) or (\ref{11}) for a constant or
piecewise constant $\mathbf{Q}\left( y\right) $.

As an exceptional option, $\mathbf{M}\left( y,y_{0}\right) =\exp \left(
\int\nolimits_{y_{0}}^{y}\mathbf{Q}\left( \varsigma \right) \mathrm{d}%
\varsigma \right) $ if the values of $\mathbf{Q}\left( y\right) $ taken at
different points in $\left[ y_{0},y_{1}\right] $ commute with each other;
however, this is practically irrelevant to the case under study. One special
context where the exponential of the integrals of eigenvalues of $\mathbf{Q}%
\left( y\right) $ comes into play is the asymptotic WKB solution (see \cite%
{K,BK} for more details).

Among the basic properties following from the matricant definition, note a
useful identity 
\begin{equation}
\mathbf{M}^{-1}\left( y,y_{0}\right) =\mathbf{M}_{-\mathbf{Q}^{+}}^{+}\left(
y,y_{0}\right) ,  \label{10.1}
\end{equation}%
where $\mathbf{M}_{-\mathbf{Q}^{+}}$ is the matricant of the ODS $\mathbf{%
\eta }^{\prime }=-\mathbf{Q}^{+}\mathbf{\eta ,}$ which is said to be
conjugate to ODS (\ref{5}) \cite{YS}. This identity may be proved directly
from (\ref{5}) or else from (\ref{10}) for any matrix $\mathbf{Q}$, i.e.
regardless of (\ref{7.2}).

\subsection{Matricant for periodic media. The plane-wave expansion\label%
{SSec2.2}}

Assume that the 1D-dependences of the density $\rho \left( y\right) $ and
stiffness coefficients $c_{ijkl}\left( y\right) $ are periodic with a period 
$T$ so that $\mathbf{Q}\left( y\right) =$ $\mathbf{Q}\left( y+T\right) $.
Then identity (\ref{9.1}) with $\widetilde{y}=y~\func{mod}T$ $(y=\widetilde{y%
}+nT)$ and $y_{0}=0$ set for convenience may be expressed in the form 
\begin{equation}
\mathbf{M}\left( y,0\right) =\mathbf{M}\left( \widetilde{y}+nT,nT\right) 
\mathbf{M}\left( nT,0\right) =\mathbf{M}\left( \widetilde{y},0\right) 
\mathbf{M}^{n}\left( T,0\right) ,  \label{12}
\end{equation}%
where the second equality is due to the periodicity. The matricant $\mathbf{M%
}\left( T,0\right) $ over a single period is called the monodromy matrix and
is customarily denoted as 
\begin{equation}
\mathbf{M}\left( T,0\right) =e^{i\mathbf{K}T}.  \label{13}
\end{equation}%
Its eigenvalues $e^{iK_{y\alpha }T}\equiv q_{\alpha },$ $\alpha =1,...,6,$
are called the multipliers, while the eigenvalues $K_{y\alpha }$ of the
matrix $\mathbf{K}$ are referred to as the Floquet-Bloch wavenumbers. Eq. (%
\ref{12}) may be continued as 
\begin{equation}
\mathbf{M}\left( y,0\right) =\mathbf{M}\left( \widetilde{y},0\right) e^{in%
\mathbf{K}T}=\mathbf{L}\left( y\right) e^{i\mathbf{K}y},  \label{14}
\end{equation}%
where $\mathbf{L}\left( y\right) =\mathbf{M}\left( \widetilde{y},0\right)
e^{-i\mathbf{K}\widetilde{y}}=\mathbf{L}\left( y+T\right) $ and $\mathbf{L}%
\left( 0\right) =\mathbf{I}$. Formula (\ref{14}) expresses the Floquet-Bloch
(or Floquet-Lyapunov) theorem.

In our case, $\mathbf{M}\left( T,0\right) $ is a function of the parameters $%
\omega $ and $k_{x},$ hence so is $\mathbf{K}.$ Note that Eqs. (\ref{13}), (%
\ref{14}) motivate the definition of a complex logarithm matrix function $i%
\mathbf{K}\left[ \omega ,k_{x}\right] T=\mathrm{Ln}\mathbf{M}\left(
T,0\right) ,$ whose appropriately defined principal branch proves
instrumental in the context of the low-frequency effective media modelling 
\cite{ShKNP, ShKN}\footnote{{\footnotesize Mind possible confusion in
interpreting the results of \cite{ShKNP}: the identity} $\mathbf{K=\mathbb{T}%
K}^{+}\mathbf{\mathbb{T}}$ {\footnotesize was rightfully attributed there to
the first Brillouin zone (see Eq. (2.11)), but this reservation was left out
later, particularly, when stating the equalities (3.9).}}.

The above matricant is associated with the 2D wave solution in the form (\ref%
{2}) and is well suited for the boundary value problems in 1D-periodic media
truncated by the plane(s) \textit{orthogonal} to the periodicity direction%
{\small . }The situation is different if the sought solution does not allow
the "pure plane-wave type" dependence in any of the two spatial coordinates,
i.e. if the medium is truncated by the plane \textit{parallel} to the
periodicity direction or if it is periodic in both coordinates. Such cases
call for the plane-wave expansion (PWE).

Consider the PWE method integrated into the Stroh formalism. Aiming at the
ODS in $y$, we assume that the material properties are periodic in $x$ and
also depend on $y$. Then the partial solutions of Eq.\ (\ref{1}) may be
sought in the Floquet-Bloch form with the periodic part expanded in the
Fourier series, namely, 
\begin{equation}
\left( 
\begin{array}{c}
\mathbf{u}\left( x,y\right) \\ 
i\mathbf{t}_{2}\left( x,y\right)%
\end{array}%
\right) =\left( 
\begin{array}{c}
\mathbf{a}\left( x,y\right) \\ 
\mathbf{b}\left( x,y\right)%
\end{array}%
\right) e^{iK_{x}x}=\sum_{n=-\infty }^{\infty }\left( 
\begin{array}{c}
\mathbf{\hat{a}}^{\left( n\right) }\left( y\right) \\ 
\mathbf{\hat{b}}^{\left( n\right) }\left( y\right)%
\end{array}%
\right) e^{ik_{n}x},  \label{15}
\end{equation}%
where the vector functions $\mathbf{a}\left( x,y\right) $ and $\mathbf{b}%
\left( x,y\right) $ are $T$-periodic in $x$ with Fourier coefficients $%
\mathbf{\hat{a}}^{\left( n\right) }\left( y\right) $ and $\mathbf{\hat{b}}%
^{\left( n\right) }\left( y\right) ,$ and 
\begin{equation}
k_{n}=K_{x}+gn,\ \left\vert K_{x}\right\vert \leq \frac{\pi }{T},\ g=\frac{%
2\pi }{T}.  \label{16}
\end{equation}%
It is understood that all functions in (\ref{15}) depend on $\omega $ and on
the wavenumber $K_{x}$ as the free parameter (unlike the wavenumbers $%
K_{y\alpha }=K_{y\alpha }\left( \omega ,k_{x}\right) $ in (\ref{13})).

In view of the practical context, we assume the series (\ref{15}) to be
truncated by the order $\pm N,$ i.e. by $M=2N+1$ terms. Inserting Eq. (\ref%
{15}) along with the similarly truncated Fourier expansion of the $x$%
-periodic functions $\rho \left( x,y\right) ,~c_{ijkl}\left( x,y\right) $%
{\small \ }in the governing equations {\small (\ref{3}), (\ref{4}) }leads to
the Stroh-like ODS \cite{NSK} 
\begin{equation}
\mathbf{\tilde{\eta}}^{\prime }\left( y\right) =\widetilde{\mathbf{Q}}\left(
y\right) \mathbf{\tilde{\eta}}\left( y\right) ,  \label{17}
\end{equation}%
where the $6M$-vector $\mathbf{\tilde{\eta}}\left( y\right) $ consists of
two $3M$-vectors $\mathbf{\tilde{a}}\left( y\right) $ and $\mathbf{\tilde{b}}%
\left( y\right) $, and these in turn are formed by $3$-vectors $\mathbf{\hat{%
a}}^{\left( n\right) }\left( y\right) $ and $\mathbf{\hat{b}}^{\left(
n\right) }\left( y\right) ,\ n=-N,...,N,$ namely, 
\begin{equation}
\mathbf{\tilde{\eta}}\left( y\right) =\left( 
\begin{array}{c}
\mathbf{\tilde{a}}\left( y\right) \\ 
\mathbf{\tilde{b}}\left( y\right)%
\end{array}%
\right) ,\ 
\begin{array}{c}
\mathbf{\tilde{a}}\left( y\right) ={\large \{}\mathbf{\hat{a}}^{\left(
n\right) }\left( y\right) {\large \}}=\left( \mathbf{\hat{a}}^{\left(
-N\right) }\left( y\right) ...\mathbf{\hat{a}}^{\left( N\right) }\left(
y\right) \right) ^{T}, \\ 
\ \mathbf{\tilde{b}}\left( y\right) ={\large \{}\mathbf{\hat{b}}^{\left(
n\right) }\left( y\right) {\large \}}=\left( \mathbf{\hat{b}}^{\left(
-N\right) }\left( y\right) ...\mathbf{\hat{b}}^{\left( N\right) }\left(
y\right) \right) ^{T}.%
\end{array}
\label{18}
\end{equation}%
Accordingly, the $6M\times 6M$ system matrix%
\begin{equation}
\widetilde{\mathbf{Q}}\left( y\right) =i\left( 
\begin{array}{cc}
-\widetilde{\mathbf{T}}^{-1}\widetilde{\mathbf{R}}^{+} & -\widetilde{\mathbf{%
T}}^{-1} \\ 
\widetilde{\mathbf{P}}-\mathbf{\widetilde{\mathbf{R}}\widetilde{\mathbf{T}}}%
^{-1}\mathbf{\widetilde{\mathbf{R}}^{+}-}\omega ^{2}\widetilde{\mathbf{\rho }%
} & -\widetilde{\mathbf{R}}\widetilde{\mathbf{T}}^{-1}%
\end{array}%
\right)  \label{19}
\end{equation}%
consists of $3M\times 3M$ "tilded" submatrices composed of $3\times 3$
"hatted" blocks and enumerated by superscripted indices: 
\begin{equation}
\begin{array}{c}
\widetilde{\mathbf{P}}={\Large \{}\widehat{\mathbf{P}}^{\left( mn\right) }%
{\Large \}},\ \mathbf{\widetilde{\mathbf{R}}}={\Large \{}\widehat{\mathbf{R}}%
^{\left( mn\right) }{\Large \}},\ \mathbf{\widetilde{\mathbf{T}}}={\Large \{}%
\widehat{\mathbf{T}}^{\left( mn\right) }{\Large \}},\ \widetilde{\mathbf{%
\rho }}={\Large \{}\hat{\rho}_{m-n}\mathbf{I}{\Large \}},{\Large \ } \\ 
\widehat{\mathbf{P}}^{\left( mn\right) }=k_{m}k_{n}\left( e_{1}e_{1}\right)
^{\left( m-n\right) }\mathbf{,\ }\widehat{\mathbf{R}}^{\left( mn\right)
}=k_{m}\left( e_{1}e_{2}\right) ^{\left( m-n\right) },\ \widehat{\mathbf{T}}%
^{\left( mn\right) }=\left( e_{2}e_{2}\right) ^{\left( m-n\right) },\ 
\end{array}
\label{20}
\end{equation}%
where $\left( e_{a}e_{b}\right) _{jk}^{\left( m-n\right) }\equiv \left( 
\mathbf{e}_{a}\right) _{i}\hat{c}_{ijkl}^{\left( m-n\right) }\left( \mathbf{e%
}_{b}\right) _{l},\ a,b=1,2,\ m,n=-N,....,N,\ $and $\hat{\rho}_{n}\left(
y\right) ,$ $\hat{c}_{ijkl}^{\left( n\right) }\left( y\right) $ are the
Fourier coefficients (cf. (\ref{4.1}); note also that the $\left( mn\right) $%
th matrix blocks (\ref{20}) incorporate the wavenumber in contrast to their
counterparts in (\ref{7})). Recall that $\hat{\rho}^{\left( -n\right) }=\hat{%
\rho}^{\left( n\right) \ast }$ and $\hat{c}_{ijkl}^{\left( -n\right) }=\hat{c%
}_{ijkl}^{\left( n\right) \ast }$ for real $\rho $ and$~c_{ijkl},$ hence the
matrix $\widetilde{\mathbf{Q}}$ satisfies%
\begin{equation}
(\mathbf{\mathbb{T}\widetilde{\mathbf{Q}})^{+}=-\mathbb{T}}\widetilde{%
\mathbf{Q}}\Leftrightarrow \widetilde{\mathbf{Q}}=-\mathbf{\mathbb{T}%
\widetilde{\mathbf{Q}}}^{+}\mathbf{\mathbb{T}},  \label{19.1}
\end{equation}%
where the $6M\times 6M$ matrix $\mathbf{\mathbb{T}}$ has a zero diagonal and
identity off-diagonal blocks (i.e. it is of the same pattern as $6\times 6$ $%
\mathbf{\mathbb{T}}$ in (\ref{7.2})). If $\rho $ and $c_{ijkl}${\small \ }%
are even functions of $x$ with respect to the midpoint of the period $\left[
x,x+T\right] ,$ then $\hat{\rho}^{\left( n\right) }$ and $\hat{c}%
_{ijkl}^{\left( n\right) }$ are real, and $\widetilde{\mathbf{Q}}$ satisfies
the identity analogous to (\ref{7.2}).

Provided that the material properties are independent of $y,$ i.e. $%
\widetilde{\mathbf{Q}}_{0}=i\widetilde{\mathbf{N}}_{0},$ the partial
solutions of Eq. (\ref{17}) are sought in the form 
\begin{equation}
\mathbf{\tilde{\eta}}_{\alpha }\left( y\right) =\mathbf{\tilde{\xi}}_{\alpha
}e^{ik_{y\alpha }y},\ \mathbf{\tilde{\xi}}_{\alpha }=\left( 
\begin{array}{c}
\mathbf{\tilde{A}}_{\alpha } \\ 
\mathbf{\tilde{B}}_{\alpha }%
\end{array}%
\right) ,\ 
\begin{array}{c}
\mathbf{\tilde{A}}_{\alpha }={\large \{}\mathbf{\hat{A}}_{\alpha }^{\left(
n\right) }{\large \}}=\left( \mathbf{\hat{A}}_{\alpha }^{\left( -N\right)
}...\mathbf{\hat{A}}_{\alpha }^{\left( N\right) }\right) ^{T} \\ 
\mathbf{\tilde{B}}_{\alpha }={\large \{}\mathbf{\hat{B}}_{\alpha }^{\left(
n\right) }{\large \}}=\left( \mathbf{\hat{B}}_{\alpha }^{\left( -N\right)
}...\mathbf{\hat{B}}_{\alpha }^{\left( N\right) }\right) ^{T}%
\end{array}%
,  \label{20.1}
\end{equation}%
where $\mathbf{\tilde{\xi}}_{\alpha }$ and $k_{y\alpha }$,\ $\alpha
=1,...,6M,$ are the eigenvectors and eigenvalues of the matrix $\widetilde{%
\mathbf{N}}_{0}$. The overall Floquet-Bloch solution (\ref{15}) in this case
reads 
\begin{equation}
\left( 
\begin{array}{c}
\mathbf{u}\left( x,y\right) \\ 
i\mathbf{t}_{2}\left( x,y\right)%
\end{array}%
\right) =\sum\limits_{\alpha =1}^{6M}\mathbf{\eta }_{\alpha }\left( x\right)
e^{ik_{y\alpha }y}=\ \sum\limits_{n=-N}^{N}\sum\limits_{\alpha
=1}^{6M}\left( 
\begin{array}{c}
\mathbf{\hat{A}}_{\alpha }^{\left( n\right) } \\ 
\mathbf{\hat{B}}_{\alpha }^{\left( n\right) }%
\end{array}%
\right) e^{i\left( k_{n}x+k_{y\alpha }y\right) },  \label{15.1}
\end{equation}%
where 
\begin{equation}
\mathbf{\eta }_{\alpha }\left( x\right) =\left( 
\begin{array}{c}
\mathbf{A}_{\alpha }\left( x\right) \\ 
\mathbf{B}_{\alpha }\left( x\right)%
\end{array}%
\right) e^{iK_{x}x}=\sum\limits_{n=-N}^{N}\left( 
\begin{array}{c}
\mathbf{\hat{A}}_{\alpha }^{\left( n\right) } \\ 
\mathbf{\hat{B}}_{\alpha }^{\left( n\right) }%
\end{array}%
\right) e^{ik_{n}x},  \label{15.2}
\end{equation}%
and $\mathbf{A}_{\alpha }\left( x\right) $ and $\mathbf{B}_{\alpha }\left(
x\right) $ are periodic functions expanded in the (truncated) Fourier series
with the coefficients $\mathbf{\hat{A}}_{\alpha }^{\left( n\right) }\mathbf{%
\ }$and $\mathbf{\hat{B}}_{\alpha }^{\left( n\right) }$. Note that each $%
\alpha $th $\mathbf{\eta }_{\alpha }\left( x\right) $ (\ref{15.2}) may
formally be defined beyond the PWE as the partial solution of the equation 
\begin{equation}
\mathbf{\eta }_{\alpha }^{\prime }\left( x\right) =\mathbf{Q}\left( x\right) 
\mathbf{\eta }_{\alpha }\left( x\right) ,\ \alpha =1,...,6M,  \label{15.3}
\end{equation}%
with the system matrix given by (\ref{8})$_{3}$ up to the replacement $%
k_{n}\rightarrow k_{y\alpha }$ and $\mathbf{e}_{1}\rightleftarrows \mathbf{e}%
_{2}$. The solution of Eq. (\ref{15.3}) can be obtained, at least
numerically, in three steps: first, $k_{y\alpha }\left( K_{x}\right) $ is
identified from the characteristic equation $\det \left( \mathbf{M}\left[
\omega ,k_{y}\right] -e^{iK_{x}T}\mathbf{I}\right) =0,$ where the monodromy
matrix $\mathbf{M}$ of (\ref{15.3}) depending on the parameters $\omega $
and $k_{y}$ is defined via (\ref{11}) or (\ref{10}); second, this $%
k_{y\alpha }\left( K_{x}\right) $ is plugged into the monodromy matrix and
the eigenvector $\mathbf{w}$ of $\mathbf{M}\left[ \omega ,k_{y\alpha }\right]
\equiv \mathbf{M}_{\alpha }$ corresponding to the eigenvalue $e^{iK_{x}T}$
is found; third, this $\mathbf{w}$ is used as an initial condition in the
formula $\mathbf{\eta }_{\alpha }\left( x\right) =\mathbf{M}_{\alpha }\left(
x,0\right) \mathbf{w}$ with the matricant containing $\omega $ and $%
k_{y\alpha }\left( K_{x}\right) .$

We conclude this section with a brief terminological remark. The "matricant" 
$\mathbf{M}\left( y_{2},y_{1}\right) $ and "monodromy matrix" $\mathbf{M}%
\left( T,0\right) $ are standard terms in mathematical courses, whereas the
alternative terms "transfer matrix" and "propagator" have been entrenched in
the literature on acoustic and electromagnetic waves. We opt for using the
former, but may occasionally invoke one of the latter, particularly in the
context of problems with fixed end points (e.g., a transfer matrix $\mathbf{M%
}\left( H,0\right) $ through a plate $\left[ 0,H\right] $).

\section{Algebraic properties of the Stroh formalism\label{Sec3}}

\subsection{Hamiltonian structure of Stroh's ODS\label{SSec3.1}}

Hereafter, unless otherwise noted, we assume that $\rho ,~c_{ijkl}$ and $%
\omega ,$ $k_{x}$ are real. Consider the governing ODS (\ref{5}) with $%
\mathbf{Q}$ in the form (\ref{8.1}). Identity (\ref{7.2}) can then be
re-written as 
\begin{equation}
\left( \mathbf{\mathbb{J}Q}\right) \mathbf{^{+}=\mathbf{\mathbb{J}}Q}\
\Leftrightarrow \mathbf{Q}^{+}=\mathbf{-\mathbf{\mathbb{J}}Q\mathbf{\mathbb{J%
}}}^{-1}\ \mathrm{with}\ \ \mathbf{\mathbb{J}}=\left( 
\begin{array}{cc}
\widehat{\mathbf{0}} & \mathbf{I} \\ 
-\mathbf{I} & \widehat{\mathbf{0}}%
\end{array}%
\right) =-\mathbf{\mathbb{J}}^{+}=-\mathbf{\mathbb{J}}^{-1},  \label{27}
\end{equation}%
which, by (\ref{10.1}), leads to the symplectic form%
\begin{equation}
\mathbf{M}^{+}\mathbf{\mathbb{J}M}=\mathbf{\mathbb{J}}  \label{27.1}
\end{equation}%
of the matricant of the ODS (\ref{5}). In turn, assuming $\mathbf{Q}$ as in (%
\ref{8}) or in any other explicit form with pure imaginary $\mathbf{Q}$
casts (\ref{7.2}) as 
\begin{equation}
\left( \mathbf{\mathbb{T}Q}\right) \mathbf{^{+}=-\mathbb{T}Q}\
\Leftrightarrow \mathbf{Q}^{+}=\mathbf{-\mathbb{T}Q\mathbb{T}}  \label{24}
\end{equation}%
and leads to 
\begin{equation}
\mathbf{M^{+}\mathbb{T}M}=\mathbf{\mathbb{T}.}  \label{26}
\end{equation}%
Matrices obeying Eqs. (\ref{24}) and (\ref{26}) are said to be,
respectively, skew $\mathbf{\mathbb{T}}$-Hermitian and $\mathbf{\mathbb{T}}$%
-unitary with $\mathbf{\mathbb{T}}$ understood as the metrics (the Gram
matrix) of an improper \cite{P} or indefinite \cite{YS} inner product.

Note that the system matrix does not have to be purely imaginary and/or meet
(\ref{7.2}) in order to satisfy (\ref{24}) and hence provide (\ref{26}). For
instance, such is the $6M\times 6M$ system matrix $\widetilde{\mathbf{Q}}$ (%
\ref{19}) of the PWE-processed ODS (\ref{17}), see Eq. (\ref{19.1}). Other
examples include the case of some non-local elasticity models, where the
matrices (\ref{4.1}) are complex Hermitian, and the case of the
acoustic-wave ODS in materials with cylindrical or certain types of
spherical anisotropy, see \cite{Sh,NSh}.

Premultiplying both sides of Stroh's ODS (\ref{5}) by $\mathbf{\mathbb{J}}$
if $\mathbf{Q}$ satisfies (\ref{27}), or by $i\mathbf{\mathbb{T}}$ if $%
\mathbf{Q}$ satisfies (\ref{24}), transforms it into the Hamiltonian
canonical pattern $\mathbf{J\dot{x}}\left( t\right) =\mathbf{H}\left(
t\right) \mathbf{x}\left( t\right) $, where $\mathbf{J}$ is a real
non-singular skew-symmetric matrix, and $\mathbf{H=H}^{+}$ is a Hermitian
matrix, both of even order (see, e.g., \S 3.1 of \cite{YS}). This
demonstrates that, regardless of the explicit formulation, the Stroh
formalism has a Hamiltonian-like nature that underlies the algebraic
properties detailed in the next Section. Further discussion involving the
link to the Lagrange formalism may be found in \cite{YZL,Fu}.

It is understood that the above equivalent choices of the system matrix $%
\mathbf{Q}$ in (\ref{5}) of the type (\ref{27}) or (\ref{24}) lead to the
same overall conclusions but through different forms of interim relations.
The subsequent text will default to the more common latter option, which
proceeds from ODS (\ref{5}) with $\mathbf{Q}$ and $\mathbf{M}$ satisfying (%
\ref{24}) and (\ref{26}). Within this framework, an additional reservation
will be made in those few cases where explicit derivation details are based
specifically on the definition (\ref{8}) of $\mathbf{Q}$.

The fundamental matrix solution $\mathbf{\mathcal{N}}\left( y\right) $
satisfies the first integral relation, known as the Poincar\'{e} invariant
in the theory of Hamiltonian systems, which stems from (\ref{5}) and (\ref%
{24}) in the form 
\begin{equation}
\mathbf{\mathcal{N}}^{+}\left( y\right) \mathbf{\mathbb{T}\mathcal{N}}\left(
y\right) =\mathcal{I}_{0},\   \label{25}
\end{equation}%
where $\mathcal{I}_{0}$ is a real symmetric constant matrix. It explicitly
expresses the energy conservation law, which in the given case implies the
constancy of the $y$-component of the energy flux density averaged over a
time period $\overline{P}_{y}=-\frac{i\omega }{4}\left( \mathbf{t}_{2}%
\mathbf{u}^{\ast }-\mathbf{t}_{2}^{\ast }\mathbf{u}\right) ,$ see (\ref{e7.0}%
) in Appendix 2. By Sylvester's law of inertia, since $\mathbf{\mathbb{T}}$ (%
\ref{7.2}) has zero signature so does the matrix $\mathcal{I}_{0},$ thus
implying an equal number of bulk modes with an upward and downward directed $%
y$-component of the flux. The matrices $\mathbf{\mathcal{N}}^{+}\left(
y\right) \mathbf{\mathbb{T}Q}^{n}\mathbf{\mathcal{N}}\left( y\right) \equiv
i^{n}\mathcal{I}_{n},$ $n\in 
\mathbb{N}
,$ are also real and symmetric; moreover, they are constant provided that so
is the system matrix $\mathbf{Q}=\mathbf{Q}_{0}.$ In particular, if the
solution of (\ref{5}) with constant $\mathbf{Q}_{0}$ satisfies the radiation
condition $\mathbf{\mathcal{N}}\left( y\right) \rightarrow 0$ at $%
y\rightarrow \infty $, then $\mathcal{I}_{n}=0$ $\forall n=0,1,2,...$. This
property was utilized to derive explicit secular equations on the speeds of
surface wave \cite{Taz} and interface wave \cite{D} in homogeneous
half-spaces.

Naturally, Stroh's ODS (\ref{5}) is no longer of the Hamiltonian type when
the viscoelasticity is taken into account. For instance, let the stiffness
tensor $c_{ijkl}$ with full symmetry in indices be generally complex (so
that the matrices $\left( e_{\alpha }e_{\beta }\right) $ (\ref{4.1}) are not
Hermitian). Then the system matrix $\mathbf{Q}$ of the form (\ref{8.1}) or (%
\ref{8}) still satisfies (\ref{7.2}) but does not satisfy (\ref{24}), and
hence identity (\ref{10.1}) leads to $\mathbf{M}^{-1}=\mathbf{\mathbb{T}M}_{-%
\mathbf{Q}}^{T}\mathbf{\mathbb{T}}$ but does not provide (\ref{26}).

\subsection{Eigenspectrum and eigenspace of $\mathbf{Q}$ and $\mathbf{M}$%
\textbf{\label{SSec3.2}}}

\subsubsection{System matrix $\mathbf{Q}\left( y\right) $\label{SSSec3.2.1}}

Consider the eigenvalue problem for the system matrix 
\begin{equation}
\mathbf{Q}\left( y\right) \mathbf{\xi }_{\alpha }\left( y\right) =i\kappa
_{\alpha }\left( y\right) \mathbf{\xi }_{\alpha }\left( y\right) ,\ \alpha
=1,...,6,  \label{28.-2}
\end{equation}%
and omit explicit mention of the variable $y$ in the following. By (\ref{7.2}%
), the matrices $\mathbf{Q}$ and $\mathbf{Q}^{T}$ are similar via the matrix 
$\mathbf{\mathbb{T}}$, hence $\mathbf{\xi }_{\alpha }$ and $\mathbf{\mathbb{T%
}\xi }_{\beta }$ are the right and left eigenvectors of $\mathbf{Q,}$ and as
such, they are mutually orthogonal. Indeed, premultiplying (\ref{28.-2}) by $%
\mathbf{\xi }_{\beta }^{T}\mathbf{\mathbb{T}}$ leads to 
\begin{equation}
\left( \kappa _{\beta }-\kappa _{\alpha }\right) \mathbf{\xi }_{\beta }^{T}%
\mathbf{\mathbb{T}\xi }_{\alpha }=0,\ \alpha ,\beta =1,...6,  \label{28.0}
\end{equation}%
\ where, for any $\mathbf{\xi }_{\alpha },$ there exists $\mathbf{\xi }%
_{\beta }$ such that $\mathbf{\xi }_{\beta }^{T}\mathbf{\mathbb{T}\xi }%
_{\alpha }\neq 0.$ Hence, if $\kappa _{\alpha }\neq \kappa _{\beta }$, then $%
\mathbf{\xi }_{\beta }^{T}\mathbf{\mathbb{T}\xi }_{\alpha }\sim \delta
_{\beta \alpha }.$ Assume $\mathbf{Q}$ semisimple (diagonalizable), i.e.
possessing a complete set of linearly independent eigenvectors $\mathbf{\xi }%
_{\alpha },$ and introduce the $6\times 6$ matrix 
\begin{equation}
\mathbf{\Xi }=\left( 
\begin{array}{cc}
\mathbf{\Xi }_{1} & \mathbf{\Xi }_{2} \\ 
\mathbf{\Xi }_{3} & \mathbf{\Xi }_{4}%
\end{array}%
\right) =\left\Vert \mathbf{\xi }_{1}...\mathbf{\xi }_{6}\right\Vert ,
\label{28-1}
\end{equation}%
whose columns are $\mathbf{\xi }_{\alpha }$ (the corresponding $3\times 3$
blocks $\mathbf{\Xi }_{1...4}$ will be extensively engaged in subsequent
developments).\ By the above, the matrix $\mathbf{\Xi }^{T}\mathbf{\mathbb{T}%
\Xi }$ is diagonal. Normalizing the self-product $\mathbf{\xi }_{\alpha }^{T}%
\mathbf{\mathbb{T}\xi }_{\alpha }$ for every $\alpha $ to 1 leads to the
orthonormality relation 
\begin{equation}
\mathbf{\Xi }^{T}\mathbf{\mathbb{T}\Xi }=\mathbf{I\ }\Leftrightarrow \ 
\mathbf{\Xi }^{-1}=\mathbf{\Xi }^{T}\mathbf{\mathbb{T\ }}\Leftrightarrow \ 
\mathbf{\Xi \Xi }^{T}=\mathbf{\mathbb{T}\ \ }\left( \det \mathbf{\Xi }=\pm
i\right) .  \label{28}
\end{equation}%
Note that Eq. (\ref{28}) rests solely on{\small \ (\ref{7.2})}, i.e. remains
valid when Stroh's ODS admits dissipation. The eigenvalue degeneracy $\kappa
_{\alpha }=\kappa _{\beta }\equiv \kappa _{\deg }$ such that keeps $\mathbf{Q%
}$ semisimple allows retaining (\ref{28}); at the same time, the 2D
eigensubspace corresponding to $\kappa _{\deg }$ contains a self-orthogonal
eigenvector $\mathbf{\xi }_{\deg }$ satisfying $\mathbf{\xi }_{\deg }^{T}%
\mathbf{\mathbb{T}\xi }_{\deg }=0$. If the eigenvalue degeneracy renders $%
\mathbf{Q}$ non-semisimple (i.e., defective or non-diagonalizable), then $%
\kappa _{\deg }$ corresponds to a single eigenvector $\mathbf{\xi }_{\deg },$
which is self-orthogonal in the above sense, and it is complemented by a
so-called generalized eigenvector. As a result, the orthonormality relation
at the degeneracy point no longer has the form (\ref{28}) and must be
appropriately modified, see \cite{T}. Note in passing that a single-mode
(one-component) wave of the form (\ref{11.01}) on a traction-free boundary
implies the eigenvector's self-orthogonality - this is why such a wave may
occur only in the case of eigenvalue degeneracy, see \cite{BCL}-\cite{WG}.

For the $\mathbf{Q}$ obeying either (\ref{27}) or (\ref{24}), i.e. in the
absence of dissipation, the set of six values $\kappa _{\alpha }~$falls into
pairs of either real or complex conjugated ones. The characteristic
polynomial of $\mathbf{Q}$ is homogeneous of degree two in $\omega ,$ $k_{x}$
and $\kappa ,$ hence the occurrence of real or complex root $\kappa _{\alpha
}$ depends on the ratio of $\omega $ to $k_{x},$ i.e. on the trace velocity $%
v=\omega /k_{x}.$ All six $\kappa _{\alpha }$'s are complex-valued at $v=0$
and hence in a certain range $0\leq v<\hat{v},$ called the subsonic interval%
\footnote{{\footnotesize It cannot reappear at any }$v>\hat{v}$%
{\footnotesize \ since the Christoffel tensor with the components }$\Gamma
_{jk}=k_{i}c_{ijkl}k_{l}${\footnotesize \ is Hermitian and hence the
slowness surface }$\mathbf{S=k}/\omega ${\footnotesize \ is a union of three
simply connected sheets with a common centre at }$\mathbf{k}=\mathbf{0}.$},
while the real pair(s) of $\kappa _{\alpha }$ appear at $v>\hat{v},$ called
the supersonic interval. The threshold $\hat{v},$ called the transonic state 
\cite{ChadS}\footnote{%
\noindent {\footnotesize Sometimes, it is specified as the \textit{first}
transonic state, while higher values of }$v\left( y\right) ,${\footnotesize %
\ at which other complex-conjugated eigenvalue pairs merge to become real,
are called the \textit{subsequent} transonic states, see \cite{Chad1,G}.}},
is thus the minimum trace velocity, at which (at least) one
complex-conjugate pair $\kappa _{\alpha }$ and $\kappa _{\alpha }^{\ast }$
merges into a double real eigenvalue $\kappa _{\deg }$ to then split into a
real pair as $v$ increases further. The matrix $\mathbf{Q}$ taken at a
transonic state is always non-semisimple. Away from transonic states, the
degeneracy between, specifically, complex eigenvalues $\kappa _{\alpha }$
and $\kappa _{\beta }\neq \kappa _{\alpha }^{\ast }$ (hence, simultaneously
between $\kappa _{\alpha }^{\ast }$ and $\kappa _{\beta }^{\ast }$)
typically renders $\mathbf{Q}$ non-semisimple, while the degeneracy between
real eigenvalues $\kappa _{\alpha }$ and $\kappa _{\beta }$ always keeps $%
\mathbf{Q}$ semisimple (if $\mathbf{Q}$ is constant, this is the case of an
acoustic axis), see \cite{Sh1} for details.

Given that $\mathbf{Q}$ is purely imaginary, i.e. of type (\ref{8}), its
eigenvectors corresponding to complex conjugated eigenvalues are also
complex conjugated. In particular, when all eigenvalues are complex (i.e.,
at $v<\hat{v}$), the eigenspace of $\mathbf{Q}$ splits into triplets such
that 
\begin{equation}
\kappa _{\alpha }=\kappa _{\alpha +3}^{\ast },\ \mathbf{\xi }_{\alpha }=%
\mathbf{\xi }_{\alpha +3}^{\ast }\ \mathrm{if\ }\func{Im}\kappa _{\alpha
}\neq 0,\ \alpha =1,2,3.  \label{28*}
\end{equation}%
When real pair(s) of values $\kappa _{\alpha }$ (at $v>\hat{v}$) exist, the
corresponding eigenvectors are scalar multiples of real vectors. Using (\ref%
{28*}), the Eucledian-type orthogonality relation (\ref{28}) can be cast
into a Hermitian-type form: $\mathbf{\xi }_{\alpha }^{+}\mathbf{\mathbb{T}%
\xi }_{\beta }=\delta _{\left\vert \alpha -\beta \right\vert ,3}$ if $\func{%
Im}\kappa _{\alpha }\neq 0$ and $\mathbf{\xi }_{\alpha }^{+}\mathbf{\mathbb{T%
}\xi }_{\beta }=\pm \delta _{\alpha \beta }$ if $\kappa _{\alpha }$ is real.
The same may be written via the eigenvector matrix $\mathbf{\Xi }$ (\ref%
{28-1}) arranged according to (\ref{28*}), namely, 
\begin{equation}
\mathbf{\Xi }^{+}\mathbf{\mathbb{T}}\mathbb{\mathbf{\Xi }}=\mathbf{\mathbb{E}%
}_{\mathbf{Q}}\mathbf{\ }\Leftrightarrow \ \mathbf{\Xi }^{-1}=\mathbf{%
\mathbb{E}}_{\mathbf{Q}}\mathbf{\Xi }^{+}\mathbf{\mathbb{T\ }}%
\Leftrightarrow \ \mathbf{\Xi \mathbb{E}}_{\mathbf{Q}}\mathbf{\Xi }^{+}=%
\mathbf{\mathbb{T}},  \label{29}
\end{equation}%
where $\mathbf{\mathbb{E}}_{\mathbf{Q}}=\mathbf{\mathbb{T}}$ if all $\kappa
_{\alpha }$ are complex or else, if there is one or several pairs of real $%
\kappa _{\alpha }$ and $\kappa _{\alpha +3},$ then $\mathbf{\mathbb{E}}_{%
\mathbf{Q}}$ differs from $\mathbf{\mathbb{T}}$ due to replacing the unit
values at the $\left( \alpha ,\alpha +3\right) $th and $\left( \alpha
+3,\alpha \right) $th positions by the values $\pm 1$ at the $\left( \alpha
\alpha \right) $th and $\left( \alpha +3,\alpha +3\right) $th diagonal
positions. Similarly to (\ref{25}), Sylvester's law of inertia applied to (%
\ref{29})$_{1}$ allows the conclusion that any matrix $\mathbf{\mathbb{E}}_{%
\mathbf{Q}}\neq \mathbf{\mathbb{T}}$ has an equal number of $1$ and $-1$
diagonal entries. Given the signs are chosen so that $-1$ and $1$ are
assigned to the $\left( \alpha \alpha \right) $th and $\left( \alpha
+3,\alpha +3\right) $th positions, respectively, normalizations (\ref{28})
and (\ref{29}) comply with one another provided $\mathbf{\xi }_{\alpha }$ is
purely imaginary and $\mathbf{\xi }_{\alpha +3}$ is real (this specifies the
statement made below (\ref{28*})). As delineated above, the transition
between two unit values on the antidiagonal of $\mathbf{\mathbb{E}}_{\mathbf{%
Q}}$ to a pair of $\pm 1$ values on its main diagonal (or vice versa) occurs
due to eigenvalue degeneracy at the transonic state, where $\mathbf{Q}$ is
non-semisimple and hence neither (\ref{28}) nor (\ref{29}) applies.

It is pertinent to remind that it is only if the system matrix is constant, $%
\mathbf{Q=Q}_{0},$ that its eigenspectrum defines the solutions $\mathbf{%
\eta }\left( y\right) $ of (\ref{5}) in the form of eigenmodes (\ref{11.01})
with $k_{y\alpha }=\kappa _{\alpha }$. In this case, the non-zero $\left(
\alpha \beta \right) $th components $\mathbf{\xi }_{\alpha }^{T}\mathbf{%
\mathbb{T}\xi }_{\beta }$ of the matrix $\mathbf{\mathbb{E}}_{\mathbf{Q}%
_{0}} $ introduced above are (negative) proportional to the $y$-components
of the time-averaged energy fluxes $\overline{\mathbf{P}};$ namely, the
diagonal entries $-1$ and $1$ correspond to the pairs of bulk modes (real $%
k_{y\alpha }$ and $k_{y,\alpha +3}$) with $\overline{\mathbf{P}}_{y\alpha
}>0 $ and $\overline{\mathbf{P}}_{y,\alpha +3}<0,$ while the off-diagonal
entries are associated with the interference of increasing/decreasing modes
(complex $k_{y\alpha }$ and $k_{y,\alpha +3}=k_{y\alpha }^{\ast }$). If $%
\mathbf{Q}$ and hence its eigenvalues and eigenvectors are varying, they are
not correlated with the solutions of (\ref{5}) (unless in the approximate
sense within the asymptotic WKB expansion). In particular, given a stack of
homogeneous layers $j=1,...,n,$ each with a constant system matrix $\mathbf{Q%
}_{0j}\left[ \omega ,k_{x}\right] ,$ the attribution of six eigenmodes as
bulk and increasing/decreasing depends on $\omega ,$ $k_{x}$ and may vary
from one layer to another (as indicated by the matrices $\mathbf{\mathbb{E}}%
_{\mathbf{Q}_{0j}}\left[ \omega ,k_{x}\right] $); at the same time, the
fundamental solution $\mathcal{N}\left( y\right) =\mathbf{M}\left(
y,y_{0}\right) \mathcal{N}\left( y_{0}\right) $ with $\mathbf{M}\left(
y,y_{0}\right) $ given in (\ref{11}) defines the invariant matrix (\ref{25})
with a constant value $\mathcal{I}_{0}$ determined by the initial condition $%
\mathcal{N}\left( y_{0}\right) .$ Physically, this implies mode conversion
at the layer interfaces, which certainly maintains the continuity of energy
flux.

\subsubsection{Matricant $\mathbf{M}\left( y,y_{0}\right) $\label{SSSec3.2.2}%
}

Consider the eigenvalue problem for the matricant $\mathbf{M}\left(
y,y_{0}\right) $%
\begin{equation}
\mathbf{M}\left( y,y_{0}\right) \mathbf{w}_{\alpha }\left( y\right)
=q_{\alpha }\left( y\right) \mathbf{w}_{\alpha }\left( y\right) ,\ \alpha
=1,...,6,  \label{30.-1}
\end{equation}%
and omit explicit mention of the variable $y$ below unless specified
otherwise. By (\ref{26}), $\mathbf{M}^{-1}$ and $\mathbf{M}^{+}$ are
similar, so they share the same eigenvalues, i.e. the set of six eigenvalues 
$\left\{ q_{\alpha }\right\} $ is equal to the sets $\left\{ q_{\alpha
}^{-1}\right\} $ and $\left\{ q_{\alpha }^{\ast }\right\} $ (recall that
equality of sets admits any ordering of their individually equated
elements). Thus, the eigenspectrum of $\mathbf{M}$ falls into pairs $%
q_{\alpha }$ and $q_{\alpha +3}$ such that 
\begin{equation}
\mathrm{either}\ \ \left\vert q_{\alpha }\right\vert =\left\vert q_{\alpha
+3}\right\vert =1\ \ \mathrm{or\ \ }q_{\alpha }=1/q_{\alpha +3}^{\ast }\ 
\mathrm{with\ }\left\vert q_{\alpha }\right\vert \neq 1.  \label{30.01}
\end{equation}%
The same may be demonstrated as in \cite{P} via premultiplying complex
conjugate of (\ref{30.-1}) by $\mathbf{\mathbb{T}w}_{\beta }$ and using (\ref%
{26}) to arrive at the relation 
\begin{equation}
\left( q_{\alpha }^{\ast }q_{\beta }-1\right) \mathbf{w}_{\alpha }^{+}%
\mathbf{\mathbb{T}w}_{\beta }=0,~\alpha ,\beta =1,...6.  \label{30.0}
\end{equation}%
It is seen that, for any $\mathbf{w}_{\alpha }$ there must exist $\mathbf{w}%
_{\sim \alpha }$ such that $\mathbf{w}_{\alpha }^{+}\mathbf{\mathbb{T}w}%
_{\sim \alpha }\neq 0$ with the index $\sim \alpha $ equal to $\alpha $ if $%
\left\vert q_{\alpha }\right\vert =1$ or to some $\beta \neq \alpha $ if $%
\left\vert q_{\alpha }\right\vert \neq 1$. Taking in the latter case $\alpha
=1,2,3$ and $\beta =\alpha +3$ leads to (\ref{30.01}). Besides, it follows
from (\ref{30.0}) that the eigenvector product $\mathbf{w}_{\alpha }^{+}%
\mathbf{\mathbb{T}w}_{\beta }$ is proportional to $\delta _{\beta ,\sim
\alpha }$ with $\sim \alpha $ defined above. Assume the generic case where
the matrix $\mathbf{M}$ is semisimple, and denote the matrix of its
eigenvectors by%
\begin{equation}
\mathbf{W}=\left( 
\begin{array}{cc}
\mathbf{W}_{1} & \mathbf{W}_{2} \\ 
\mathbf{W}_{3} & \mathbf{W}_{4}%
\end{array}%
\right) =\left\Vert \mathbf{w}_{1}...\mathbf{w}_{6}\right\Vert .
\label{30.00}
\end{equation}%
Then adding the normalization condition to the aforementioned eigenvector
orthogonality yields the identity 
\begin{equation}
\mathbf{W}^{+}\mathbf{\mathbb{T}W}=\mathbf{\mathbb{E}}_{\mathbf{M}}\mathbf{\ 
}\Leftrightarrow \ \mathbf{W}^{-1}=\mathbf{\mathbb{E}}_{\mathbf{M}}\mathbf{W}%
^{+}\mathbf{\mathbb{T\ }}\Leftrightarrow \ \mathbf{W\mathbb{E}}_{\mathbf{M}}%
\mathbf{W}^{+}=\mathbf{\mathbb{T}\ \ }\left( \left\vert \det \mathbf{W}%
\right\vert ^{2}=1\right) .  \label{30}
\end{equation}%
It is alike (\ref{29}) up to the replacement of $\mathbf{\mathbb{E}}_{%
\mathbf{Q}}$ with $\mathbf{\mathbb{E}}_{\mathbf{M}},$ which is equal to $%
\mathbf{\mathbb{T}}$ if all $\left\vert q_{\alpha }\right\vert \neq 1,$ and
otherwise has $\pm 1$ diagonal entries at the $\left( \alpha \alpha \right) $%
th and $\left( \alpha +3,\alpha +3\right) $th positions associated with the
pair $\left\vert q_{\alpha }\right\vert =\left\vert q_{\alpha +3}\right\vert
=1$. In the event of eigenvalue degeneracy $q_{\alpha }=q_{\beta }\equiv
q_{\deg }$ such that renders the matrix $\mathbf{M}\left( y,y_{0}\right) $
non-semisimple, the eigenvector $\mathbf{w}_{\deg }$ corresponding to $%
q_{\deg }$ satisfies $\mathbf{w}_{\deg }^{+}\mathbf{\mathbb{T}w}_{\deg }=0$.

In the case of the $T$-periodic system matrix $\mathbf{Q}\left( y\right) $
outlined in \S \ref{SSec2.2}, the solution (\ref{9}) taken at $y=\widetilde{y%
}+nT$ ($\widetilde{y}<T$) can be expressed as a superposition of six
eigenmodes $\mathbf{\eta }_{\alpha }\left( y\right) =q_{\alpha }^{n}\mathbf{M%
}\left( \widetilde{y},0\right) \mathbf{w}_{\alpha },$ where $q_{\alpha
}=q_{\alpha }\left[ \omega ,k_{x}\right] $ ($\equiv e^{iK_{y\alpha }T}$) and 
$\mathbf{w}_{\alpha }=\mathbf{w}_{\alpha }\left[ \omega ,k_{x}\right] $ are
the eigenvalues and eigenvectors of the monodromy matrix $\mathbf{M}\left(
T,0\right) ,$ and $\mathbf{M}\left( \widetilde{y},0\right) $ is bounded (see
(\ref{12}), (\ref{13})). Thus, the wave evolution at large propagation
distance is governed by the eigenvalues of $\mathbf{M}\left( T,0\right) $
according to their placement relative to the unit circle $C_{\left\vert
q\right\vert =1}$ in the $\left( \func{Re}q,\func{Im}q\right) $-complex
plane. Specifically, the pairs of $q_{\alpha }$ (\ref{30.01})$_{1}$ lying on
this circle identify \textit{propagating} (or \textit{bulk}) modes, while
each "reciprocal" pair (\ref{30.01})$_{2}$, which is symmetric relative to
this circle, corresponds to a decreasing mode and an increasing one. As $%
\omega ~$and $k_{x}$ vary, the pairs $q_{\alpha }$ either rotate on the unit
circle or move concertedly on opposite sides of the circle towards or away
from each other. This behavior can also be described via partitioning the $%
\left( \omega ,k_{x}\right) $-plane into areas, termed (full) stopbands if
no pairs (\ref{30.01})$_{1}$ of unit absolute value are admitted therein%
\footnote{%
Given $\omega ,k_{x}$ in a stopband, it may be useful to split the matricant 
$\mathbf{M}\left( nT,0\right) $ into the decreasing and increasing parts.
Employing (\ref{30}) with $\mathbf{\mathbb{E}}_{\mathbf{M}}=\mathbf{\mathbb{T%
}}$ yields%
\begin{equation}
\mathbf{M}\left( nT,0\right) =\left( 
\begin{array}{c}
\mathbf{W}_{1} \\ 
\mathbf{W}_{3}%
\end{array}%
\right) \mathrm{diag}\left( q_{\alpha }^{n}\right) \left( \mathbf{W}_{4}^{+}%
\mathbf{W}_{2}^{+}\right) +\left( 
\begin{array}{c}
\mathbf{W}_{2} \\ 
\mathbf{W}_{4}%
\end{array}%
\right) \mathrm{diag}\left( \frac{1}{q_{\alpha }^{n}}\right) ^{\ast }\left( 
\mathbf{W}_{3}^{+}\mathbf{W}_{1}^{+}\right) .  \label{30a}
\end{equation}%
where the first term may be discarded at $n\gg 1.$}, and passbands
otherwise. A band edge, i.e. the curve separating $\left( \omega
,k_{x}\right) $-areas with different numbers of such eigenvalue pairs$,$
indicates that (at least) one pair merges at the unit circle $C_{\left\vert
q\right\vert =1}$, thus forming a double eigenvalue of unit absolute value: $%
q_{\alpha }=q_{\alpha +3}\equiv q_{\deg }$ with $\left\vert q_{\deg
}\right\vert $ $=1$. Such a degeneracy renders the matrix $\mathbf{M}\left(
T,0\right) $ non-semisimple (except in the case of a "zero-width stopband",
see \cite{KSPN} and \S \ref{SSSec4.5.3}){\small . }By contrast, a degeneracy 
$q_{\alpha }=q_{\beta }\equiv q_{\deg }$ ($\beta \neq \alpha +3$) and hence $%
q_{\alpha +3}=q_{\beta +3}=q_{\deg }^{\ast }$ with $\left\vert q_{\deg
}\right\vert =1,$ involving two pairs of eigenvalues on the circle $%
C_{\left\vert q\right\vert =1},$ i.e. two pairs of propagating modes, always
keeps $\mathbf{M}\left( T,0\right) $ semisimple (otherwise one of the modes
of the degenerate pair would grow proportionally to $K_{y\alpha }T$). On the
other hand, a degeneracy $q_{\alpha }=q_{\beta }\equiv q_{\deg }$ and hence $%
q_{\alpha +3}=q_{\beta +3}=1/q_{\deg }^{\ast }$ with $\left\vert q_{\deg
}\right\vert \neq 1$ between the two decreasing and hence between the two
increasing modes usually leads to a non-semisimple $\mathbf{M}\left(
T,0\right) $.

A comment is in order concerning the propagating modes with $\left\vert
q_{\alpha }\right\vert =1,$ whose parameters $\omega ,~k_{x}$ lie either
inside the passbands or on the band edges in the $\left( \omega
,k_{x}\right) $-plane. They are the Lagrange-stable partial solutions of ODS
(\ref{5}) in the sense that they remain bounded as the variable $y$ tends to
infinity; at the same time, some of them may pairwise turn into
unstable/evanescent couples with $\left\vert q_{\alpha }\right\vert \neq 1$
under a small perturbation of the parameters $\omega $ and $k_{x}$ or
material coefficients. It is evident that such "wobbly" solutions occur,
specifically, on the band edges in the $\left( \omega ,k_{x}\right) $-plane.
The question arises as to how to distinguish them among the solutions with $%
\left\vert q_{\alpha }\right\vert =1$ on general grounds, that is, without
appeal to the $\omega ,~k_{x}$ parametrization. Obviously, the primary
prerequisite is that the sought solutions must correspond to the degenerate
eigenvalue $\left\vert q_{\deg }\right\vert =1$. A more subtle condition is
that, as was mentioned above, a degenerate eigenvalue $\left\vert q_{\deg
}\right\vert =1$ may get off the unit circle $C_{\left\vert q\right\vert =1}$
due to a small perturbation iff (if and only if) this degeneracy renders the
monodromy matrix $\mathbf{M}\left( T,0\right) $ non-semisimple. Strict proof
of this statement is due to Krein, see \cite{YS}.

\subsubsection{PWE matrices\label{SSSec3.2.3}}

Consider briefly the case of $6M\times 6M$ ODS (\ref{17}) generated by the
PWE approach. According to \S \ref{SSSec3.2.3}, the system matrix $%
\widetilde{\mathbf{Q}}\left( y\right) $ of (\ref{17}) satisfies identity (%
\ref{19.1}) (the scaled version of (\ref{24})), hence its eigenvalues $i%
\tilde{\kappa}_{\alpha }\left( y\right) $ and eigenvectors $\mathbf{\tilde{%
\xi}}_{\alpha }\left( y\right) $ fulfil the relation 
\begin{equation}
\left( \tilde{\kappa}_{\alpha }^{\ast }-\tilde{\kappa}_{\beta }\right) 
\mathbf{\tilde{\xi}}_{\alpha }^{+}\mathbf{\mathbb{T}\tilde{\xi}}_{\beta }=0,
\label{30*}
\end{equation}%
where $\alpha ,\beta =1,...,6M$ if $\widetilde{\mathbf{Q}}$ is semisimple.
The set of $\tilde{\kappa}_{\alpha }$'s can be split into two subsets
satisfying $\tilde{\kappa}_{\alpha }=\tilde{\kappa}_{\alpha +3M}^{\ast },$ $%
\alpha =1,...,3M,$ if $\func{Im}\tilde{\kappa}_{\alpha }\neq 0,$ but the
corresponding $\mathbf{\tilde{\xi}}_{\alpha }$ and $\mathbf{\tilde{\xi}}%
_{\alpha +3M}$ are generally not complex conjugated (cf. (\ref{28*})).
According to (\ref{30*}), the eigenvector matrix $\widetilde{\mathbf{\Xi }}%
\left( y\right) ={\large ||}\mathbf{\tilde{\xi}}_{1}...\mathbf{\tilde{\xi}}%
_{6M}{\large ||}$ satisfies the $6M\times 6M$ analogue of (\ref{29}).
Similarly, the matrix $\widetilde{\mathbf{W}}\left( y\right) =\left\Vert 
\mathbf{\tilde{w}}_{1}...\mathbf{\tilde{w}}_{6M}\right\Vert $ of normalized
eigenvectors of the matricant $\widetilde{\mathbf{M}}\left( y,y_{0}\right) $
of (\ref{17}) satisfies the analogue of (\ref{30}).

\subsubsection{Symmetric inhomogeneity profile\label{SSSec3.2.4}}

We shall call the inhomogeneity profile symmetric if the material
coefficients $\rho \left( y\right) ,$ $c_{ijkl}\left( y\right) $ and hence
the system matrix $\mathbf{Q}\left( y\right) $ are even functions about some
point, which may be taken for convenience as $y=0$ so that $\mathbf{Q}\left(
y\right) =\mathbf{Q}\left( -y\right) $. Then, it may be spotted from Eq. (%
\ref{11}) and rigorously proved using identity (\ref{10.1}) and (\ref{26})%
\footnote{%
\noindent {\footnotesize Let }$\mathbf{M}_{\mathbf{Q}\left( y\right) }\left(
y,-y\right) \equiv \mathbf{M}${\footnotesize ; then }$\mathbf{M}=\mathbf{M}%
_{-\mathbf{Q}\left( -y\right) }^{-1}=\mathbf{M}_{-\mathbf{Q}\left( y\right)
}^{-1}=\mathbf{M}_{\mathbf{Q}^{+}\left( y\right) }^{+}=\mathbf{M}_{\mathbf{Q}%
^{T}\left( y\right) }^{T}=\mathbf{TM}^{T}\mathbf{T}${\footnotesize , q.e.d.
If }$\mathbf{Q}^{\ast }=-\mathbf{Q}${\footnotesize , then also }$\mathbf{M}=%
\mathbf{M}_{-\mathbf{Q}\left( y\right) }^{-1}=\mathbf{M}_{\mathbf{Q}^{\ast
}\left( y\right) }^{-1}=\mathbf{M}^{\ast -1}.$} that 
\begin{equation}
\mathbf{M}\left( y,-y\right) =\mathbf{\mathbb{T}M}^{T}\left( y,-y\right) 
\mathbf{\mathbb{T}},\ \ \mathbf{M}^{-1}\left( y,-y\right) =\mathbf{M}^{\ast
}\left( y,-y\right) ,  \label{28.2}
\end{equation}%
where the latter holds if $\mathbf{Q}$ is purely imaginary.

The above identities are the same as if the matricant were given by Eq. (\ref%
{16}), i.e. the medium were homogeneous within the considered interval $%
\left[ -y,y\right] $. Accordingly, provided $\mathbf{M}\left( y,-y\right) $
is diagonalizable, the matrix $\mathbf{W}$ of its (normalized) eigenvectors
satisfies the relation%
\begin{equation}
\mathbf{W}^{T}\mathbf{\mathbb{T}W}=\mathbf{I\ }\Leftrightarrow \ \mathbf{W}%
^{-1}=\mathbf{W}^{T}\mathbf{\mathbb{T\ }}\Leftrightarrow \ \mathbf{WW}^{T}=%
\mathbf{\mathbb{T}},  \label{28.1}
\end{equation}%
which is the same as (\ref{28}) for the matrix of eigenvectors $\mathbf{\Xi }
$ of (arbitrarily varying) $\mathbf{Q}\left( y\right) $. In particular, the
conjunction of (\ref{30}) with (\ref{28.1}) necessitates a pairwise
partitioning of the eigenspace of $\mathbf{M}$ similar to that of $\mathbf{Q}
$ (see the discussion around Eqs. (\ref{28*}), (\ref{29})). It reads as
follows: 
\begin{equation}
\mathbf{w}_{\alpha }=\mathbf{w}_{\alpha +3}^{\ast }\ \mathrm{if}\ q_{\alpha
}=1/q_{\alpha +3}^{\ast };\ \mathbf{w}_{\alpha }=-\mathbf{w}_{\alpha }^{\ast
},\ \mathbf{w}_{\alpha +3}=\mathbf{w}_{\alpha +3}^{\ast }\ \mathrm{if}\
\left\vert q_{\alpha }\right\vert =\left\vert q_{\alpha +3}\right\vert =1,
\label{28.3}
\end{equation}%
where choosing $\mathbf{w}_{\alpha }$ as purely imaginary and $\mathbf{w}%
_{\alpha +3}$ as real replicates the similar choice for the plane modes (\ref%
{11.01}), which renders the modal energy flux along the $Y$-axis positive
for $\alpha $ and negative for $\alpha +3$ (see the end of \S \ref%
{SSSec3.2.1}).

Note that if the profile is defined on an infinite axis $Y$ and hence the
choice of the reference point $y=0$ is optional, it is called symmetric iff $%
y=0$ \textit{can} be chosen so that $\mathbf{Q}\left( y\right) $ is even. In
particular, an infinite periodic profile $\mathbf{Q}\left( y\right) =\mathbf{%
Q}\left( y+T\right) $ is symmetric iff it permits fixing the period frame $%
\left[ 0,T\right] $ so that $\mathbf{Q}\left( y\right) $ is even or,
equivalently, $\mathbf{Q}\left( \widetilde{y}\right) $ with $\widetilde{y}=y-%
\frac{1}{2}T$ is even. A standard example is an infinite periodically
bilayered structure $...A/B/A/B...,$ which has an even profile over a period 
$\frac{1}{2}A/B/\frac{1}{2}A$ and thus manifests as a symmetric structure.

\subsection{Impact of symmetry planes\label{SSec3.3}}

\subsubsection{Symmetry plane orthogonal to $\mathbf{e}_{1}$\ or $\mathbf{e}%
_{2}$\label{SSSec3.3.1}}

Assume that the medium is monoclinic, i.e. its stiffness tensor $\mathbf{c}%
=\left\{ c_{ijkl}\right\} $ possesses a single plane of symmetry, and let
this plane be orthogonal to either vectors $\mathbf{e}_{1}$ or $\mathbf{e}%
_{2}$. This means that the components $c_{ijkl}$ are invariant under the
orthogonal transformation $\mathbf{g}_{1}=\mathbf{I}-2\mathbf{e}_{1}^{T}%
\mathbf{e}_{1}$ or $\mathbf{g}_{2}=\mathbf{I}-2\mathbf{e}_{2}^{T}\mathbf{e}%
_{2},$ which inverts the sign of the vectors $\mathbf{e}_{1}~$or $\mathbf{e}%
_{2},$ respectively. Hence, in both cases, the Stroh matrix (\ref{7}) and
therefore the system matrix $\mathbf{Q}\left( y\right) $ of any explicit
form (\ref{8.1}) or (\ref{8}) satisfy the identity \cite{BH} 
\begin{equation}
\mathbf{Q}\left( y\right) =-\mathbf{\mathbb{G}Q}\left( y\right) \mathbf{%
\mathbb{G},\ \ \mathbb{G}}=\left( 
\begin{array}{cc}
\mathbf{g}_{1,2} & \widehat{\mathbf{0}} \\ 
\widehat{\mathbf{0}} & -\mathbf{g}_{1,2}%
\end{array}%
\right) \left( =\mathbf{\mathbb{G}}^{-1}=\mathbf{\mathbb{G}}^{T}\right) .
\label{31}
\end{equation}%
Noteworthy that the matrix $\mathbf{Q}$ (\ref{31}) can be transformed into a
form with zero diagonal blocks, see \cite{XN}.

Recall that the eigenvalues $i\kappa _{\alpha }\left( y\right) $ of $\mathbf{%
Q}$ occur in pairs with either real or complex conjugated values of $\kappa
_{\alpha }$ (see \S \ref{SSSec3.2.1}). On top of that, by (\ref{31}), the
matrices $\mathbf{Q}$ and $\mathbf{-Q}$ in the present case are similar;
thus, the sets $\left\{ \kappa _{\alpha }\right\} ,\ \left\{ \kappa _{\alpha
}^{\ast }\right\} $ and $\left\{ -\kappa _{\alpha }\right\} $ are equal.
Accordingly, the characteristic polynomial for $\mathbf{Q}$ contains only
even powers of $\kappa .$ The six values of $\kappa _{\alpha }$ may be split
into two pairwise-connected triplets 
\begin{equation}
\kappa _{\alpha }=-\kappa _{\beta },\ \ \alpha \in \left\{ 1,2,3\right\} ,\
\ \beta \in \left\{ 4,5,6\right\} .  \label{31.1}
\end{equation}%
Matching $\beta $ to the numbering rule adopted in and below (\ref{29})
implies that $\beta =\alpha +3$ in (\ref{31.1}) unless possibly when four or
all six $\kappa $'s are complex. In such cases, $\beta $ may not equal $%
\alpha +3$ for two pairs, say,%
\begin{equation}
\begin{array}{c}
\kappa _{1}=-\kappa _{4}\ (\func{real}\ \mathrm{or\ c.c.}),\ \kappa
_{2}=\kappa _{2}^{\prime }+i\kappa _{2}^{\prime \prime }=-\kappa _{3}^{\ast
}=\kappa _{5}^{\ast }=-\kappa _{6} \\ 
\Leftrightarrow \ \kappa _{3}=-\kappa _{2}^{\prime }+i\kappa _{2}^{\prime
\prime }=-\kappa _{2}^{\ast }=\kappa _{6}^{\ast }=-\kappa _{5}\ \left(
\kappa _{2}^{\prime }\neq 0\right) ,%
\end{array}
\label{31.2}
\end{equation}%
where c.c. denotes complex conjugate, $\kappa ^{\prime }\equiv \func{Re}%
\kappa ,\ \kappa ^{\prime \prime }\equiv \func{Im}\kappa $ and $\kappa
_{2}^{\prime }\neq 0$. For homogeneous media (hence $\kappa \equiv k_{y}$),
a sufficient condition for the occurrence of (\ref{31.2}) is the concavity
of the slowness surface in the direction of the $X$-axis.

Provided that $\mathbf{Q}$ obeys (\ref{31}) and hence its eigenvalues are
linked according to (\ref{31.1}), the corresponding eigenvectors $\mathbf{%
\xi }_{\alpha }\left( y\right) $ satisfy%
\begin{equation}
\mathbf{\xi }_{\alpha }=i\mathbf{\mathbb{G}\xi }_{\beta },\   \label{31.3}
\end{equation}%
where $\beta =\alpha +3$ for $\alpha =1,2,3$ unless the option of the type (%
\ref{31.2}). The factor "$i$" ensures that Eq. (\ref{31.3}) conforms with
the normalization adopted in (\ref{28}) and (\ref{29}). Their conjunction
with (\ref{31.3}) can be expressed in the form 
\begin{equation}
\mathbf{\Xi }^{T}\mathbf{\mathbb{Y}\Xi }=i\widetilde{\mathbf{\mathbb{J}}},
\label{31.4}
\end{equation}%
where $\mathbf{\mathbb{Y}}\ =\mathbb{\mathbf{GT}}\left( =-\mathbf{\mathbb{Y}}%
^{T}\ =-\mathbf{\mathbb{Y}}^{-1}\right) $ and $\mathbf{\mathbb{J}}$ is the
matrix with all entries being zero except for $1$ and $-1$ at the $\left(
\alpha \beta \right) $th and $\left( \beta \alpha \right) $th positions,$\ $%
respectively, $\alpha $ and $\beta $ being the indices linked by (\ref{31.3}%
). Specifically, if $\beta =\alpha +3$ for all $\alpha =1,2,3,$ then $%
\widetilde{\mathbf{\mathbb{J}}}$ is equal to the matrix $\mathbf{\mathbb{J}}$
that appeared previously in (\ref{27}), while if the option (\ref{31.2})
comes about, then $\widetilde{\mathbf{\mathbb{J}}}$ contains $1$ in
positions $14,\ 26,\ 35$ and $-1$ in positions $41,53,\ 62$.

Now consider the impact of symmetry planes orthogonal to $\mathbf{e}_{1}$ or 
$\mathbf{e}_{2}$ on the matricant $\mathbf{M}\left( y,y_{0}\right) \equiv 
\mathbf{M}$ of (\ref{5}). By (\ref{31}), $\mathbf{Q}\left( y\right) $ has
zero trace and hence, by Liouville's formula, $\det \mathbf{M}=1$ (cf.
remark under (\ref{9.1})). Moreover, combining (\ref{10.1}) with (\ref{31})
yields the identity $\mathbf{M}^{-1}=\mathbf{\mathbb{G}M}_{\mathbf{Q}%
^{T}}^{T}\mathbf{\mathbb{G},}$ which may be further developed, using (\ref%
{7.2}) and (\ref{26}), to the form%
\begin{equation}
\mathbf{M}^{T}\mathbf{\mathbb{Y}M=\mathbb{Y}}\ \Leftrightarrow \ \mathbf{M}=%
\mathbf{\mathbb{G}M}^{\ast }\mathbf{\mathbb{G}}.  \label{32}
\end{equation}%
The same may be observed from (\ref{11}) with (\ref{31}), see \cite{BH}. By (%
\ref{32}), $\mathbf{M}^{-1},$ $\mathbf{M}^{T}$ (hence, $\mathbf{M}$) and $%
\mathbf{M}^{\ast }$ are similar, so they share the same eigenspectrum, i.e.
the set of eigenvalues $\left\{ q_{\alpha }\right\} $ of $\mathbf{M}$ is
equal to the sets $\left\{ q_{\alpha }^{-1}\right\} $ and $\left\{ q_{\alpha
}^{\ast }\right\} $. Therefore it may be partitioned as follows:%
\begin{equation}
q_{\alpha }=q_{\beta }^{-1},\ \mathrm{\ }\alpha \in \left\{ 1,2,3\right\} ,\
\beta \in \left\{ 4,5,6\right\} .  \label{32.1}
\end{equation}%
By (\ref{32.1}), the characteristic polynomial for $\mathbf{M}$ is a
self-reciprocal one and hence may be reduced to a third-degree polynomial in
the variable $q+q^{-1}$ \cite{BH,Pod}\footnote{{\footnotesize Note an
instructive demonstration of the Hamiltonian framework for this case
provided in \cite{Pod}.}}. Similarly to (\ref{31.1}), the pairing in Eq. (%
\ref{32.1}) consistent with that adopted in (\ref{30.01}) implies $\beta
=\alpha +3,$ i.e. 
\begin{equation}
q_{\alpha }=q_{\alpha +3}^{\ast }\text{ }\mathrm{if~}\left\vert q_{\alpha
}\right\vert =1,\ q_{\alpha }=q_{\alpha }^{\ast }\text{ }\mathrm{if~}%
\left\vert q_{\alpha }\right\vert \neq 1,  \label{32a}
\end{equation}%
unless possibly in the case of three or two pairs (\ref{32.1}) of $q$'s with
not a unit absolute value. This case allows for $\beta \neq \alpha +3$ for
two pairs, say, $\alpha =2,3,$ namely, 
\begin{equation}
q_{1}=1/q_{4}\ \left( =q_{4}^{\ast }\ \mathrm{or\ }\ q_{1}^{\ast }\right) ,\
q_{2}=1/q_{5}^{\ast }=1/q_{6}=q_{3}^{\ast }\ \Leftrightarrow \
q_{3}=1/q_{6}^{\ast }=1/q_{5}=q_{2}^{\ast }.  \label{32.2}
\end{equation}%
Note that if the medium is periodic and $\mathbf{M=M}\left( T,0\right) $
with $q_{\alpha }=e^{iK_{\alpha }T}$ (see (\ref{13})), then reformulation of
(\ref{32.2}) in terms of $K_{\alpha }$ is precisely the same as (\ref{31.2})
in terms $\kappa _{\alpha }$.

According to Eqs. (\ref{32}) and (\ref{32.1}), the eigenvectors of
(diagonalizable) $\mathbf{M}$ are linked as%
\begin{equation}
\mathbf{w}_{\alpha }=i\mathbf{Gw}_{\beta }^{\ast }\ \mathrm{if}\ \
\left\vert q_{\alpha }\right\vert =\left\vert q_{\beta }\right\vert =1;\ 
\mathbf{w}_{\alpha }=i\mathbf{Gw}_{\alpha }^{\ast },\ \mathbf{w}_{\beta }=-i%
\mathbf{Gw}_{\beta }^{\ast }\ \mathrm{if}\ \ \left\vert q_{\alpha
}\right\vert ,\left\vert q_{\beta }\right\vert \neq 1,  \label{32.3}
\end{equation}%
where $\beta =\alpha +3$ for all $\alpha =1,2,3$, except in the case (\ref%
{32.2}) when $\mathbf{w}_{2}=i\mathbf{Gw}_{3}^{\ast },\ \mathbf{w}_{5}=-i%
\mathbf{Gw}_{6}^{\ast }$. Inserting equalities (\ref{32.3}) into (\ref{30})
verifies their consistency with the above-adopted normalization and provides
an additional identity 
\begin{equation}
\mathbf{W}^{T}\mathbf{\mathbb{Y}W}=i\widetilde{\mathbf{\mathbb{J}}},
\label{32.4}
\end{equation}%
where $\widetilde{\mathbf{\mathbb{J}}}$ is the same as in Eq. (\ref{30}).

Now, we suppose that the symmetry plane orthogonal to the vector $\mathbf{e}%
_{1}$ or $\mathbf{e}_{2}$ \textit{coexists with the symmetric profile} of
inhomogeneity, i.e. the system matrix $\mathbf{Q}\left( y\right) $ satisfies
(\ref{31}) and is an even function on $\left[ -y,y\right] $ Then, in view of
(\ref{28.2})$_{1},$ Eq. (\ref{32}) applied to $\mathbf{M=M}\left(
y,-y\right) $ can be brought into the form%
\begin{equation}
\mathbf{M}\left( y,-y\right) \mathbf{\mathbb{G}M}\left( y,-y\right) =\mathbf{%
\mathbb{G}.}  \label{33}
\end{equation}%
Consequently, the triplets of (normalized) eigenvectors $\mathbf{w}_{\alpha
}\left( y\right) $ and $\mathbf{w}_{\beta }\left( y\right) $ of $\mathbf{M}%
\left( y,-y\right) $ corresponding to the mutually inverse eigenvalues (\ref%
{32.1}) are related as 
\begin{equation}
\mathbf{w}_{\alpha }=i\mathbf{\mathbb{G}w}_{\beta },\ \mathrm{\ }\alpha \in
\left\{ 1,2,3\right\} ,\ \beta \in \left\{ 4,5,6\right\} ,  \label{33.1}
\end{equation}%
where the pairing of $\alpha $ and $\beta $ is the same as described below (%
\ref{32.1}). In parallel, due to the profile symmetry, the eigenvectors
possess property (\ref{28.3}), which ensures the compatibility of relation (%
\ref{33.1}) with (\ref{32.3}) and (\ref{32.4}).

Let us mention one more feature of the displacement and traction solutions $%
\mathbf{u}$ and $\mathbf{t}_{i}=\mathbf{e}_{i}^{T}\mathbf{\sigma }$ of the
wave equation (\ref{1}) in the media with the plane(s) of crystallographic
symmetry and a symmetric dependence of the material coefficients $\rho $ and 
$c_{ijkl}$. This property concerns the general 3D-inhomogeneity setting, in
which $\rho ,$ $c_{ijkl}$ and hence $\mathbf{u}$ and $\mathbf{t}_{i}$ depend
on all three coordinates $x,~y,~z$. Assume that the medium possesses a
symmetry plane orthogonal to the vector $\mathbf{e}_{1}\parallel X,$ and $%
\rho $ and $c_{ijkl}$ are even functions $f\left( x,\cdot \right) =f\left(
-x,\cdot \right) $ of $x$ (here $\cdot $ implies $y,~z$). Then two linearly
independent partial solutions of Eq. (\ref{1}) may be cast to the form where
their vector components are either even or odd in $x.$ Specifically, they
fall under one of the following two options (indicated by the superscripts $%
^{\left( 1\right) }$ and $^{\left( 2\right) }$):%
\begin{equation}
\begin{array}{c}
\mathbf{u}^{\left( 1\right) }\left( x,\cdot \right) =-\mathbf{g}_{1}\mathbf{u%
}^{\left( 1\right) }\left( -x,\cdot \right) ,\ \mathbf{t}_{1}^{\left(
1\right) }\left( x,\cdot \right) =\mathbf{g}_{1}\mathbf{t}_{1}^{\left(
1\right) }\left( -x,\cdot \right) ,\ \mathbf{t}_{2,3}^{\left( 1\right)
}\left( x,\cdot \right) =-\mathbf{g}_{1}\mathbf{t}_{2,3}^{\left( 1\right)
}\left( -x,\cdot \right) , \\ 
\mathbf{u}^{\left( 2\right) }\left( x,\cdot \right) =\mathbf{g}_{1}\mathbf{u}%
^{\left( 2\right) }\left( -x,\cdot \right) ,\ \mathbf{t}_{1}^{\left(
2\right) }\left( x,\cdot \right) =-\mathbf{g}_{1}\mathbf{t}_{1}^{\left(
2\right) }\left( -x,\cdot \right) ,\ \mathbf{t}_{2,3}^{\left( 2\right)
}\left( x,\cdot \right) =\mathbf{g}_{1}\mathbf{t}_{2,3}^{\left( 2\right)
}\left( -x,\cdot \right) .%
\end{array}
\label{34}
\end{equation}%
Accordingly, if there is a symmetry plane orthogonal to the vector $\mathbf{e%
}_{2}\parallel Y$ and/or $\mathbf{e}_{3}\parallel Z$ and $\rho ,$ $c_{ijkl}$
are even functions of $y$ and/or$~z$, then Eq. (\ref{34}) holds with $%
\mathbf{g}_{2}=\mathbf{I}-2\mathbf{e}_{2}^{T}\mathbf{e}_{2}$ and/or $\mathbf{%
g}_{3}=\mathbf{I}-2\mathbf{e}_{3}^{T}\mathbf{e}_{3}$, respectively. It is
noteworthy that if the medium with a symmetry plane orthogonal to $\mathbf{e}%
_{1}\parallel X$ is 2D-inhomogeneous, i.e. characterized with $\rho \left(
x,y\right) $ and $c_{ijkl}\left( x,y\right) $, then the evenness of $\rho $
and $c_{ijkl}$ in $x$ leads not only to relation (\ref{34}), but also to a
similar relation between the displacements and tractions taken at $\left(
x,y\right) $ and $\left( x,-y\right) $ (the same statement may be reworded
with $x$ and $y$ swapped). The above partitioning into symmetric and
antisymmetric families appears helpful in solving the boundary-value problem
in waveguides.

\subsubsection{Symmetry plane orthogonal to $\mathbf{e}_{3}:$ uncoupling of
the SH modes\label{SSSec3.3.2}}

If the sagittal plane $\left( \mathbf{e}_{1},\mathbf{e}_{2}\right) $ is the
symmetry plane, then the system of three equations (\ref{1}) applied to the
2D wave field $\mathbf{u}\left( x,y\right) $ splits into a system of two
equations and a single uncoupled equation. The former system defines the
in-plane or sagittally polarized vector waves with the displacement $\mathbf{%
u}$ and the tractions $\mathbf{t}_{1,2}$ lying in the plane $\left( \mathbf{e%
}_{1},\mathbf{e}_{2}\right) $ (the traction $\mathbf{t}_{3}$ is parallel to $%
\mathbf{e}_{3}$). The latter uncoupled equation defines the out-of-plane or
shear horizontally (SH) polarized scalar waves with $\mathbf{u}$ and $%
\mathbf{t}_{1,2}$ orthogonal to $\left( \mathbf{e}_{1},\mathbf{e}_{2}\right) 
$ (the traction $\mathbf{t}_{3}$ is orthogonal to $\mathbf{e}_{3}$).

Consider the SH\ waves in the most general setup, which is when $\left( 
\mathbf{e}_{1},\mathbf{e}_{2}\right) $ is the symmetry plane but other
coordinate planes are not, as is the case in monoclinic and trigonal
materials. The equation following from (\ref{1}) for SH displacement $%
u_{3}\left( y\right) =$ $\mathbf{u}\left( y\right) \cdot \mathbf{e}_{3}$
reads 
\begin{equation}
\left( c_{44}u_{3}^{\prime }+ik_{x}c_{45}u_{3}\right) ^{\prime
}+ik_{x}c_{45}u_{3}^{\prime }+\left( \rho \omega ^{2}-k_{x}^{2}c_{55}\right)
u_{3}=0,  \label{35}
\end{equation}%
where the stiffness coefficients are written in Voight's notations and
referred to the monoclinic basis $\left\{ \mathbf{e}_{1},\mathbf{e}_{2},%
\mathbf{e}_{3}\right\} .$ Interestingly, substituting the replacement $\ $ 
\begin{equation}
u_{3}\left( y\right) =w\left( y\right) e^{-i\Phi \left( y\right) },\ \Phi
\left( y\right) =k_{x}\int_{0}^{y}\frac{c_{45}\left( s\right) }{c_{44}\left(
s\right) }ds  \label{36}
\end{equation}%
casts Eq. (\ref{35}) into the canonical Sturm-Liouville form of equation on $%
w\left( y\right) $: 
\begin{equation}
\left( c_{44}w^{\prime }\right) ^{\prime }+\left( \rho \omega
^{2}-k_{x}^{2}C_{55}\right) w=0,  \label{37}
\end{equation}%
where $C_{55}=c_{55}-\dfrac{c_{45}^{2}}{c_{44}}.$ Equation (\ref{37})
emulates the SH wave equation with $c_{45}=0$, which is the case in
orthorhombic materials with symmetry planes along all three coordinate
planes. Replacement (\ref{36}) does not involve differentiation of material
parameters and hence remains valid in the case of their piecewise continuous
coordinate dependence, such as occurs in layered media.

Equivalent equations (\ref{35}) and (\ref{37}) may be brought in the form of
the 1st-order ODS $\mathbf{\eta }_{u}^{\prime }=\mathbf{Q}_{u}\mathbf{\eta }%
_{u}$ and $\mathbf{\eta }_{w}^{\prime }=\mathbf{Q}_{w}\mathbf{\eta }_{w}$
representing the SH subsystem uncoupled from the general Stroh's ODS (\ref{5}%
). Using, say, explicit format (\ref{8})$_{3}$ yields $\mathbf{\eta }%
_{u}=\left( u_{3}\ i\sigma _{23}\right) ^{T}=\mathbf{\eta }_{w}e^{-i\Phi
\left( y\right) }$ and 
\begin{equation}
\mathbf{Q}_{u}=-ik_{x}\dfrac{c_{45}}{c_{44}}\mathbf{I+Q}_{w},\ \mathbf{Q}%
_{w}=i\left( 
\begin{array}{cc}
0 & -c_{44}^{-1} \\ 
C_{55}k_{x}^{2}-\rho \omega ^{2} & 0%
\end{array}%
\right) ,  \label{38}
\end{equation}%
whence the corresponding $2\times 2\ $matricants are related as $\mathbf{M}%
_{u}=e^{-i\Phi }\mathbf{M}_{w}$. Due to (\ref{26}), the diagonal and
off-diagonal elements of $\mathbf{M}_{w}$ are real and complex-valued,
respectively\footnote{{\footnotesize Alternative use of the state vector }$%
\eta =\left( u_{3}\ \sigma _{23}\right) ^{T}${\footnotesize \ converts (\ref%
{37}) to the ODS with a purely real system matrix and a matricant satisfying
(\ref{27}) and (\ref{27.1}).}}. The matrix $\mathbf{Q}_{w}$ is traceless,
hence $\mathbf{M}_{w}$ has a unit determinant and the eigenvalues $q$ and $%
q^{-1}$ which are either real or have unit absolute value. The eigenvectors $%
\mathbf{w}_{\alpha },$ $\alpha =1,2,$ of (diagonalizable) $\mathbf{M}_{w}$
satisfy%
\begin{equation}
\mathbf{w}_{1}=i\widehat{\mathbf{h}}\mathbf{w}_{2}^{\ast }\ \mathrm{if}\
\left\vert q\right\vert =1,\ \mathbf{w}_{1}=i\widehat{\mathbf{h}}\mathbf{w}%
_{1}^{\ast },\ \mathbf{w}_{2}=-i\mathbf{\widehat{\mathbf{h}}w}_{2}^{\ast }\ 
\mathrm{if}\ q\ \mathrm{is}\ \func{real}\left( \neq \pm 1\right) ,\ \mathrm{%
where\ }\widehat{\mathbf{h}}=\mathrm{diag}\left( 1,-1\right) ,  \label{39}
\end{equation}%
which is the "$2\times 2$ reduction" of (\ref{32.3}). In view of (\ref{39}), 
$\det \left\Vert \mathbf{w}_{1}\mathbf{w}_{2}\right\Vert =\pm i$. If the
inhomogeneity profile is symmetric, then, by (\ref{28.2}) and (\ref{28.1}),
the diagonal elements of $\mathbf{M}_{w}$ are equal to each other and so $%
\mathbf{w}_{1}=\pm i\mathbf{\widehat{\mathbf{h}}w}_{2}$ and $u_{1}v_{1}=%
\frac{1}{2}$ at any $q.$

A detailed survey of the SH wave formalism in periodic media may be found in 
\cite{KSPN}.

\subsubsection{Orthorhombic symmetry: separation of variables solution\label%
{SSSec3.3.3}}

Let all three coordinate planes be the symmetry planes for the stiffness
tensor, i.e. the medium be of orthorhombic or higher symmetry. In this case,
the sagittal and SH solutions of (\ref{1}) admit the separation of
variables, leading to a form more general than (\ref{2}). In particular, the
sagittal solutions may be sought as 
\begin{equation}
\mathbf{u}\left( x,y\right) =\left( 
\begin{array}{c}
X^{\prime }Y_{1} \\ 
ik_{x}XY_{2}%
\end{array}%
\right) ,\ \mathbf{t}_{1}\left( x,y\right) =\left( 
\begin{array}{c}
ik_{x}X\tau _{11} \\ 
X^{\prime }\tau _{12}%
\end{array}%
\right) ,\ \mathbf{t}_{2}\left( x,y\right) =\left( 
\begin{array}{c}
X^{\prime }\tau _{21} \\ 
ik_{x}X\tau _{22}%
\end{array}%
\right) ,  \label{40}
\end{equation}%
with $X=X\left( x\right) ,$ $\mathbf{Y}=\left( Y_{1}\ Y_{2}\right) ^{T}=%
\mathbf{Y}\left( y\right) $ and $\mathbf{\tau }_{i}=\left( \tau _{i1}\ \tau
_{i2}\right) ^{T}=\mathbf{\tau }_{i}\left( y\right) ,$ $i=1,2$. The function 
$X\left( x\right) $ satisfies the equation $X^{\prime \prime }+k_{x}^{2}X=0$
with an arbitrary constant $k_{x},$ while the vector functions $\mathbf{Y}%
\left( y\right) $ and $\mathbf{\tau }_{2}\left( y\right) $ are defined by
the $4\times 4$ ODS of the form (\ref{5}) and, say, (\ref{8})$_{3}$ with the
entries $\mathbf{\eta }\left( y\right) =\left( \mathbf{Y\ }i\mathbf{\tau }%
_{2}\right) ^{T}$ and 
\begin{equation}
\mathbf{N}_{1}=-\left( 
\begin{array}{cc}
0 & 1 \\ 
c_{12}c_{22}^{-1} & 0%
\end{array}%
\right) ,\ \mathbf{N}_{2}=-\left( 
\begin{array}{cc}
c_{66}^{-1} & 0 \\ 
0 & c_{22}^{-1}%
\end{array}%
\right) ,\ \mathbf{N}_{3}=\left( 
\begin{array}{cc}
c_{11}-c_{12}^{2}c_{22}^{-1} & 0 \\ 
0 & 0%
\end{array}%
\right) ,  \label{41}
\end{equation}%
where the stiffness coefficients are referred to the basis $\left\{ \mathbf{e%
}_{1},\mathbf{e}_{2}\right\} $. The remaining vector function $\mathbf{\tau }%
_{1}\left( y\right) $ follows from the relation $\mathbf{\tau }_{1}=ik_{x}%
\mathbf{N}_{3}\mathbf{Y}-\mathbf{N}_{1}^{T}\mathbf{\tau }_{2}$. In turn, the
SH modes admit the form 
\begin{equation}
u_{3}\left( x,y\right) =X\left( x\right) Y\left( y\right) ,  \label{42}
\end{equation}%
where $X\left( x\right) $ fulfils $X^{\prime \prime }+k_{x}^{2}X=0$ and $%
Y\left( y\right) $ is defined by (\ref{37}) with $Y$ and $c_{55}$ in place
of $w$ and $C_{55}$. Note that the SH solution cannot take the form (\ref{42}%
) in the case of the symmetry plane in a monoclinic or trigonal media
considered in \S \ref{SSSec3.3.2}.

The sagittal solution in the separation of variables form (\ref{40}) has
been discussed in \cite{Ach,Pa1} for the transversely isotropic homogeneous
and 1D-inhomogeneous materials, respectively. By admitting a trigonometric
form of $x$-dependence, such solutions are well-suited to wave problems on a
2D-bounded domain truncated in both the $\mathbf{e}_{1}$ and $\mathbf{e}_{2}$
directions. They are particularly advantageous for dealing with
semi-infinite or rectangular plates (strips) subjected to the mixed
homogeneous boundary conditions at the faces $x=x_{1},$ $x_{2}$ orthogonal
to the $X$-axis: $u_{i}\left( x_{1},y\right) =0,\ t_{1j}\left(
x_{2},y\right) =0$\ $\forall y,\ i\neq j,$ which are complemented by, say,
homogeneous boundary conditions at the faces $y=y_{1},y_{2}$ or at $y=y_{1}$
plus the radiation condition at $y\rightarrow \infty $. By (\ref{40}), such
a boundary-value problem retains the decoupling of the $x$- and $y$%
-dependencies so that the dispersion spectrum represents a discrete set of
frequencies $\omega _{n,m}=\omega _{n}(k_{x}^{\left( m\right) }),$ in which
the set of arguments $\{k_{x}^{\left( m\right) }\}$ is prescribed by the
boundary condition on $X\left( x\right) ,$ and the functions $\omega
_{n}=\omega _{n}\left( k_{x}\right) $ follow from the solvability condition
of Eq. (\ref{41}) under the constraint on $Y\left( y\right) $. This idea,
which goes back to the classical Mindlin's development \cite{Mind}, was
implemented in \cite{WK} for isotropic homogeneous plates; as shown above,
it equally applies to the orthorhombic 1D-inhomogeneous materials. The same
way the separation of variables approach (\ref{42}) readily provides the
solutions for the SH guided waves in laterally inhomogeneous finite or
semi-infinite strips with free or clamped faces, see e.g. \cite{ShPG}.

Another significant aspect of (\ref{42}) is that, when the material
properties depend only on $y$, the uncoupled scalar function may be taken as
a function $X\left( x,z\right) $ of two variables $x$ and $z$ (see remark to
(\ref{2})) which is defined by the 2D Helmholtz (reduced membrane) equation $%
\nabla ^{2}X+k^{2}X=0.$ Such an extension allows modelling surface wave
fields with a particular spatial structure in the $XZ$-plane \cite{PK}-\cite%
{KP}.

\section{Reflection/transmission problem\label{Sec4}}

\subsection{Preamble\label{SSec4.1}}

Let us precede derivations with a general methodological remark comparing
the statement of the reflection/transmission problem with those of the
initial value and boundary value problems. The initial value problem posed
in \S \ref{Sec2} amounts to calculating the matricant with no need to stitch
the wave field whenever material properties undergo a jump at interfaces. In
turn, the boundary-value problem involves auxiliary conditions at two points
(or the radiation condition at infinity), which bind the equation parameters
such as $\omega $ and $k_{x}$, see Part II. By contrast, the
reflection/transmission problem proceeds from an "incomplete" initial value,
which is the incident wave without the sought reflected complement, and the
condition that the sought transmitted wave must consist only of the
outflowing or decreasing modes (the attribution of modes is unambiguous once
they propagate in a homogeneous non-dissipative medium). This framework
reduces the problem to an algebraic linear inhomogeneous system on the
partial amplitudes of the reflected and transmitted modes, solvable without
restricting the parameters $\omega $ and $k_{x}$. Note a similarity to the
Green's function problem that is evident upon replacing the incident wave
with the source in the wavenumber domain.

The present section intends to set up the reflection/transmission problem in
the context of the Stroh formalism. With this purpose, we consider a planar
transversely inhomogeneous layer $\left[ y_{1},y_{2}\right] $ embedded
between two homogeneous half-spaces $y\leq y_{1}$ and $y\geq y_{2}$ referred
to below as substrates 1 and 2 (the axis $Y$ is therefore directed from the
former to the latter). The substrates are characterized by constant system
matrices $\mathbf{Q}_{0}^{\left( 1\right) }$ and $\mathbf{Q}_{0}^{\left(
2\right) }$, respectively. Let the eigenvectors $\mathbf{\xi }_{\alpha
}^{\left( 1,2\right) }$ of either of $\mathbf{Q}_{0}^{\left( 1,2\right) }$
be numbered so that $\alpha =1,2,3$ correspond to the modes, which
exponentially decay with growing $y$ if $\func{Im}k_{y\alpha }^{\left(
1,2\right) }\neq 0$ or have a positive $y$-component of the time-averaged
energy flux density $\overline{P}_{y\alpha }$ if $\func{Im}k_{y\alpha
}^{\left( 1,2\right) }=0$. The layer-substrate interfaces maintain rigid
contact, implying that the jump of the system matrix elements is finite (see
the discussion under Eq. (\ref{25})). We emphasize that the results of \S \S %
\ref{SSec4.2}\ and \ref{SSec4.3} are based solely on the Stroh matrix
symmetry (\ref{7.2}), i.e. they are equally valid for the case of a
viscoelastic layer. The assumption of no dissipation is reinstated from \S %
\ref{SSec4.4}. The problem of reflection/transmission from an immersed plate
is addressed separately in \S \ref{SSSec9.3.3}.

\subsection{Definition and properties\label{SSec4.2}}

An incident wave $\mathbf{\eta }_{\mathrm{inc}}^{\left( 1\right) }\left(
y\right) $ propagating in substrate 1 generates the reflected wave $\mathbf{%
\eta }_{\mathrm{ref}}^{\left( 1\right) }\left( y\right) $ in substrate 1,
the 6-partial wave packet in the layer, and the transmitted wave $\mathbf{%
\eta }_{\mathrm{tran}}^{\left( 2\right) }\left( y\right) $ in substrate 2.
The waves in the (homogeneous) substrates represent a packet of three plane
modes (\ref{11.01}). According to the above-adopted numbering convention and
the direction of the $Y$-axis chosen from substrate 1 to substrate 2, they
are written as follows:%
\begin{equation}
\begin{array}{l}
\mathbf{\eta }_{\mathrm{inc}}^{\left( 1\right) }\left( y\right) =\left( 
\begin{array}{c}
\mathbf{\Xi }_{1}^{\left( 1\right) } \\ 
\mathbf{\Xi }_{3}^{\left( 1\right) }%
\end{array}%
\right) \mathrm{diag}\left( e^{ik_{y\alpha }^{\left( 1\right) }y}\right) 
\mathbf{c},\ \mathbf{\eta }_{\mathrm{ref}}^{\left( 1\right) }\left( y\right)
=\left( 
\begin{array}{c}
\mathbf{\Xi }_{2}^{\left( 1\right) } \\ 
\mathbf{\Xi }_{4}^{\left( 1\right) }%
\end{array}%
\right) \mathrm{diag}\left( e^{ik_{y,\alpha +3}^{\left( 1\right) }y}\right) 
\mathbf{R}^{\left( 11\right) }\mathbf{c}, \\ 
\mathbf{\eta }_{\mathrm{tran}}^{\left( 2\right) }\left( y\right) =\left( 
\begin{array}{c}
\mathbf{\Xi }_{1}^{\left( 1\right) } \\ 
\mathbf{\Xi }_{3}^{\left( 1\right) }%
\end{array}%
\right) \mathrm{diag}\left( e^{ik_{y\alpha }^{\left( 2\right) }y}\right) 
\mathbf{T}^{\left( 12\right) }\mathbf{c},\ \alpha =1,2,3,%
\end{array}
\label{48.00}
\end{equation}%
where $\mathbf{\Xi }_{1...4}^{\left( 1,2\right) }$ are the $3\times 3$
blocks (see (\ref{0})) of the $6\times 6$ eigenvector matrices $\mathbf{\Xi }%
^{\left( 1\right) }$ and $\mathbf{\Xi }^{\left( 2\right) }$ in substrates 1
and 2, $\mathbf{R}^{\left( 11\right) }$ and $\mathbf{T}^{\left( 12\right) }$
are the reflection and transmission matrices, $\mathbf{c}=\left(
c_{1}~c_{2}~c_{3}\right) ^{T}$ is a vector of arbitrary scalar factors. The
matrix elements $R_{\alpha \beta }^{\left( 11\right) }$ and $T_{\alpha \beta
}^{\left( 12\right) }$ of $\mathbf{R}^{\left( 11\right) }$ and $\mathbf{T}%
^{\left( 12\right) }$ ($\alpha ,\beta =1,2,3$) identify the amplitudes of
the $\left( \alpha +3\right) $th reflected and $\alpha $th transmitted plane
modes, respectively, produced by the $\beta $th incident mode. A similar
definition applies when the incident wave propagates in substrate 2.

Using continuity conditions for the wave fields taken with the same $k_{x}$
at the interfaces $y_{1}$ and $y_{2},$ the two above reflection/transmission
problems associated with counter-propagating incident waves can be
formulated in matrix form as%
\begin{equation}
\begin{array}{l}
\mathbf{M}\left( y_{2},y_{1}\right) \mathbf{\Xi }^{\left( 1\right) }\left( 
\begin{array}{c}
\mathbf{I} \\ 
\mathbf{R}^{\left( 11\right) }%
\end{array}%
\right) =\mathbf{\Xi }^{\left( 2\right) }\left( 
\begin{array}{c}
\mathbf{T}^{\left( 12\right) } \\ 
\widehat{\mathbf{0}}%
\end{array}%
\right) , \\ 
\mathbf{M}\left( y_{2},y_{1}\right) \mathbf{\Xi }^{\left( 1\right) }\left( 
\begin{array}{c}
\widehat{\mathbf{0}} \\ 
\mathbf{T}^{\left( 21\right) }%
\end{array}%
\right) =\mathbf{\Xi }^{\left( 2\right) }\left( 
\begin{array}{c}
\mathbf{R}^{\left( 22\right) } \\ 
\mathbf{I}%
\end{array}%
\right) ,%
\end{array}
\label{48.0}
\end{equation}%
where $\mathbf{M}\left( y_{2},y_{1}\right) $ is the matricant through the
layer. It depends on $\omega $ and $k_{x},$\ hence so do the matrices $%
\mathbf{R}$ and $\mathbf{T}$ (unless the layer is replaced with a planar
interface, in which case the dependence is on $v=\omega /k_{x}$). The sought
reflection and transmission matrices for both directions of wave incidence
may be incorporated into the $6\times 6$ matrix 
\begin{equation}
\mathbf{D}\equiv \left( 
\begin{array}{cc}
i\mathbf{R}^{\left( 11\right) } & \mathbf{T}^{\left( 21\right) } \\ 
\mathbf{T}^{\left( 12\right) } & -i\mathbf{R}^{\left( 22\right) }%
\end{array}%
\right) ,  \label{48.D}
\end{equation}%
where the factors $\pm i$ are added for convenience (they render $\mathbf{D}$
orthogonal at $\mathbf{M}=\mathbf{I}$, see below). Combining Eqs. (\ref{48.0}%
) in a blockwise form determines $\mathbf{D}$ as 
\begin{equation}
\mathbf{D}=\left( 
\begin{array}{cc}
i\mathbf{\Psi }_{2}^{\left( 1\right) } & \mathbf{\Xi }_{1}^{\left( 2\right) }
\\ 
i\mathbf{\Psi }_{4}^{\left( 1\right) } & \mathbf{\Xi }_{3}^{\left( 2\right) }%
\end{array}%
\right) ^{-1}\left( 
\begin{array}{cc}
\mathbf{\Psi }_{1}^{\left( 1\right) } & i\mathbf{\Xi }_{2}^{\left( 2\right) }
\\ 
\mathbf{\Psi }_{3}^{\left( 1\right) } & i\mathbf{\Xi }_{4}^{\left( 2\right) }%
\end{array}%
\right) ,  \label{48.D1}
\end{equation}%
where the auxiliary notation $\mathbf{\Psi }^{\left( 1\right) }=\mathbf{M}%
\left( y_{2},y_{1}\right) \mathbf{\Xi }^{\left( 1\right) }$ is used for
brevity, and the $3\times 3$ blocks of the $6\times 6$ matrices $\mathbf{%
\Psi }^{\left( 1\right) }$ and $\mathbf{\Xi }^{\left( 2\right) }$ are
numbered according to convention (\ref{0}).

Alternatively, the reflection/transmission problem may be recast in terms of
the $6\times 6$ scattering matrix. It can be introduced in either of the
forms%
\begin{equation}
\mathbf{S}^{\left( 12\right) }=\mathbf{\Xi }^{\left( 1\right) T}\mathbf{%
\mathbb{T}M}\left( y_{1},y_{2}\right) \mathbf{\Xi }^{\left( 2\right) },\ 
\mathbf{S}^{\left( 21\right) }=\mathbf{\Xi }^{\left( 2\right) T}\mathbf{%
\mathbb{T}M}\left( y_{2},y_{1}\right) \mathbf{\Xi }^{\left( 1\right) }\left(
=\mathbf{S}^{\left( 12\right) -1}\right) ,  \label{48.S0}
\end{equation}%
where the orthonormalization relation $\mathbf{\Xi }^{-1}=\mathbf{\Xi }^{T}%
\mathbf{\mathbb{T}}$ and the identity $\mathbf{M}^{-1}\left(
y_{2},y_{1}\right) =\mathbf{M}\left( y_{1},y_{2}\right) $ are implied.
Using, say, (\ref{48.S0})$_{1}$ allows collecting Eqs. (\ref{48.0}) into a
single equation with the unknowns separated on one side: 
\begin{equation}
\mathbf{S}^{\left( 12\right) }=\left( 
\begin{array}{cc}
\mathbf{T}^{\left( 12\right) -1} & -\mathbf{T}^{\left( 12\right) -1}\mathbf{R%
}^{\left( 22\right) } \\ 
\mathbf{R}^{\left( 11\right) }\mathbf{T}^{\left( 12\right) -1} & \mathbf{T}%
^{\left( 21\right) -1}-\mathbf{R}^{\left( 11\right) }\mathbf{T}^{\left(
12\right) -1}\mathbf{R}^{\left( 22\right) }%
\end{array}%
\right) \ \ (\det \mathbf{S}^{\left( 12\right) }=\frac{\det \mathbf{T}%
^{\left( 21\right) }}{\det \mathbf{T}^{\left( 12\right) }}),  \label{48.S}
\end{equation}%
where the matrices $\mathbf{T}^{\left( 21\right) -1}$ and $\mathbf{T}%
^{\left( 12\right) -1}$ are well defined since the matricant continuity
rules out a zero transmitted field. Hence follows the expression for the
matrix $\mathbf{D}$ (\ref{48.D}), and thereby for the reflection and
transmission matrices, via the blocks of $\mathbf{S}^{\left( 12\right) }$,
namely,%
\begin{equation}
\mathbf{D}=\left( 
\begin{array}{cc}
i\mathbf{S}_{3}^{\left( 12\right) }\mathbf{S}_{1}^{\left( 12\right) -1} & 
\mathbf{S}_{4}^{\left( 12\right) }-\mathbf{S}_{3}^{\left( 12\right) }\mathbf{%
S}_{1}^{\left( 12\right) -1}\mathbf{S}_{2}^{\left( 12\right) } \\ 
\mathbf{S}_{1}^{\left( 12\right) -1} & -i\mathbf{S}_{2}^{\left( 12\right) }%
\mathbf{S}_{4}^{\left( 12\right) -1}%
\end{array}%
\right) \ \ (\det \mathbf{D}=-\frac{\det \mathbf{S}_{4}^{\left( 12\right) }}{%
\det \mathbf{S}_{1}^{\left( 12\right) }}).  \label{48.D2}
\end{equation}%
This representation of $\mathbf{D}$ is equivalent to (\ref{48.D1}). The
expressions involving $\mathbf{S}^{\left( 21\right) }$ are obtainable from (%
\ref{48.S}) and (\ref{48.D2}) by swapping the superscripts $%
1\rightleftarrows 2$ and the (abridged) block indices $1\rightleftarrows
4,2\rightleftarrows 3.\ $

If the two substrates are identical ($\mathbf{\Xi }^{\left( 1\right) }=%
\mathbf{\Xi }^{\left( 2\right) }\equiv \mathbf{\Xi ,}$ hence $\det \mathbf{S}%
=\det \mathbf{M}$) and the intermediate layer has a symmetric profile of
inhomogeneity (see \S \ref{SSSec3.2.4}), then the scattering matrices (\ref%
{48.S0}) are symmetric and hence 
\begin{equation}
\mathbf{R}^{\left( 11\right) }=-\mathbf{R}^{\left( 22\right) T},\ \mathbf{T}%
^{\left( 12\right) }=\mathbf{T}^{\left( 12\right) T},\ \mathbf{T}^{\left(
21\right) }=\mathbf{T}^{\left( 21\right) T}.  \label{48.021}
\end{equation}

Let us also mention the case of two substrates in direct contact via a
planar interface $y=0$. Then the reflection/transmission and scattering
matrices (\ref{48.D1}) and (\ref{48.S0}) with $\mathbf{M=I}$ are orthogonal
matrices. In consequence of $\mathbf{DD}^{T}=1$, 
\begin{equation}
\begin{array}{c}
\mathbf{I}+\mathbf{R}^{\left( 11\right) T}\mathbf{R}^{\left( 11\right) }=%
\mathbf{T}^{\left( 12\right) T}\mathbf{T}^{\left( 12\right) },\ \mathbf{R}%
^{\left( 11\right) T}\mathbf{T}^{\left( 21\right) }=\mathbf{T}^{\left(
12\right) T}\mathbf{R}^{\left( 22\right) }; \\ 
\mathbf{I}+\mathbf{R}^{\left( 11\right) }\mathbf{R}^{\left( 11\right) T}=%
\mathbf{T}^{\left( 21\right) }\mathbf{T}^{\left( 21\right) T}\mathbf{,\ R}%
^{\left( 11\right) }\mathbf{T}^{\left( 12\right) T}=\mathbf{T}^{\left(
21\right) }\mathbf{R}^{\left( 22\right) T}%
\end{array}
\label{48.022}
\end{equation}%
and the same identities hold with swapped subscripts $1\rightleftarrows 2$.
Assuming incidence, say, from substrate 1 ($y<0$) and applying (\ref{28})
allows specifying the components of the reflection and transmission matrices
in the form%
\begin{equation}
\begin{array}{c}
R_{\alpha \beta }^{\left( 11\right) }=c_{\beta }\det \left\Vert \mathbf{\xi }%
_{\alpha }^{\left( 1\right) }\mathbf{\xi }_{\gamma }^{\left( 1\right) }%
\mathbf{\xi }_{\delta }^{\left( 1\right) }\mathbf{\xi }_{4}^{\left( 2\right)
}\mathbf{\xi }_{5}^{\left( 2\right) }\mathbf{\xi }_{6}^{\left( 2\right)
}\right\Vert ,\ T_{\alpha \beta }^{\left( 12\right) }=c_{\beta }\det
\left\Vert \mathbf{\xi }_{\alpha }^{\left( 2\right) }\mathbf{\xi }_{\gamma
}^{\left( 1\right) }\mathbf{\xi }_{\delta }^{\left( 1\right) }\mathbf{\xi }%
_{4}^{\left( 2\right) }\mathbf{\xi }_{5}^{\left( 2\right) }\mathbf{\xi }%
_{6}^{\left( 2\right) }\right\Vert \\ 
\mathrm{with}\ \ 1/c_{\beta }=\det \left\Vert \mathbf{\xi }_{\beta }^{\left(
1\right) }\mathbf{\xi }_{\gamma }^{\left( 1\right) }\mathbf{\xi }_{\delta
}^{\left( 1\right) }\mathbf{\xi }_{4}^{\left( 2\right) }\mathbf{\xi }%
_{5}^{\left( 2\right) }\mathbf{\xi }_{6}^{\left( 2\right) }\right\Vert ,%
\end{array}
\label{48.22}
\end{equation}%
where $\left\Vert ...\right\Vert $ denotes a matrix whose enclosed columns
are the eigenvectors $\mathbf{\xi }_{1...6}^{\left( 1,2\right) }$ of the
substrates' system matrices $\mathbf{Q}_{0}^{\left( 1,2\right) },$ and $%
\alpha ,\beta ,\gamma ,\delta =1,2,3,\ \gamma ,\delta \neq \beta ,$ see \cite%
{ADS}.

\subsection{Reciprocity between $k_{x}$ and $-k_{x}$\label{SSec4.3}}

Swapping the incident and reflected modes propagating in the same half-space
corresponds to changing $k_{x}$ to $-k_{x}$. Consider the link between these
two settings. For definiteness, let us take the system matrix of the Stroh
ODS (\ref{5}) in the form (\ref{8})$_{3}$ and introduce its abridged
notation $\mathbf{Q}\left[ k_{x},\omega ^{2}\right] \equiv \mathbf{Q}%
_{k_{x}} $ so that%
\begin{equation}
\mathbf{Q}_{k_{x}}=-\mathbf{\mathbb{H}Q}_{-k_{x}}\mathbf{\mathbb{H\ }}%
\mathrm{with}\ \mathbf{\mathbb{H}}=\left( 
\begin{array}{cc}
\mathbf{I} & \widehat{\mathbf{0}} \\ 
\widehat{\mathbf{0}} & -\mathbf{I}%
\end{array}%
\right) .  \label{48.06}
\end{equation}%
Denote the matricant and fundamental solutions of (\ref{5}) with $\mathbf{Q}%
_{\pm k_{x}}\left( y\right) $ by $\mathbf{M}_{\pm k_{x}}\left(
y,y_{0}\right) $ and $\mathbf{\mathcal{N}}_{\pm k_{x}}\left( y\right) ;$ then%
\begin{equation}
\mathbf{M}_{k_{x}}^{T}\mathbf{\mathbb{J}M}_{-k_{x}}=\mathbf{\mathbb{J}},\ \
\left( \mathbf{\mathcal{N}}_{k_{x}}\mathbf{\mathbb{J}\mathcal{N}}%
_{-k_{x}}\right) ^{\prime }=\widehat{\mathbf{0}},  \label{48.07}
\end{equation}%
where $\mathbf{\mathbb{J}}=\mathbf{\mathbb{HT\ }}$is the same matrix as
defined in (\ref{27}), and the prime means derivative with respect to $y$.
Similarly, denote the set of six eigenvalues and the matrix of normalized
eigenvectors of (semisimple) $\mathbf{Q}_{\pm k_{x}}$ by $i\left\{ \kappa
_{\pm k_{x}}\right\} $ and $\mathbf{\Xi }_{\pm k_{x}}$ (see \S \ref%
{SSSec3.2.1}). It follows that 
\begin{equation}
\left\{ \kappa _{k_{x}}\right\} =\left\{ -\kappa _{-k_{x}}\right\} ,\ 
\mathbf{\Xi }_{k_{x}}=\mathbf{\mathbb{H}\Xi }_{-k_{x}}\ \left( \Rightarrow
\pm i\mathbf{\Xi }_{k_{x}}^{T}\mathbf{\mathbb{J}\Xi }_{-k_{x}}=\mathbf{I}%
\right) .  \label{48.08}
\end{equation}

Using the above notations, introduce two reciprocal reflection/transmission
problems for which the incident wave propagates in the same substrate (say,
substrate 1) but with mutually inverse $k_{x}$ or $-k_{x}$, namely, 
\begin{equation}
\begin{array}{l}
\mathbf{M}_{k_{x}}\left( y_{2},y_{1}\right) \mathbf{\Xi }_{k_{x}}^{\left(
1\right) }\left( 
\begin{array}{c}
\mathbf{I} \\ 
\mathbf{R}_{k_{x}}^{\left( 11\right) }%
\end{array}%
\right) =\mathbf{\Xi }_{k_{x}}^{\left( 2\right) }\left( 
\begin{array}{c}
\mathbf{T}_{k_{x}}^{\left( 12\right) } \\ 
\widehat{\mathbf{0}}%
\end{array}%
\right) , \\ 
\mathbf{M}_{-k_{x}}\left( y_{2},y_{1}\right) \mathbf{\Xi }_{-k_{x}}^{\left(
1\right) }\left( 
\begin{array}{c}
\mathbf{R}_{-k_{x}}^{\left( 11\right) } \\ 
\mathbf{I}%
\end{array}%
\right) =\mathbf{\Xi }_{-k_{x}}^{\left( 2\right) }\left( 
\begin{array}{c}
\widehat{\mathbf{0}} \\ 
\mathbf{T}_{-k_{x}}^{\left( 12\right) }%
\end{array}%
\right) ,%
\end{array}
\label{48.09}
\end{equation}%
where the interchange of the upgoing/decreasing and downgoing/increasing
triplets of partial modes due to (\ref{48.08})$_{1}$ is taken into account.
The product of the transposed first relation with the $\mathbf{\mathbb{J}}$
times the second yields the reflection reciprocity identity%
\begin{equation}
\mathbf{R}_{k_{x}}^{\left( 11\right) }=-\mathbf{R}_{-k_{x}}^{\left(
11\right) T}.  \label{48.10}
\end{equation}%
Now let the incident mode with $-k_{x}$ propagate in substrate 2 so that 
\begin{equation}
\mathbf{\Xi }_{-k_{x}}^{\left( 1\right) }\left( 
\begin{array}{c}
\mathbf{T}_{-k_{x}}^{\left( 21\right) } \\ 
\widehat{\mathbf{0}}%
\end{array}%
\right) =\mathbf{M}_{-k_{x}}\left( y_{1},y_{2}\right) \mathbf{\Xi }%
_{-k_{x}}^{\left( 2\right) }\left( 
\begin{array}{c}
\mathbf{I} \\ 
\mathbf{R}_{-k_{x}}^{\left( 22\right) }%
\end{array}%
\right) .  \label{48.11}
\end{equation}%
The product of the transposed (\ref{48.09})$_{1}$ with the $\mathbf{\mathbb{J%
}}$ times (\ref{48.11}) provides the transmission reciprocity identity%
\begin{equation}
\mathbf{T}_{k_{x}}^{\left( 12\right) }=\mathbf{T}_{-k_{x}}^{\left( 21\right)
T}.  \label{48.12}
\end{equation}%
Finally, applying (\ref{48.07}) and (\ref{48.08}) to definition (\ref{48.S0}%
) or else inserting (\ref{48.10}) and (\ref{48.12}) into (\ref{48.S})
reveals the reciprocity property for the scattering matrix in the form%
\begin{equation}
\mathbf{S}_{k_{x}}^{\left( 12\right) }=-\mathbf{S}_{-k_{x}}^{\left(
21\right) T}.  \label{48.13}
\end{equation}%
Note that the resulting identities (\ref{48.10}), (\ref{48.12}) and (\ref%
{48.13}) do not depend on the choice between the explicit definitions of the
system matrix $\mathbf{Q}$ introduced in \S \ref{Sec1}.

The above proof of reflection and transmission reciprocities basically
follows that of \cite{Chap}. The derived equalities between the Fourier
harmonics with wavenumbers $k_{x}$ and $-k_{x}$ in the rectangular basis
correspond to the symmetry relations\ for the Hankel modes with radial
wavenumber $k_{r}\left( >0\right) $ established in \cite{K}.

\subsection{Case of non-dissipative layer\label{SSec4.4}}

So far, we have assumed the absence of dissipation in the substrates (to
gain clear attribution of reflected and transmitted modes), but not
necessarily in the intermediate layer. Now let us extend this assumption to
the layer material. Then, using Eqs. (\ref{26}) and (\ref{29}), it is
immediate to verify that the scattering matrix $\mathbf{S}^{\left( 21\right)
}$ (\ref{48.S0}) corresponding to the incidence from substrate 1 satisfies
the identity 
\begin{equation}
\mathbf{S}^{\left( 21\right) +}\mathbf{\mathbb{E}}_{\mathbf{Q}_{02}}\mathbf{S%
}^{\left( 21\right) }=\mathbf{\mathbb{E}}_{\mathbf{Q}_{01}},  \label{48.03}
\end{equation}%
where $\mathbf{\mathbb{E}}_{\mathbf{Q}_{01}}$ and $\mathbf{\mathbb{E}}_{%
\mathbf{Q}_{02}}$ are the matrices described below (\ref{29}). An equivalent
identity following from (\ref{48.0})$_{1}$ and (\ref{26}), (\ref{29}) has
the form 
\begin{equation}
\left( 
\begin{array}{cc}
\mathbf{I} & \mathbf{R}^{\left( 11\right) +}%
\end{array}%
\right) \mathbf{\mathbb{E}}_{\mathbf{Q}_{01}}\left( 
\begin{array}{c}
\mathbf{I} \\ 
\mathbf{R}^{\left( 11\right) }%
\end{array}%
\right) =\left( 
\begin{array}{cc}
\mathbf{T}^{\left( 12\right) +} & \widehat{\mathbf{0}}%
\end{array}%
\right) \mathbf{\mathbb{E}}_{\mathbf{Q}_{02}}\left( 
\begin{array}{c}
\mathbf{T}^{\left( 12\right) } \\ 
\widehat{\mathbf{0}}%
\end{array}%
\right) .  \label{48.04}
\end{equation}%
Diagonal $\left( \beta \beta \right) $th elements of matrix identity (\ref%
{48.04}) express the continuity of the normal energy flux at the incidence
of the $\left( \beta +3\right) $th mode, namely,%
\begin{equation}
\sum\nolimits_{\alpha }\left\vert R_{\alpha \beta }^{\left( 11\right)
}\right\vert ^{2}+\sum\nolimits_{\gamma }\left\vert T_{\gamma \beta
}^{\left( 12\right) }\right\vert ^{2}=1,  \label{48.05}
\end{equation}%
where the indices $\alpha $ and $\gamma $ (both $\leq 3$) enumerate,
specifically, the bulk (propagating) reflected and transmitted modes
generated in substrates 1 and 2 by the given $\left( \beta +3\right) $th
incident mode. Equations (\ref{48.04}) and (\ref{48.05}) remain valid under
the replacement of substrates' subscripts $1\rightleftarrows 2.$

One more identity is provided by the energy balance (\ref{e7'}) obtained in
Appendix 2. For brevity, assume that all three reflected modes generated by
the incident mode are bulk modes and denote their reflection coefficients by 
$R_{\alpha },\ \alpha =1,2,3$. Also, denote by $\theta _{\mathrm{inc}}$ and $%
\theta _{\mathrm{ref}}^{\left( \alpha \right) }$ the incidence and
reflection angles made by the (real) wave vectors with the $Y$-axis normal
to the junction surface. Then Eq. (\ref{e7'}) yields%
\begin{equation}
k_{x}\overline{P}_{x}+\overline{p}_{y\left( \mathrm{inc}\right) }\cot \theta
_{_{\mathrm{inc}}}+\sum\nolimits_{\alpha =1}^{3}\left\vert R_{\alpha
}\right\vert ^{2}\overline{p}_{y\left( \mathrm{ref}\right) }^{\left( \alpha
\right) }\cot \theta _{\mathrm{ref}}^{\left( \alpha \right) }=\omega \left( 
\overline{\mathcal{K}}+\overline{\mathcal{W}}\right) ,  \label{48.051}
\end{equation}%
where $k_{x}$ is the common tangential wavenumber; $\overline{p}_{y\left( 
\mathrm{inc}\right) }$ and $\overline{p}_{y\left( \mathrm{ref}\right)
}^{\left( \alpha \right) }$ are the $y$-components of the time-averaged
energy fluxes of the similarly normalized incident and reflected partial
modes; $\overline{P}_{x},$ $\overline{\mathcal{K}}$ and $\overline{\mathcal{W%
}}$ are the time-averaged tangential flux, kinetic energy and stored energy
of the overall wave field (see Appendix 2 for details).

\subsection{Poles and zeros\label{SSec4.5}}

\subsubsection{Reflection/transmission poles\label{SSSec4.5.1}}

The reflection/transmission poles are related to the exceptional possibility
that the waves in the layer are precisely matched by the reflected and
transmitted waves, i.e. the boundary conditions are satisfied without the
incident wave. Indeed, it is seen from either of (\ref{48.D1})-(\ref{48.D2})
that the common denominator of the reflection/transmission matrix $\mathbf{D}
$ equals zero under either of the following conditions: 
\begin{equation}
\det \left( 
\begin{array}{cc}
\mathbf{\Psi }_{2}^{\left( 1\right) } & \mathbf{\Xi }_{1}^{\left( 2\right) }
\\ 
\mathbf{\Psi }_{4}^{\left( 1\right) } & \mathbf{\Xi }_{3}^{\left( 2\right) }%
\end{array}%
\right) =0\Leftrightarrow \det \mathbf{S}_{1}^{\left( 12\right)
}=0\Leftrightarrow \det \mathbf{S}_{4}^{\left( 21\right) }=0,  \label{R1}
\end{equation}%
whose equivalence may be demonstrated using Schur's formula and identity (%
\ref{28})$.$ The former $6\times 6$ matrix 
\begin{equation}
\left( 
\begin{array}{cc}
\mathbf{\Psi }_{2}^{\left( 1\right) } & \mathbf{\Xi }_{1}^{\left( 2\right) }
\\ 
\mathbf{\Psi }_{4}^{\left( 1\right) } & \mathbf{\Xi }_{3}^{\left( 2\right) }%
\end{array}%
\right) \equiv \left\Vert \mathbf{\psi }_{4}^{\left( 1\right) }\mathbf{\psi }%
_{5}^{\left( 1\right) }\mathbf{\psi }_{6}^{\left( 1\right) }\mathbf{\xi }%
_{1}^{\left( 2\right) }\mathbf{\xi }_{2}^{\left( 2\right) }\mathbf{\xi }%
_{3}^{\left( 2\right) }\right\Vert  \label{R2}
\end{equation}%
is composed of $6$-component displacement-traction vectors $\mathbf{\psi }%
_{\alpha }^{\left( 1\right) }$ and $\mathbf{\xi }_{\beta }^{\left( 2\right)
} $ ($\alpha =4,5,6$ and $\beta =1,2,3$) of two triplets of partial modes
propagating in substrates 1 and 2 and decreasing away from the layer when $%
\omega $ and $k_{x}$ are real and subsonic, i.e. at $\omega /k_{x}<\min
\left( \hat{v}^{\left( 1\right) },\hat{v}^{\left( 2\right) }\right) $, see 
\S \ref{SSSec3.2.1}. Hence, Eq. (\ref{R1}) is the dispersion equation for a
layer-localized wave, as expected. Similarly, if the intermediate layer is
replaced by a planar interface so that $\mathbf{M}=\mathbf{I}$ and $\mathbf{%
\psi }_{\alpha }^{\left( 1\right) }\equiv \mathbf{\xi }_{\alpha }^{\left(
1\right) },$ then (\ref{R1}) is the dispersion equation for the Stoneley
wave (cf. Eq. (\ref{48*}) in \S \ref{SSec6.5}). At the same time, it is
physically clear that a bulk incident wave real $\omega $ and $k_{x}$ cannot
give rise to infinite reflection/transmission. Let us look into the formal
side.

Typically, the above localized wave solutions are confined to the subsonic
range and, therefore, are irrelevant to the reflection and transmission
generated by a bulk (i.e. necessarily supersonic) incident mode\footnote{%
{\footnotesize Formally, the reflection/transmission problem may be posed in
the subsonic domain with an incident mode having complex }$k_{y}.$%
{\footnotesize \ Then, in contrast to the supersonic case, the reflection
and transmission coefficients tend to infinity when approaching the subsonic
solution for a localized wave. This behavior is perfectly "physical" as it
just corresponds to the pole of the Green's function in the }$k_{x}$%
{\footnotesize \ space (this observation is owed to Olivier Poncelet). }}.
However, as a restricted possibility, Eq. (\ref{R1}) may also hold in the
supersonic domain $\omega /k_{x}>\min \left( \hat{v}^{\left( 1\right) },\hat{%
v}^{\left( 2\right) }\right) $ due to the linear dependence of fewer than
six vectors (\ref{R2}), which signals a localized wave and thus a zero
denominator of the reflection and transmission coefficients. The point is
that when this occurs, their numerators turn to zero as well. In other
words, the governing system of (generally) six equations (\ref{48.0})
becomes of rank four and thus has two types of uncoupled solutions: one is
the solution of Eq. (\ref{R1}) for the layer-localized supersonic wave, and
the other describes reflection/transmission but differs from (\ref{48.D1}).
The general proof is essentially similar to that developed for the case of a
free surface of an anisotropic elastic half-space, where a solution for the
supersonic Rayleigh wave with $v_{\mathrm{R}}>\hat{v}$ coexists alongside
the solution for a "pure reflection" involving bulk incident and reflected
modes but no surface ones \cite{AL81}. In more elaborate cases, the two
packets constituting the localized wave and the reflection/transmission
solution "share" partial modes.

An arbitrary perturbation $\left\vert \varepsilon \right\vert \ll 1$ leads
to the hybridization of the (formerly uncoupled) localized-wave and
reflection/transmission solutions. The former "shifts" to the complex plane,
i.e. transforms into the so-called leaky wave with complex $\omega $ and/or $%
k_{x},$ while its coupling with the reflection/transmission solution entails
the strong generation of the accompanying guided-wave field at real $\omega $
and $k_{x}$. It is significant that the numerator and denominator of the
reflection and transmission coefficients, both of which are small near the
bifurcation point, may be of different orders in $\varepsilon $ allowing
their ratio to exhibit resonant-type behavior with a potential for fine
tuning. Examples of such effects, where the perturbation $\varepsilon $ is
realized through a slight turn of the propagation geometry or by loading a
solid half-space with a relatively light fluid or a thin layer, may be found
in \cite{ADS,DW,EM}.

Resonance of the plane-wave reflection/transmission causes the non-specular
phenomena for bounded beams, such as the classical Schoch effect \cite{BG}
and its analogues \cite{N1,ADS2} on the fluid-solid interface. In terms of
the boundary-value problem, the supersonic localized wave is the bound state
in the radiation continuum, and the above resonant behavior of
reflection/transmission is in line with special features of the response
(Green's) function, see e.g. \cite{ME1,EM1}. Note in conclusion that some
fundamental issues related to the poles of the reflection and transmission
coefficients in the complex plane are discussed in \cite{BG}.

\subsubsection{Reflection zeros\label{SSSec4.5.2}}

The vanishing of one of the partial modes within the reflected wave packet
(an acoustic analogue of Brewster's effect) is a fairly common feature. At
the same time, it is clear that the simultaneous vanishing of all elements
of the $6\times 6$ reflection matrix $\mathbf{R},$ which would mean
"transparency" of the interface or layer for any incident (vector) wave, is
a heavily overdetermined problem. We shall be interested in a different
phenomenon, namely, the possibility of such transparency for a particular
incident wave with real $\omega $ and $k_{x}$. This wave may be referred to
as "reflectionless" by analogy with a similar type of guided waves, see \cite%
{BCP}. Given the incidence, say, from substrate 1, the necessary and
sufficient condition for its existence is that the matrix $\mathbf{R}%
^{\left( 11\right) }$ is singular, 
\begin{equation}
\ \det \mathbf{R}^{\left( 11\right) }\left( \omega ,k_{x}\right) =0,
\label{R3}
\end{equation}%
and hence admits a null vector $\mathbf{c}_{0}$ defined as $\mathbf{R}%
^{\left( 11\right) }\mathbf{c}_{0}=\mathbf{0}$, which identifies the
amplitudes of the partial modes of the reflectionless incident wave, see (%
\ref{48.00}). Generally, the determinant of $\mathbf{R}^{\left( 11\right) }$
is complex with linearly independent real and imaginary parts, so its zeros
are restricted to isolated values of (real) $\omega $ and $k_{x}.$ Let us
specify the (sufficient) conditions under which Eq. (\ref{R3}) is equivalent
to a real equation.

First, we note that the zeros of $\det \mathbf{R}^{\left( 11\right) }$
coincide with those of the determinant of the left off-diagonal block $%
\mathbf{S}_{3}^{\left( 12\right) }$ of the scattering matrix, as it is seen
from Eq. (\ref{48.S}) and the non-singularity of the transmission matrices
(see below). Next, we assume that the substrates are identical and that the
inhomogeneity profile through the layer is symmetric, so that $\mathbf{S}%
^{\left( 12\right) }$ is a symmetric matrix (see \S \ref{SSec4.2}).
Therefore, its spectral decomposition is of the form $\mathbf{S}^{\left(
12\right) }=\mathbf{\Theta }\mathrm{diag}\left( q_{1}...q_{6}\right) \mathbf{%
\Theta }^{T}$ where the eigenvalues $q_{1...6}$ of $\mathbf{M}\left(
y_{1},y_{2}\right) $ are also those of $\mathbf{S}^{\left( 12\right) }$ and
the matrix of eigenvectors $\mathbf{\Theta =\Xi }^{T}\mathbf{\mathbb{T}W}$
is orthogonal. Besides, we assume that the stiffness tensors of the
substrates and the layer possess a symmetry plane orthogonal to $\mathbf{e}%
_{1}$ or $\mathbf{e}_{2},$ and that the corresponding Eqs. (\ref{31.3}) and (%
\ref{33.1}) hold with $\beta =\alpha +3.$ Consequently, the blocks $\mathbf{%
\Theta }_{1...4}$ of $\mathbf{\Theta }$ satisfy $\mathbf{\Theta }_{1}=-%
\mathbf{\Theta }_{3}$ and $\mathbf{\Theta }_{4}=\mathbf{\Theta }_{1},$ which
leads to 
\begin{equation}
\mathbf{S}_{3}^{\left( 12\right) }=\mathbf{\Theta }_{3}\mathrm{diag}\left(
q_{\alpha }\right) \mathbf{\Theta }_{1}^{T}-\mathbf{\Theta }_{1}\mathrm{diag}%
\left( q_{\alpha +3}\right) \mathbf{\Theta }_{3}^{T}.\   \label{R6}
\end{equation}%
If all partial modes in the substrates and the layer are propagating, i.e. $%
\mathbf{\xi }_{\alpha }$ and $\mathbf{w}_{\alpha }$ are purely imaginary, $%
\mathbf{\xi }_{\alpha +3}$ and $\mathbf{w}_{\alpha +3}$ are real, and $%
q_{\alpha }=q_{\alpha +3}^{\ast }$ (see (\ref{32a})$_{1}$), then (\ref{R6})
yields $\mathbf{S}_{3}=\mathbf{S}_{3}^{+},$ so that $\det \mathbf{S}_{3}$ is
real. Thus, the above conditions reduce Eq. (\ref{R3}) to a real equation
and hence suggest the existence of reflectionless waves on 1D manifolds
(curves) $\omega \left( k_{x}\right) $. The case of the uncoupled SH waves,
which somewhat stands out, is considered in the next subsection.

As mentioned in \S \ref{SSec4.2}, the transmission field cannot totally
vanish and hence the transmission matrices are never singular due to the
wave field continuity at the welded interfaces\footnote{{\footnotesize This
is certainly apart from the exponentially asymptotic decay of the
transmission }$\mathbf{T}${\footnotesize \ at large frequency-thickness
values if the wave packet in the plate does not include propagating modes.
Such a trend may be observed via splitting the scattering matrix }$\mathbf{S}
$ {\footnotesize (\ref{48.S0}) into two parts similarly to (\ref{30a}) and
noting that the triplet of growing eigenvalues }$q_{\alpha }${\footnotesize %
\ causes the blocks of }$\mathbf{S}$ {\footnotesize to grow and hence }$%
\mathbf{T}${\footnotesize \ to decrease.}}. At the same time, the matricant
is no longer continuous if two solids are put in sliding contact, which
therefore admits zero transmission. Interestingly, a sliding-contact
interface may serve as a "universal sonic mirror" in the sense that the
condition ensuring transmission cutoff at certain fixed values of $v=\omega
/k_{x}$ depends, apart from $v$, only on the "host" substrate containing the
incident wave but not at all on the adjusted medium \cite{ShG1}. Another
implication of this idea is the reflection/transmission in a periodic
structure of layers in sliding contact, where the coupling between the Bragg
phenomenon and the above effect of zero transmission leads to some unusual
spectral features \cite{ShG2,HBDRA}.

Zero reflection and transmission on the solid/fluid interfaces is commented
on in \S \ref{SSSec9.3.3}.

\subsubsection{Zero reflection of SH waves\label{SSSec4.5.3}}

The reflection/transmission of SH waves (see \S \ref{SSSec3.3.2}) is much
more amenable than that of the vector waves. Let the SH bulk incident wave
propagate in substrate 1 and impinge on the layer $\left[ 0,H\right] $ at an
angle of incidence less than "critical" so that the SH wave transmitted into
substrate 2 is also bulk. The absolute value of the SH reflection
coefficient reads \cite{K,BG,BSP} 
\begin{equation}
\left\vert R^{\left( 11\right) }\right\vert ^{2}=\frac{A^{2}}{A^{2}+4}\ \ 
\mathrm{with}\ \ A^{2}=\frac{1}{Z_{1}Z_{2}}\left[ \left(
Z_{2}M_{1}-Z_{1}M_{4}\right) ^{2}+\left( Z_{1}Z_{2}\func{Im}M_{3}-\func{Im}%
M_{2}\right) ^{2}\right] ,  \label{R7}
\end{equation}%
where $M_{1},M_{4}$ and $M_{2},M_{3}$ are real diagonal and purely imaginary
off-diagonal elements of the matricant $\mathbf{M}_{w}\left( H,0\right) $ of
(\ref{37}), and 
\begin{equation}
Z_{i}=\sqrt{c_{44}^{\left( i\right) }\left( \rho _{i}-\frac{k_{x}^{2}}{%
\omega ^{2}}C_{55}^{\left( i\right) }\right) },\ \ i=1,2,  \label{R8}
\end{equation}%
are the "dynamic" impedances\footnote{{\footnotesize They reduce to static
values }$Z_{i}=\sqrt{c_{44}^{\left( i\right) }\rho _{i}}${\footnotesize \ at 
}$k_{x}=0,${\footnotesize \ i.e. at the normal incidence.}} of the
substrates, both real as assumed above. Obviously, $\left\vert T^{\left(
12\right) }\right\vert ^{2}=1-\left\vert R^{\left( 11\right) }\right\vert
^{2}$ ($\left\vert R^{\left( 11\right) }\right\vert ^{2}=1$ if $Z_{2}$ is
purely imaginary). According to (\ref{R7}), equation $R^{\left( 11\right)
}=0 $ is equivalent to a system of two real equations%
\begin{equation}
Z_{2}M_{1}=Z_{1}M_{4},\ Z_{1}Z_{2}M_{3}=M_{2}  \label{R9}
\end{equation}%
in unknowns $\omega $ and $k_{x}$. Its (real) solutions, if they exist,
define isolated points of zero reflection on the $\left( \omega
,k_{x}\right) $-plane, which depend on the material parameters of the layer
and also of both substrates. In the case of a symmetric profile (hence $%
M_{1}=M_{4}$, see (\ref{28.2})) and equal substrate impedances $Z_{1}=Z_{2}$%
, condition (\ref{R9}) reduces to one real equation which may be satisfied
on curves $\omega \left( k_{x}\right) $.

Within the set of solutions of (\ref{R9}), there may exist a subset of
points $\left( \omega ,k_{x}\right) $ that render $\mathbf{M}_{w}\left(
H,0\right) $ a scalar matrix, and hence plus or minus $\mathbf{I}$: 
\begin{equation}
\mathbf{M}_{w}\left( H,0\right) =\pm \mathbf{I\ }\Leftrightarrow \ \left\{ 
\begin{array}{c}
M_{1}=M_{4}\left( =\pm 1\right) , \\ 
M_{2}=0,\ M_{3}=0.%
\end{array}%
\right. \Leftrightarrow \left\{ 
\begin{array}{c}
\mathrm{trace}\mathbf{M}_{w}\left( H,0\right) \equiv \Delta \left( \omega
^{2},k_{x}^{2}\right) =\pm 2 \\ 
\partial \Delta /\partial \omega ^{2}=0\ \mathrm{or}\ \partial \Delta
/\partial k_{x}^{2}=0%
\end{array}%
\right. .  \label{R10}
\end{equation}%
By definition of the matricant or, equally, by insertion into (\ref{R9}),
Eq. (\ref{R10}) signifies that the layer is "transparent" for any SH
incident mode and thus provides zero reflection between any substrates with
equal impedances $Z_{1}=Z_{2}$. Besides, a matricant equal to a scalar
matrix admits eigenvectors with zero first (displacement) or second
(traction) component, i.e. Eq. (\ref{R10}) allows for either clamped or
traction-free conditions at the layer faces. To this end, recall the
textbook case of homogeneous layers, for which Eq. (\ref{R10}) reduces to
the "half-lambda" equality $k_{y}\left( \omega ,k_{x}\right) H=\pi n,$ $n\in 
\mathbb{Z}
_{>0}$ that defines a set of dispersion branches realizing simultaneously
the transparency and the traction-free or clamped boundary conditions%
\footnote{{\footnotesize Such coincidence of the transparency and the free
or clamped boundary conditions, sometimes referred to as a "layer
resonance", is a fundamental feature for scalar waves, but is irrelevant to
the vector waves, see \S \ref{SSSec4.5.2}\ .}}. {\footnotesize \ }Once the
plate is inhomogeneous, each dispersion branch splits into a pair of
traction-free and clamped branches, and the transparency condition (\ref{R10}%
) "survives" only at their occasional intersection points. As a theoretical
possibility, the branches of the two types may pairwise merge into "joint"
ones satisfying Eq. (\ref{R10}) and thereby emulating those in a homogeneous
plate (see \cite{KSPN}, where such a branch was termed "zero-width stopband"
or ZWS).

It is instructive to interpret the above through the behavior of the
eigenvalues $q$ and $q^{-1}$ of $\mathbf{M}_{w}\left( H,0\right) $ ($\det 
\mathbf{M}_{w}=1$). According to \S \ref{SSSec3.2.2}, the $\left( \omega
,k_{x}\right) $-plane is partitioned into the passband and stopband areas
where $q$ is complex and $\left\vert q\right\vert =1$ and where $q$ is real
and $\left\vert q\right\vert \neq 1$. These areas are separated by the band
edge curves, at which the eigenvalue degeneracy $q=q^{-1}=\pm 1$ renders $%
\mathbf{M}$ similar to the Jordan block\footnote{{\footnotesize Note that
plugging the spectral decomposition }$\mathbf{M}_{w}=q\mathbf{w}_{1}\mathbf{w%
}_{1}^{T}+q^{-1}\mathbf{w}_{2}\mathbf{w}_{2}^{T}${\footnotesize \ into the
SH }$2\times 2${\footnotesize \ scattering matrix }$\mathbf{S}^{\left(
12\right) }${\footnotesize \ or }$\mathbf{S}^{\left( 12\right) }$%
{\footnotesize \ yields the off-diagonal elements proportional to the
difference }$\left( q-q^{-1}\right) ,${\footnotesize \ but this does not
mean their vanishing at any degeneracy }$q=q^{-1}${\footnotesize \ since the
spectral decomposition in the above form does not apply to }$\mathbf{M}_{w}$%
{\footnotesize \ similar to the Jordan box. }}. There is one traction-free
branch and one clamped branch within any stopband\ or else along its edges
if the profile is symmetric. The edges of the \textit{same }stopband may
occasionally intersect, thus causing the traction-free and one clamped
dispersion branches to intersect too, and the intersection point is where $%
\mathbf{M}$ assumes the scalar matrix form (\ref{R10}).

The layer transparency to SH waves becomes a much more frequent phenomenon
when the layer is periodic. Suppose it consists of $N$ periods $T$. Then,
the eigenvalues of $\mathbf{M}_{w}\left( H,0\right) =\mathbf{M}%
_{w}^{N}\left( T,0\right) $ are equal to $q^{N}$ and $q^{-N}$, where $%
q,~q^{-1}\equiv e^{\pm iKT}$ are the eigenvalues of the monodromy matrix $%
\mathbf{M}_{w}\left( T,0\right) $. It is seen that the eigenvalue degeneracy 
$q^{N}=q^{-N}\left( =\pm 1\right) $ comes about when $q=e^{i\frac{\pi n}{N}%
},~q^{-1}=e^{-i\frac{\pi n}{N}}$ ($\left\vert q\right\vert =1$) at any $%
n=1,2...$ If $n$ is not a multiple of $N,$ then $q\neq q^{-1},$ i.e. $%
\mathbf{M}_{w}\left( T,0\right) $ has distinct eigenvalues and so cannot be
similar to a Jordan block; hence neither can its matrix power function $%
\mathbf{M}_{w}\left( H,0\right) $. Therefore, $\mathbf{M}_{w}\left(
H,0\right) $ is guaranteed to be a scalar matrix satisfying (\ref{R10}) at
any eigenvalue degeneracy $q^{N}=q^{-N}$ unless $q=q^{-1},\ $i.e. at 
\begin{equation}
K\left( \omega ,k_{x}\right) T=\frac{\pi n}{N},\ n=1,2,3...\neq N,
\label{R10a}
\end{equation}%
This may be called "half-\textit{quasi}lambda" condition, which ensures that
the layer is transparent but not that its faces are traction-free or
clamped. The values of $n$ that are multiples of $N$ imply the eigenvalue
degeneracy $q=q^{-1}$ of the matrix $\mathbf{M}_{w}\left( T,0\right) ,$
which only exceptionally may match (\ref{R10}) (the above ZWS option), but
usually becomes similar to the Jordan block; hence so does $\mathbf{M}%
_{w}\left( H,0\right) $ and therefore band edge curves $KT=\pi \mathbb{Z}%
_{>0}$ usually do not support zero reflection.

Thus, the layer periodicity brings in transparency curves defined by Eq. (%
\ref{R10a}) and arising as a set of $\left( N-1\right) $ ones between the
edges $\sin KT=\pm 1$ of each passband. These curves tend to form a dense
cluster with the narrowing of passbands (areas of stable solutions), which
is the case for commensurately large values of $\omega $ and $k_{x}$ (see \S %
19 of \cite{Dem} and Fig. S1 in the Supplemental Material of \cite{ShPG}).

The above speculation on the periodic case may be illustrated explicitly by
the identity for the $N$th power of a $2\times 2$ matrix $\mathbf{C}$ with
unit determinant:%
\begin{equation}
\mathbf{C}^{N}=\frac{\sin NKT}{\sin KT}\mathbf{C}-\frac{\sin \left(
N-1\right) KT}{\sin KT}\mathbf{I,}  \label{R11}
\end{equation}%
where $2\cos KT=\mathrm{trace}\mathbf{C}$. Applying this to Eq. (\ref{R7})
with $\mathbf{M}_{w}\left( H,0\right) =\mathbf{M}_{w}^{N}\left( T,0\right) $
say, for the case of identical substrates ($Z_{1}=Z_{2}\equiv Z$) and
denoting the elements of $\mathbf{M}_{w}\left( T,0\right) $ by $m_{1...4}$
yields 
\begin{equation}
\left\vert R^{\left( 11\right) }\right\vert ^{2}=\dfrac{a^{2}}{a^{2}+4\dfrac{%
\sin ^{2}KT}{\sin ^{2}NKT}}\ \mathrm{with}\ \ 
\begin{array}{c}
a^{2}=\left( Z\func{Im}m_{3}-Z^{-1}\func{Im}m_{2}\right) ^{2}+\left(
m_{1}-m_{4}\right) ^{2}= \\ 
=\left( Z\func{Im}m_{3}+Z^{-1}\func{Im}m_{2}\right) ^{2}-4\sin ^{2}KT,%
\end{array}
\label{R12}
\end{equation}%
where $m_{1},m_{4}$ are real and $m_{2},m_{3}$ are imaginary. It is observed
that $R^{\left( 11\right) }=0$ at $\sin NKT=0,$ i.e. along the $N$-dependent
curves $\omega \left( k_{x}\right) $ (\ref{R10a}), and also at $a=0,$ which
may occur at isolated points $\left( \omega ,k_{x}\right) $ either due to
the vanishing of each of the two perfect squares in the first formula
equivalent to (\ref{R9}) (recall that $m_{1}=m_{4}$ if the profile is
symmetric) or due to the vanishing of $\sin KT$ along with $m_{2}$ or $m_{3}$
(ZWS option, independent of the substrate impedance $Z$).

We conclude with a remark, similar to that at the end of \S \ref{SSSec3.3.3}%
, which now concerns the reflection/transmission of SH waves on a laterally
inhomogeneous obstacle in free strips, see e.g. \cite{BCP}. It is that the
separation of variables approach (\ref{42}) can be used in this problem to
obtain the set of zero-reflection points $\omega _{n}^{\left( m\right) }$
defined by Eq. (\ref{R10a}) with $k_{x}^{\left( m\right) }d=\pi m$ (here $d$
is the strip thickness). Finally, the above analysis is equally relevant to
the reflection/transmission of the (scalar) electromagnetic waves, see \cite%
{YY}.

\pagebreak

\part{Impedance matrix\label{Part2}}

\section{Preamble\label{Sec5}}

Analytical and numerical treatment of the boundary-value problems associated
with Stroh's ODS (\ref{5}) is much facilitated by using the impedance
matrix, sometimes called the stiffness matrix. Generally speaking, the
dynamic stiffness matrix relates the multidimensional vector of
displacements to the corresponding vector of forces, each associated with
the wave field at the given set of coordinates, and, when appropriate, takes
into account the radiation condition (a.k.a. the limiting absorption
principle). It is one of the key concepts of structural vibration analysis,
see e.g. \cite{Ban}. The impedance matrix may be seen as its particular
realization, which is explicitly equipped with far-reaching algebraic and
analytical properties stemming from the Hamiltonian structure of the Stroh
formalism and the link to the energy parameters. These attributes of the
impedance lend direct access to the core aspects of the problem, such as the
existence/non-existence of the wave solutions and some of their fundamental
characteristics, which are hardly reachable through explicit derivations.
The impedance approach also fosters efficient computation and allows
circumventing the exponential dichotomy problem inherent to the transfer
matrix at a large frequency-thickness product.

The surface impedance matrix underpinning the theory of surface (Rayleigh)
waves in a homogeneous half-space was introduced in the seminal papers \cite%
{IT,LB} and honed to completion in \cite{BL85}. We revisit these classical
results in \S \ref{Sec6}, where some original derivations are streamlined
with a view to facilitate their further generalization. Indeed, it turns out
that the above Barnett and Lothe's methodology can be fruitfully adapted to
studying surface waves in transversely and laterally periodic half-spaces 
\cite{DSh18, ShWM}. This extension of the surface impedance method is
described in \S \S \ref{Sec7} and \ref{Sec8}.

While the surface (half-space) impedance must keep the surface-localized
modes and discard divergent ones, i.e. it requires "dismantling" the
fundamental matrix solution, the plate (two-point) impedance involves all
modes and thus admits a relatively straightforward definition linked to the
matricant of (\ref{5}). Its properties and application to analyzing the wave
dispersion spectra in homogeneous and transversely inhomogeneous plates are
considered in \S \ref{Sec9}.

Note also the interesting implications of the two-point impedance in the
cylindrical coordinates \cite{NSh2}{\small \ }and a somewhat different
impedance approach resting on the normal modes in laterally inhomogeneous
plates \cite{PM}. A rigorous mathematical analysis of the two-point boundary
problem for the acoustic-wave equation can be found in \cite{SK}.

Throughout Part II, the material characteristics $\rho ,~c_{ijkl}$ and the
parameters $\omega ,~k_{x}$ are assumed to be real. We recall the notation $%
\mathbf{\eta }=\left( \mathbf{a}\ \mathbf{b}\right) ^{T}$ embracing
different explicit forms (\ref{8}) of the displacement-traction state vector
of Eq. (\ref{5}). The same letters $\mathbf{Z}$ and$\mathbf{~Y=Z}^{-1}$ will
denote explicitly different impedance and admittance matrices in various
settings, including homogeneous or periodic half-spaces and a plate (the
specific context of discussion must preclude confusion).

\section{Impedance of a homogeneous half-space\label{Sec6}}

\subsection{Definition\label{SSec6.1}}

Given an anisotropic homogeneous medium, consider ODS (\ref{5}) with a
constant $6\times 6$ system matrix $\mathbf{Q}_{0}$ and hence with partial
solutions $\mathbf{\eta }_{\alpha }\left( y\right) =\left( \mathbf{a}%
_{\alpha }\ \mathbf{b}_{\alpha }\right) ^{T}=\mathbf{\xi }_{\alpha
}e^{ik_{y\alpha }y},$ where $\mathbf{\xi }_{\alpha }=\left( \mathbf{A}%
_{\alpha }\ \mathbf{B}_{\alpha }\right) ^{T}$ and $ik_{y\alpha }$ are the
eigenvectors and eigenvalues of $\mathbf{Q}_{0}$ (see (\ref{11.01})). The
velocity $v=\omega /k_{x}$ will be restricted to the subsonic interval $v<%
\hat{v},$ in which all $k_{y\alpha }$'s are complex and hence pairwise
complex conjugated. For definiteness, assume $\mathbf{Q}_{0}$ in pure
imaginary form (\ref{8}) ensuring that the corresponding eigenvectors $%
\mathbf{\xi }_{\alpha }$ are also pairwise complex conjugated. Let them be
arranged in triplets (\ref{28*}) where $\kappa _{\alpha }\equiv k_{y\alpha }$
and $\func{Im}k_{y\alpha }>0,\ \alpha =1,2,3.$ Recall (see \S \ref%
{SSSec3.2.1}) that the matrix $\mathbf{Q}_{0}$ at $v=\hat{v}$ is
non-semisimple and has a self-orthogonal eigenvector satisfying $\mathbf{\xi 
}_{\deg }^{T}\mathbf{\mathbb{T}\xi }_{\deg }=2\mathbf{A}_{\deg }^{T}\mathbf{B%
}_{\deg }=0$; if incidentally $\mathbf{B}_{\deg }=\mathbf{0}$, i.e. if the
so-called limiting wave $\mathbf{\xi }_{\deg }=\mathbf{\xi }_{\deg
}e^{ik_{y,\deg }y}$ fulfils the traction-free boundary condition $\mathbf{t}%
_{2}=\mathbf{0}$, then both this wave and the transonic state $v=\hat{v}$
are called exceptional \cite{ChadS}.

The impedance associated with the governing wave equation (\ref{3}) is
broadly understood as a matrix linking the elastic displacement $\mathbf{u}$
and the exerted traction $\mathbf{t}_{2}$. Let us look at this concept more
closely. According to Eq. (\ref{4})$_{2}$ applied to a single partial mode $%
\mathbf{\eta }_{\alpha }\left( y\right) $ (\ref{11.01}), the amplitudes of
its traction $\mathbf{b}_{\alpha }$ and displacement $\mathbf{a}_{\alpha }$
are related by a (constant) matrix $k_{y\alpha }\left( e_{2}e_{2}\right)
+k_{x}\left( e_{2}e_{1}\right) ,$ but it depends on $\alpha $, i.e. is
different for different $\alpha $th modes, and thereby unsuitable for
describing a wave packet. In turn, traction and displacement of an arbitrary
superposition of several partial modes (\ref{11.01}) generally (for sure if
the number of modes exceeds three) cannot be related by a constant matrix.
This perspective allows us to better appreciate the concept of the impedance
matrix introduced by Ingebrigtsen and Tonning \cite{IT} and further
developed by Lothe and Barnett \cite{LB}. It is concerned, specifically,
with a subsonic wave field travelling along the surface of a half-space and
represented by an arbitrary superposition of three exponentially decreasing
(evanescent) partial modes. Assuming the half-space $y\geq 0$, this
surface-localized wave field is defined as 
\begin{equation}
\mathbf{\eta }\left( y\right) =\left( 
\begin{array}{c}
\mathbf{a}\left( y\right) \\ 
\mathbf{b}\left( y\right)%
\end{array}%
\right) =\sum\nolimits_{\alpha =1}^{3}c_{\alpha }\left( 
\begin{array}{c}
\mathbf{A}_{\alpha } \\ 
\mathbf{B}_{\alpha }%
\end{array}%
\right) e^{ik_{y\alpha }y}=\left( 
\begin{array}{c}
\mathbf{\Xi }_{1} \\ 
\mathbf{\Xi }_{3}%
\end{array}%
\right) \mathrm{diag}\left( e^{ik_{y\alpha }y}\right) \mathbf{c,\ \ }
\label{43}
\end{equation}%
where the $3\times 3$ matrices $\mathbf{\Xi }_{1}=\left\Vert \mathbf{A}_{1}%
\mathbf{A}_{2}\mathbf{A}_{3}\right\Vert $ and $\mathbf{\Xi }_{3}=\left\Vert 
\mathbf{B}_{1}\mathbf{B}_{2}\mathbf{B}_{3}\right\Vert $ are the left
diagonal and off-diagonal blocks of the $6\times 6$ matrix of eigenvectors $%
\mathbf{\Xi }$ (\ref{28-1}) (see (\ref{0})), $\func{Im}k_{y\alpha }>0$ for $%
\alpha =1,2,3,$ and $\mathbf{c}$ is a vector of disposable constants $%
c_{\alpha }.$ The sought impedance $\mathbf{Z}$ is uniquely determined by
either of the equivalent definitions \cite{IT,LB}%
\begin{equation}
\mathbf{b}\left( y\right) =-i\mathbf{Za}\left( y\right) \ \Leftrightarrow \ 
\mathbf{B}_{\alpha }=-i\mathbf{ZA}_{\alpha },\ \alpha =1,2,3.  \label{44.0}
\end{equation}%
By this definition, $\mathbf{Z}$ is independent of $y$ and of the modal
index $\alpha $. From (\ref{43}) and (\ref{44.0}), $\mathbf{Z}$ and its
inverse, the admittance $\mathbf{Z}^{-1}\mathbf{\equiv Y},$ are expressed in
the form 
\begin{equation}
\mathbf{Z}=i\mathbf{\Xi }_{3}\mathbf{\Xi }_{1}^{-1}\left( =\mathbf{Z}%
^{+}\right) ,\ \mathbf{Y}=-i\mathbf{\Xi }_{1}\mathbf{\Xi }_{3}^{-1}\left( =%
\mathbf{Y}^{+}\right) ,  \label{44}
\end{equation}%
where the Hermiticity follows from identity (\ref{29}) taken with $\mathbf{%
\mathbb{E}}_{\mathbf{Q}}=\mathbf{\mathbb{T}}$. The vectors $\mathbf{B}%
_{\alpha }$ and $\mathbf{A}_{\alpha }$ are homogeneous functions of $\omega $
and $k_{x}$, hence so are $\mathbf{Z}$ and $\mathbf{Y}$ (\ref{44}) \cite{IT}%
. They may always be chosen to be of degree zero so that $\mathbf{\Xi =\Xi }%
\left[ v\right] ,$ $\mathbf{Z=Z}\left[ v\right] $ and $\mathbf{Y=Y}\left[ v%
\right] $ where $v=\omega /k_{x}$. This is understood below by default,
unless explicitly specified.

Complex conjugating Eq. (\ref{44.0}) and taking into account (\ref{28*})
shows that the traction and displacement parts of the exponentially
divergent wave field $\mathbf{\eta }\left( y\right) =\sum\nolimits_{\alpha
=1}^{3}c_{\alpha +3}\mathbf{\xi }_{\alpha }^{\ast }e^{ik_{y\alpha }^{\ast
}y} $ are related by the matrices 
\begin{equation}
\mathbf{Z}^{\prime }=-i\mathbf{\Xi }_{4}\mathbf{\Xi }_{2}^{-1}=\mathbf{Z}%
^{\ast },\ \mathbf{Y}^{\prime }=i\mathbf{\Xi }_{2}\mathbf{\Xi }_{4}^{-1}=%
\mathbf{Y}^{\ast },  \label{46}
\end{equation}%
where $\mathbf{\Xi }_{2,4}=\mathbf{\Xi }_{1,3}^{\ast }$ are the right-side
blocks of $\mathbf{\Xi }$, and the primes are not to be confused with
differentiation. Clearly, the matrices $\mathbf{Z}^{\prime }$ and $\mathbf{Y}%
^{\prime }$ are also Hermitian. They may be called "non-physical" impedance
and admittance relative to the half-space $y\geq 0$ in the sense that they
connect the divergent modes; however, their implication jointly with
"physical" ones (\ref{44}) enables one to fully exploit the properties of
the Stroh formalism when analyzing surface waves, see below. It is also
understood that the above-specified attribution of the quantities (\ref{44})
and (\ref{46}) as "physical" and "non-physical" swaps when they are
considered in the half-space $y\leq 0.$

\subsection{Properties\label{SSec6.2}}

Diagonal and off-diagonal blocks of the orthonormality relation (\ref{28})
yield the equalities 
\begin{equation}
\mathbf{Z}^{\prime }=\mathbf{Z}^{T},\ \ \mathbf{Y}^{\prime }=\mathbf{Y}^{T}
\label{48.1}
\end{equation}%
and 
\begin{equation}
\mathbf{Z}+\mathbf{Z}^{T}=i\left( \mathbf{\Xi }_{1}\mathbf{\Xi }%
_{1}^{T}\right) ^{-1},\ \mathbf{Y}+\mathbf{Y}^{T}=-i\left( \mathbf{\Xi }_{3}%
\mathbf{\Xi }_{3}^{T}\right) ^{-1}.  \label{48.2}
\end{equation}%
Note that Eq. {\small (\ref{48.1}) }is also evident from (\ref{46}) taken in
the subsonic interval where the impedances are Hermitian; however, both (\ref%
{48.1}) and (\ref{48.2}) with entries (\ref{44}) are not restricted to the
subsonic velocity, since neither is their root cause (\ref{28}).

Introduce the matrix formed by the dyadic products of the eigenvectors of $%
\mathbf{Q}_{0}$ as follows: 
\begin{equation}
\mathbf{\Upsilon }\left[ v\right] =i\mathbf{\Xi }\mathrm{diag}\left( \mathbf{%
I,}-\mathbf{I}\right) \mathbf{\Xi }^{-1}=i\left( 
\begin{array}{cc}
-\mathbf{I}+2\mathbf{\Xi }_{1}\mathbf{\Xi }_{3}^{T} & 2\mathbf{\Xi }_{1}%
\mathbf{\Xi }_{1}^{T} \\ 
2\mathbf{\Xi }_{3}\mathbf{\Xi }_{3}^{T} & -\mathbf{I}+2\mathbf{\Xi }_{3}%
\mathbf{\Xi }_{1}^{T}%
\end{array}%
\right) \equiv \left( 
\begin{array}{cc}
\mathbf{\Upsilon }_{1} & \mathbf{\Upsilon }_{2} \\ 
\mathbf{\Upsilon }_{3} & \mathbf{\Upsilon }_{1}^{T}%
\end{array}%
\right) .  \label{50}
\end{equation}%
By construction, $\mathbf{\Upsilon }$ has zero trace and its off-diagonal
blocks $\mathbf{\Upsilon }_{2},\ \mathbf{\Upsilon }_{3}$ are symmetric;
moreover, the latter are real in the subsonic interval due to (\ref{28})
with $\mathbf{\Xi }_{2,4}=\mathbf{\Xi }_{1,3}^{\ast }.$ Combining (\ref{44})
and (\ref{50}) yields 
\begin{equation}
\mathbf{Z}=-\mathbf{\Upsilon }_{2}^{-1}\left( \mathbf{I}+i\mathbf{\Upsilon }%
_{1}\right) ,\ \mathbf{Y}=\mathbf{\Upsilon }_{3}^{-1}\left( \mathbf{I}+i%
\mathbf{\Upsilon }_{1}^{T}\right) .  \label{53}
\end{equation}%
In particular, it is seen from (\ref{53}) that 
\begin{equation}
\func{Re}\mathbf{Z}=-\mathbf{\Upsilon }_{2}^{-1},\ \func{Re}\mathbf{Y}=%
\mathbf{\Upsilon }_{3}^{-1},  \label{48.21}
\end{equation}%
which agrees with (\ref{48.2}) specified to the subsonic interval. It can be
shown that the components of $\mathbf{\Upsilon }_{2}$ and $\mathbf{\Upsilon }%
_{3}$ are finite\footnote{{\footnotesize For example, a proof by
contradiction proceeds from the equalities }$\mathbf{\Upsilon }_{2}=-2\left( 
\mathbf{Z}+\mathbf{Z}^{\prime }\right) ^{-1},${\footnotesize \ }$\mathbf{%
\Upsilon }_{3}=2\left( \mathbf{Y}+\mathbf{Y}^{\prime }\right) ^{-1}$%
{\footnotesize \ (see (\ref{48.1})-(\ref{50})) and observes that the
matrices }$\mathbf{Z}+\mathbf{Z}^{\prime }${\footnotesize \ and }$\mathbf{Y}+%
\mathbf{Y}^{\prime }${\footnotesize \ cannot be singular at }$v<\hat{v}$%
{\footnotesize \ as the opposite would lead to a senseless conclusion of the
existence of a wave solution localized on an arbitrary plane }$y=const$%
{\footnotesize \ inside the infinite space (see (\ref{48*})}$_{2}$%
{\footnotesize \ in \S 6.5). The same consideration applies to the cases of
transversely and laterally periodic half-spaces, see \S \S 7,8.}} at $v<\hat{%
v}$ but diverge at $v=\hat{v},$ except for $\mathbf{\Upsilon }_{3}$ if $\det 
\mathbf{\Xi }_{3}\left[ \hat{v}\right] =0$. The pole of $\mathbf{\Upsilon }%
_{2}$ and $\mathbf{\Upsilon }_{3}$ at $v=\hat{v}$ is due to the
non-semisimple form of the system matrix $\mathbf{Q}_{0}\left[ \hat{v}\right]
$ (though $\mathbf{\Upsilon }_{2}$ and $\mathbf{\Upsilon }_{3}$ remain
finite if non-semisimple $\mathbf{Q}_{0}$ occurs at $v\neq \hat{v},$ see the
details in \cite{T}).

An essential attribute of the above impedance concept is its link to the
energy quantities, which results in a specific sign-definiteness of the
(Hermitian) matrices $\mathbf{Z}\left[ v\right] $ and $\mathbf{Y}\left[ v%
\right] ,$ established in \cite{IT,LB,BL85} and detailed in Appendix 2. Let
us formulate it with respect to the system matrix $\mathbf{Q}$ defined as in
(\ref{8})$_{1}$ or (\ref{8})$_{3}$ (the use of (\ref{8})$_{2}$ leads to all
signs being opposite). It follows that

\begin{equation}
\begin{array}{c}
\mathbf{Z}\ \mathrm{is}\ \func{positive}\ \mathrm{definite\ at}\ v=0,\ 
\dfrac{\mathrm{d}\mathbf{Z}}{\mathrm{d}v}\ \mathrm{is}\ \func{negative}\ 
\mathrm{definite;} \\ 
\mathbf{Y}\ \mathrm{is}\ \func{positive}\ \mathrm{definite\ at}\ v=0,\ 
\dfrac{\mathrm{d}\mathbf{Y}}{\mathrm{d}v}\ \mathrm{is}\ \func{positive}\ 
\mathrm{definite.}%
\end{array}
\label{47.1}
\end{equation}%
In consequence, by (\ref{48.21}), the real symmetric matrices $\mathbf{%
\Upsilon }_{2}=2i\mathbf{\Xi }_{1}\mathbf{\Xi }_{1}^{T}$ and $\mathbf{%
\Upsilon }_{3}=2i\mathbf{\Xi }_{3}\mathbf{\Xi }_{3}^{T}$ satisfy the
following properties: 
\begin{equation}
\begin{array}{c}
\mathbf{\Upsilon }_{2}~\mathrm{is}\ \func{negative}\ \mathrm{definite\ at}\
v=0,\ \dfrac{\mathrm{d}\mathbf{\Upsilon }_{2}}{\mathrm{d}v}\ \mathrm{is}\ 
\func{negative}\ \mathrm{definite;} \\ 
\mathbf{\Upsilon }_{3}~\mathrm{is}\ \func{positive}\ \mathrm{definite\ at}\
v=0,\ \dfrac{\mathrm{d}\mathbf{\Upsilon }_{3}}{\mathrm{d}v}\ \mathrm{is}\ 
\func{negative}\ \mathrm{definite.}%
\end{array}
\label{47.2}
\end{equation}%
From (\ref{47.2}) and the finiteness of $\mathbf{\Upsilon }_{2}$ and$\ 
\mathbf{\Upsilon }_{3}$ at $v<\hat{v}$, \ one concludes that their
eigenvalues are \textit{continuously} decreasing functions of $v<\hat{v},$
and so $\mathbf{\Upsilon }_{2}$ is negative definite throughout the subsonic
interval. Two principal corollaries follow. First, by (\ref{48.21}), the
matrix $\func{Re}\mathbf{Z}$ is positive definite at $v<\hat{v}$. Second, $%
\mathbf{\Xi }_{1}$ and hence $\mathbf{Y}$ are non-singular and therefore, by
(\ref{47.1})$_{1}$, the eigenvalues of $\mathbf{Z}$ are \textit{continuously}
decreasing functions of subsonic $v$.

\subsection{Direct evaluation of the impedance\label{SSec6.3}}

Whereas evaluating the impedance $\mathbf{Z}$ from Eq. (\ref{44}) requires
the preliminary step of finding the eigenvectors of the system matrix $%
\mathbf{Q}_{0},$ it also appears possible to calculate $\mathbf{Z}$
directly, that is, by an explicit formula expressed in material constants of
the medium. The first way to achieve this is through the integral method of
Barnett and Lothe \cite{BL73}-\cite{ChadS}. Based on the system matrix in
the form $\mathbf{Q}_{0}=ik\mathbf{N}_{0}\left[ v\right] $ (\ref{8})$_{1},$
the method proceeds from the angular average 
\begin{equation}
\left\langle \mathbf{N}_{\varphi }\left[ v\right] \right\rangle =\frac{1}{%
\pi }\int_{0}^{\pi }\mathbf{N}_{\varphi }\left[ v\right] \mathrm{d}\varphi
\label{51}
\end{equation}%
of the matrix $\mathbf{N}_{\varphi }\left[ v\right] $ defined similarly to $%
\mathbf{N}_{0}\left[ v\right] $, except that the frame of vectors $\left( 
\mathbf{e}_{1},\mathbf{e}_{2}\right) $ is not fixed but rotates by the angle 
$\varphi $ within the fixed sagittal plane. By construction, the blocks of $%
\left\langle \mathbf{N}_{\varphi }\right\rangle $ are $3\times 3$ real
matrices that are finite at $v<\hat{v}$ but diverge at $v\rightarrow \hat{v}%
; $ also, the left and right off-diagonal blocks are, respectively, positive
and negative definite at $v=0$. Numerical integration in (\ref{51}) may be
realized iteratively, see \cite{GL,CK}.

Remarkably, the matrix $\left\langle \mathbf{N}_{\varphi }\right\rangle $ (%
\ref{51}) considered in the subsonic velocity interval $v<\hat{v}$ satisfies
the eigenrelation of the form 
\begin{equation}
\left\langle \mathbf{N}_{\varphi }\right\rangle \mathbf{\Xi =\Xi }\mathrm{%
diag}\left( i\mathbf{I,}-i\mathbf{I}\right) ,  \label{52}
\end{equation}%
and hence equals the matrix $\mathbf{\Upsilon }$ (\ref{50}) consisting of
dyads of the eigenvectors of $\mathbf{Q}_{0}$, i.e. 
\begin{equation}
\left\langle \mathbf{N}_{\varphi }\right\rangle =\mathbf{\Upsilon }.
\label{49}
\end{equation}%
The equivalence of the eigenvector and integral representations of the same
matrix $\left\langle \mathbf{N}_{\varphi }\right\rangle $ is a key point of
the Barnett-Lothe development. In particular, plugging Eq. (\ref{49}) into (%
\ref{53}) expresses the impedance $\mathbf{Z}$ via the blocks of $%
\left\langle \mathbf{N}_{\varphi }\right\rangle $ and thus enables its
numerical evaluation directly from the given material data. Furthermore, the
above sign properties of $\left\langle \mathbf{N}_{\varphi }\right\rangle $
provide an alternative proof (actually, the original Barnett-Lothe's one) of
statements (\ref{47.1}), (\ref{47.2}) and below them.

Another approach to finding the impedance directly from the system matrix is
to numerically solve the Riccati matrix equation in $\mathbf{Z}$, which
follows from differentiating (\ref{44.0}) and invoking (\ref{5}) \cite{Bir}.
In the case of a constant system matrix $\mathbf{Q}_{0},$ this equation
takes an algebraic form%
\begin{equation}
\mathbf{ZQ}_{02}\mathbf{Z}+i\mathbf{ZQ}_{01}-i\mathbf{Q}_{01}^{T}\mathbf{Z}+%
\mathbf{Q}_{03}=\mathbf{0,}  \label{54}
\end{equation}%
where $\mathbf{Q}_{0i}$ are the blocks of $\mathbf{Q}_{0}$. Fu and Mielke 
\cite{FM} have shown that Eq. (\ref{54}) has a unique Hermitian solution for 
$\mathbf{Z}\left[ v\right] $ that is positive definite at $v<v_{\mathrm{R}%
}\left( <\hat{v}\right) $. The link of this approach to Barnett and Lothe's
integral method was established in \cite{MF}.

More recently, a somewhat reconciling look at the problem was taken in \cite%
{NSK} by proceeding from the sign function of the matrix $i\mathbf{N}_{0}.$
It satisfies the same eigenrelation as that for $i\left\langle \mathbf{N}%
_{\varphi }\right\rangle ,$ see (\ref{52}); hence, $\mathrm{sign}\left( i%
\mathbf{N}_{0}\right) $ is independently equal to $i\left\langle \mathbf{N}%
_{\varphi }\right\rangle $ and $i\mathbf{\Upsilon ,}$ thus recovering (\ref%
{49}). This perspective also provides an analytical proof that $\mathbf{Z}$
given by (\ref{53})$_{1}$ is a root of Eq. (\ref{54}).

\subsection{Rayleigh wave\label{SSec6.4}}

Consider the half-space $y\geq 0$ with a traction-free surface $y=0$. The
surface, or Rayleigh, wave (RW) is described by Eq. (\ref{43}) subjected to
the boundary condition $\mathbf{t}_{2}\left( 0\right) =\mathbf{0}$, i.e. $%
\mathbf{b}\left( 0\right) =\mathbf{\Xi }_{3}\left[ v\right] \mathbf{c}=%
\mathbf{0.}$ The resulting dispersion equation defining the RW velocity $v_{%
\mathrm{R}}$ can be expressed via the impedance (\ref{44})$_{1}$ as 
\begin{equation}
\det \mathbf{Z}\left[ v\right] =0,  \label{45}
\end{equation}%
where the left-hand side is irrational function real at $v<\hat{v}$ (due to $%
\mathbf{Z=Z}^{+}$). It can be evaluated for each given set of material
constants by one or another method outlined in the previous subsection;
however, the analytical solution of Eq. (\ref{45}) is generally out of
reach. All the more powerful are Barnett-Lothe theorems of the existence and
uniqueness of the subsonic RW with $v_{\mathrm{R}}<\hat{v}$ in an arbitrary
anisotropic material. The main points of their proof, based on the
sign-definite properties of the impedance recapped in \S \ref{SSec6.2}, are
as follows.

By (\ref{45}), the subsonic RW comes about due to the vanishing of any
eigenvalue of $\mathbf{Z}\left[ v\right] $ at $v<\hat{v}.$ The primary point
is that, by (\ref{47.1}), all three eigenvalues are positive at $v=0$ and
decrease continuously at $v\leq \hat{v};$ hence, it suffices to examine
their signs at the transonic state $v=\hat{v}$. The self-orthogonality
relation $\mathbf{\xi }_{\deg }^{T}\mathbf{\mathbb{T}\xi }_{\deg }=0$ (see 
\S \ref{SSec6.1}) implies $\mathbf{A}_{\deg }^{T}\mathbf{Z}\left[ \hat{v}%
\right] \mathbf{A}_{\deg }=0$ with real $\mathbf{A}_{\deg }$, hence $\mathbf{%
Z}\left[ \hat{v}\right] $ can neither be strictly negative nor strictly
positive. The former bars the occurrence of three subsonic surface waves
(which is yet far too weak a statement, see below); the latter allows
claiming that at least one wave exists unless the limiting wave solution $%
\mathbf{\xi }_{\deg }$ is exceptional (i.e. unless $\mathbf{B}_{\deg }=%
\mathbf{0}$ hence $\det \mathbf{Z}\left[ \hat{v}\right] =0,$ in which case $%
\mathbf{Z}\left[ \hat{v}\right] $ may not be positive definite even in the
absence of zero eigenvalues at $v<\hat{v}$). Next, by using the positive
definiteness of the matrix $\func{Re}\mathbf{Z}$ (see below (\ref{47.2})),
it may be shown \cite{LB} that in fact no more than one (subsonic) RW is
possible. A slightly modified speculation resting on the properties (\ref%
{47.2}) of the matrix $\mathbf{\Upsilon }_{3}\left[ v\right] $ and the
observation that its zero eigenvalue is always double was developed in \cite%
{BL74,ChadS}. Later on, the study was extended to all possible types of
transonic states $v=\hat{v}$ with more than one degenerate eigenvalue $%
k_{y,\deg }$ and hence more than one limiting wave. Within this broader
context, a comprehensive statement of the existence theorem established in 
\cite{BL85} (see also \cite{B,BLG}) reads that the RW must exist if the
transonic state $v=\hat{v}$ is \textit{normal} in the sense that it admits
at least one limiting wave which is not exceptional.

It is of interest to mention the essentially different proofs of the RW
uniqueness theorem established in \cite{MF} and \cite{KK}.

\subsection{Related boundary-value problems\label{SSec6.5}}

Note first that the clamped boundary condition, which implies zero
displacement $\mathbf{a}\left( 0\right) =\mathbf{\Xi }_{1}\left[ v\right] 
\mathbf{c}=\mathbf{0}$ (see (\ref{43})), rules out the existence of subsonic
surface waves since, in view of (\ref{47.2}), $\mathbf{\Xi }_{1}$ is
non-singular at $v\leq \hat{v}.$

Further, assume two rigidly bonded homogeneous materials, labelled 1 and 2,
which occupy the half-spaces $y\geq 0$ and $y\leq 0$, respectively. The
velocity $v_{\mathrm{St}}<\min \left( \hat{v}^{\left( 1\right) },\hat{v}%
^{\left( 2\right) }\right) \equiv \hat{v}^{\left( 12\right) }$ of the
interfacial (Stoneley) wave vanishing at $y\rightarrow \pm \infty $ must
match the equality of the interface values $\mathbf{\eta }^{\left( 1\right)
}\left( 0\right) =\mathbf{\eta }^{\left( 2\right) }\left( 0\right) $ of the
two surface-localized wave fields built of the triplets of modes with $\func{%
Im}k_{y\alpha }^{\left( 1\right) }>0$ and $\func{Im}k_{y,\alpha +3}^{\left(
2\right) }<0$, respectively. Hence, the dispersion equation may be written
in the form \cite{BLGM} 
\begin{equation}
\det \left( 
\begin{array}{cc}
\mathbf{\Xi }_{2}^{\left( 1\right) } & \mathbf{\Xi }_{1}^{\left( 2\right) }
\\ 
\mathbf{\Xi }_{4}^{\left( 1\right) } & \mathbf{\Xi }_{3}^{\left( 2\right) }%
\end{array}%
\right) =0\Leftrightarrow \det \left( \mathbf{Z}^{\left( 1\right) T}\left[ v%
\right] +\mathbf{Z}^{\left( 2\right) }\left[ v\right] \right) =0,
\label{48*}
\end{equation}%
where the equality $\mathbf{Z}^{\left( 1\right) \prime }=\mathbf{Z}^{\left(
1\right) T}$ is used (see (\ref{48.1})$_{1}$). By (\ref{48*})$_{2}$ and the
positive-definiteness of $\mathbf{Z}\left[ v\right] $ at $v<v_{\mathrm{R}}$,
if the Stoneley wave exists, it is unique and its velocity $v_{\mathrm{St}}$
is greater than the least of the Rayleigh velocities $v_{\mathrm{R}}^{\left(
1\right) }$ and $v_{\mathrm{R}}^{\left( 2\right) }$ in the adjacent
half-spaces; accordingly, no (subsonic) Stoneley wave exists if both $v_{%
\mathrm{R}}^{\left( 1\right) }$ and $v_{\mathrm{R}}^{\left( 2\right) }$ are
greater than $\hat{v}^{\left( 12\right) }$ \cite{BLGM}. It may be added
that, by Weyl's inequality, the sufficient condition for the Stoneley wave
existence is the negativeness of the sum of the greatest eigenvalue of
either one of the impedances $\mathbf{Z}^{\left( 1\right) },$ $\mathbf{Z}%
^{\left( 2\right) }$ and the least eigenvalue of the other, both taken at $%
\hat{v}^{\left( 12\right) }.$ This inequality appears to be fairly
restrictive; in fact, unlike the Rayleigh wave case, "allowed Stoneley wave
propagation is usually the exception and not the rule" \cite{B}.

A similar boundary-value problem at the sliding-contact interface of two
homogeneous half-spaces was studied in \cite{BGL}. Break of continuity
across such an interface turns out to be fairly consequential: the existence
of localized (slip) waves was shown to be a much more general feature than
that of the Stoneley wave; moreover, two such waves at the sliding-contact
interface are admitted (see \cite{WL}). As merely a by-product (!), the
paper \cite{BGL} also proves that the interface between homogeneous solid
and fluid half-spaces always supports at least one, and possibly two,
localized wave solutions (the Scholte waves). Remarkably, the above
fundamental results were obtained in \cite{BLGM,BGL} by solely appealing to
the general properties of the impedance matrix without any additional
calculations.

In conclusion, we reiterate that the discussion in this Section, based on
the surface impedance properties, is related to the subsonic range $v<\hat{v}
$ where none of the partial modes is bulk (propagating). The supersonic
surface or interface waves have, so to say, a reduced number of evanescent
modes available for matching the boundary condition. Hence, apart from the
case of the SH uncoupling in the symmetry plane (see \S \ref{SSSec3.3.2}),
they waves come about as relatively rare \textit{secluded} occasions (in
terms of spectral theory, they imply embedded eigenvalues in the continuous
spectrum). General conditions for the existence of supersonic waves were
analyzed in \cite{GWL,DALLS} and specialized for cubic crystals in \cite%
{ME,E}. Their implication in the reflection/transmission resonant phenomena
was mentioned in \S \ref{SSSec4.5.1}.

\section{Impedance of a transversely periodic half-space\label{Sec7}}

\subsection{Definition\label{SSec7.1}}

Consider a functionally graded and/or layered half-space $y\geq 0$ such that
the variation of its elastic properties $\rho \left( y\right) $ and $%
c_{ijkl}\left( y\right) $ is periodic with a period $T$ and hence so is the
system matrix $\mathbf{Q}\left( y\right) =\mathbf{Q}\left( y+T\right) $ of
Eq. (\ref{5}). Let $\mathbf{Q}$ have any explicit form satisfying (\ref{24}%
). The values of $\omega $ and $k_{x}$ will be assumed to lie in the (full)
stopbands, i.e. such areas of the $\left( \omega ,k_{x}\right) $-plane,
where the set of eigenvalues of the monodromy matrix $\mathbf{M}\left(
T,0\right) $ (\ref{13}) splits into two triplets (\ref{30.01})$_{2}$
satisfying $\left\vert q_{\alpha }\right\vert =\left\vert q_{\alpha
+3}^{-1}\right\vert <1$,\ $\alpha =1,2,3$.

For the wave field $\mathbf{\eta }\left( y\right) $ to asymptotically vanish
at $y\rightarrow \infty $, its value at the surface $y=0$ must be of the
form $\mathbf{\eta }\left( 0\right) =\sum\nolimits_{\alpha =1}^{3}c_{\alpha }%
\mathbf{w}_{\alpha }q_{\alpha },$ where $\mathbf{w}_{\alpha }\equiv \left( 
\mathbf{u}_{\alpha }\mathbf{\ v}_{\alpha }\right) ^{T}$ are eigenvectors of $%
\mathbf{M}\left( T,0\right) $ corresponding to its eigenvalues $\left\vert
q_{\alpha }\right\vert <1,$ $\alpha =1,2,3.$ In consequence, the
surface-localized wave field evaluated at the period edges $y=nT$ is%
\begin{equation}
\mathbf{\eta }\left( nT\right) =\left( 
\begin{array}{c}
\mathbf{a}\left( nT\right) \\ 
\mathbf{b}\left( nT\right)%
\end{array}%
\right) =\mathbf{M}^{n}\left( T,0\right) \mathbf{\eta }\left( 0\right)
=\left( 
\begin{array}{c}
\mathbf{W}_{1} \\ 
\mathbf{W}_{3}%
\end{array}%
\right) \mathrm{diag}\left( q_{\alpha }^{n}\right) \mathbf{c,}  \label{55}
\end{equation}%
where $\mathbf{W}_{1}=\left\Vert \mathbf{u}_{1}\mathbf{u}_{2}\mathbf{u}%
_{3}\right\Vert $ and $\mathbf{W}_{3}=\left\Vert \mathbf{v}_{1}\mathbf{v}_{2}%
\mathbf{v}_{3}\right\Vert $ are the left diagonal and off-diagonal blocks of
the eigenvector matrix (\ref{30.00}), and $\mathbf{c}$ is the vector of
disposable coefficients $c_{\alpha }$. The impedance $\mathbf{Z}\left[
\omega ,k_{x}\right] $ and admittance $\mathbf{Z}^{-1}\equiv \mathbf{Y}\left[
\omega ,k_{x}\right] $ of a transversely periodic half-space $y\geq 0$ are
defined as the matrices relating the traction and displacement parts of the
vector $\mathbf{\eta }\left( nT\right) $ (\ref{55}) so that%
\begin{equation}
\mathbf{b}\left( nT\right) =-i\mathbf{Za}\left( nT\right) \Leftrightarrow \ 
\mathbf{v}_{\alpha }=-i\mathbf{Zu}_{\alpha },  \label{55.0}
\end{equation}%
and therefore%
\begin{equation}
\mathbf{Z}=i\mathbf{W}_{3}\mathbf{W}_{1}^{-1}\left( =\mathbf{Z}^{+}\right)
,\ \mathbf{Y}=-i\mathbf{W}_{1}\mathbf{W}_{3}^{-1}\left( =\mathbf{Y}%
^{+}\right) ,  \label{55.1}
\end{equation}%
where the Hermiticity follows from (\ref{30}) with $\mathbf{\mathbb{E}}_{%
\mathbf{M}}=\mathbf{\mathbb{T}}.$

Also introduced are the matrices $\mathbf{Z}^{\prime }\left[ \omega ,k_{x}%
\right] $ and $\mathbf{Z}^{\prime -1}=\mathbf{Y}^{\prime }\left[ \omega
,k_{x}\right] $ built from the "complementary" triplet of eigenvectors $%
\mathbf{w}_{\alpha +3}$ corresponding to the eigenvalues $\left\vert
q_{\alpha +3}\right\vert >1,$ $\alpha =1,2,3,$ so that 
\begin{equation}
\mathbf{Z}^{\prime }=-i\mathbf{W}_{4}\mathbf{W}_{2}^{-1}\left( =\mathbf{Z}%
^{\prime +}\right) ,\ \mathbf{Y}^{\prime }=i\mathbf{W}_{2}\mathbf{W}%
_{4}^{-1}\left( =\mathbf{Y}^{\prime +}\right) ,  \label{55.2}
\end{equation}%
where $\mathbf{W}_{2}$ and $\mathbf{W}_{4}$ are the right off-diagonal and
diagonal blocks of $\mathbf{W}$ (see (\ref{0})). The wave field defined via
these blocks similarly to (\ref{55}) infinitely grows at $y=nT\rightarrow
\infty $, in which sense $\mathbf{Z}^{\prime }$ and $\mathbf{Y}^{\prime }$
are "non-physical" impedance and admittance relative to the half-space $%
y\geq 0$ like their analogue (\ref{46}) for a homogeneous half-space.
However, the difference with the latter case is that the matrices $\mathbf{Z}%
^{\prime }$ and $\mathbf{Y}^{\prime }$ (\ref{55.2}) are generally not
complex conjugates of $\mathbf{Z}$ and $\mathbf{Y}$ (\ref{55.1}) (unless the
particular case of an even function $\mathbf{Q}\left( y\right) ,$ see
below). Note the identity%
\begin{equation}
\mathbf{M}\left( T,0\right) =\mathbf{M}^{-1}\left( -T,0\right) =\mathbf{M}_{%
\mathbf{Q}^{\ast }\left( -y\right) }^{T}\left( T,0\right) ,  \label{56}
\end{equation}%
where the former equality is due to periodicity, and the latter follows from
the successive use of (\ref{10.1}), (\ref{24}) and (\ref{26}). According to (%
\ref{56}), the eigenvalues of the monodromy matrix $\mathbf{M}\left(
T,0\right) $ are inverse of those of $\mathbf{M}\left( -T,0\right) $ and of $%
\mathbf{M}_{\mathbf{Q}^{\ast }\left( -y\right) }\left( T,0\right) $. Hence, $%
\mathbf{Z}^{\prime }$ and $\mathbf{Y}^{\prime }$ are the "physical"
impedance and admittance for the surface-localized waves in the half-space $%
y\leq 0$ with $\mathbf{Q}\left( y\right) $ (i.e. with $\rho \left( y\right) $
and $c_{ijkl}\left( y\right) $), while $\mathbf{Z}^{\prime \ast }\left( =%
\mathbf{Z}^{\prime T}\right) $ and $\mathbf{Y}^{\prime \ast }\left( =\mathbf{%
Y}^{\prime T}\right) $ are those in the half-space $y\geq 0$ with $\mathbf{Q}%
\left( -y\right) $ (i.e. with $\rho \left( -y\right) $ and $c_{ijkl}\left(
-y\right) $. Let us refer to any of these two periodic half-space
configurations as "\textit{inverted"} relative to the given half-space $%
y\geq 0$ with $\mathbf{Q}\left( y\right) $. Note also that, by (\ref{56}),
the stopband/passband partitioning of the $\left( \omega ,k_{x}\right) $%
-plane related to any given half-space and its "inverted" counterpart is the
same, and hence so is the area of the definition of the impedances $\mathbf{Z%
}\left[ \omega ,k_{x}\right] $ and $\mathbf{Z}^{\prime }\left[ \omega ,k_{x}%
\right] .$

\subsection{Properties\label{SSec7.2}}

Apart from proving the Hermiticity in (\ref{55.1}) and (\ref{55.2}),
orthonormality relation (\ref{30}) taken with $\mathbf{\mathbb{E}}_{\mathbf{M%
}}=\mathbf{\mathbb{T}}$ also yields the equalities%
\begin{equation}
\mathbf{Z+Z}^{\prime }=-2\mathbf{\Upsilon }_{2}^{-1},\ \mathbf{Y+Y}^{\prime
}=2\mathbf{\Upsilon }_{3}^{-1},  \label{55.3}
\end{equation}%
where 
\begin{equation}
\mathbf{\Upsilon }_{2}=2i\mathbf{W}_{1}\mathbf{W}_{2}^{+}\left( =\mathbf{%
\Upsilon }_{2}^{+}\right) ,\ \mathbf{\Upsilon }_{3}=2i\mathbf{W}_{3}\mathbf{W%
}_{4}^{+}\left( =\mathbf{\Upsilon }_{3}^{+}\right) .  \label{55.4}
\end{equation}%
The latter notation (\ref{55.4}) is motivated by the semblance to the case
of homogeneous half-space, see (\ref{48.2}). Following this analogy, $%
\mathbf{\Upsilon }_{2}$ and $\mathbf{\Upsilon }_{3}$ defined in (\ref{55.4})
may be viewed as the off-diagonal blocks of the $6\times 6$ dyadic matrix $%
\mathbf{\Upsilon }$ satisfying the eigenrelation $\mathbf{\Upsilon W}=%
\mathbf{W}\mathrm{diag}\left( i\mathbf{I,}-i\mathbf{I}\right) ,$ which
replicates (\ref{50}) and thus provides expression (\ref{53}) of $\mathbf{Z}$
and $\mathbf{Y}$ via the blocks of $\mathbf{\Upsilon }$; however, such $%
\mathbf{\Upsilon }$ has no integral representation similar to (\ref{49}).
Note aside that there is a formal analogy between the function $\mathrm{sign}%
\left( i\mathbf{N}_{0}\right) $ and the sign function of $\mathrm{Ln}\mathbf{%
M}\left( T,0\right) ,$ mentioned at the end of \S \ref{SSec6.3}\ and below
Eq. (\ref{14}), respectively.

As demonstrated in Appendix 2, the above impedances and admittances $\mathbf{%
Z,}$ $\mathbf{Z}^{\prime }$ and $\mathbf{Y,~Y}^{\prime }$ along with the
matrices $\mathbf{\Upsilon }_{2}$ and $\mathbf{\Upsilon }_{3}$ taken at a
fixed $k_{x}$ satisfy the sign-definiteness properties (\ref{47.1}) and (\ref%
{47.2}) with $v$ replaced by $\omega $ (the signs are referred to
definitions (\ref{8})$_{1,3}$ and must be inverted if (\ref{8})$_{2}$ is
used). It can also be shown that the components of matrices $\mathbf{%
\Upsilon }_{2}$ and $\mathbf{\Upsilon }_{3}$ are finite within the stopbands
and generally diverge at the band edges $\hat{\omega},$ unless exceptional
cases where, respectively, $\mathbf{W}_{2}$ or $\mathbf{W}_{3}$ at $\hat{%
\omega}$ is singular due to vanishing components $\mathbf{u}_{\deg }=\mathbf{%
0}\ $or $\mathbf{v}_{\deg }=\mathbf{0}$ of the eigenvector $\mathbf{w}_{\deg
}$ corresponding to the degenerate eigenvalue. The above properties emulate
those of the counterpart matrices of a homogeneous material discussed in \S %
\ref{Sec6}.

At the same time, transverse periodicity brings in essential
dissimilarities. First is that, unlike the case of a homogeneous medium, the
left-hand side in (\ref{55.3})$_{1}$ is not the (twice) real part of $%
\mathbf{Z}$ as in (\ref{48.2})$_{1},$ which makes (\ref{48.21}) irrelevant
and hence $\func{Re}\mathbf{Z}$ not positive definite. Secondly, there is a
subtle yet far-reaching difference in the impedance behavior at $\omega =%
\hat{\omega}$ depending on whether it is the transonic state for a
homogeneous half-space or the stopband edge for a periodic one. In the
former case, the $3\times 3$ impedance $\mathbf{Z}$ taken at $\hat{\omega}$
satisfies $\mathbf{A}_{\deg }^{T}\mathbf{ZA}_{\deg }=0$ with real $\mathbf{A}%
_{\deg }$ and hence can be neither positive nor negative definite (see \S %
\ref{SSec6.2}), while this statement does not apply in the latter case.
Consequently, the uniqueness theorem of the Rayleigh wave does not hold;
that is why the transversely periodic half-space with a generic (asymmetric)
arrangement over the period can support more than one surface wave, see
below.

\subsection{Surface waves\label{SSec7.3}}

Consider the surface waves in the transversely periodic half-space $y\geq 0$
with traction-free surface $y=0$. The surface wave must vanish at $%
y\rightarrow \infty ,$ i.e. fit (\ref{55}), hence the boundary condition $%
\mathbf{t}_{2}\left( 0\right) =\mathbf{0}$ leads to the equation $\mathbf{b}%
\left( 0\right) =\mathbf{W}_{3}\mathbf{c=0}$ and further, via (\ref{55.1}),
to the dispersion equation%
\begin{equation}
\det \mathbf{Z}\left[ \omega ,k_{x}\right] =0,  \label{56.1}
\end{equation}%
which defines the surface wave branches $\omega \left( k_{x}\right) $ in the
stopband areas of the $\left( \omega ,k_{x}\right) $-plane. In turn, the
boundary condition for the surface waves in the traction-free "inverted"
half-space reads $\mathbf{W}_{4}\mathbf{c}^{\prime }=\mathbf{0}$ (or its
complex conjugate), which provides, via (\ref{55.2}), the dispersion
equation in the form 
\begin{equation}
\det \mathbf{Z}^{\prime }\left[ \omega ,k_{x}\right] =0.  \label{56.2}
\end{equation}%
The roots of Eqs. (\ref{56.1}) and (\ref{56.2}) are zeros of the eigenvalues
of $3\times 3$ matrices $\mathbf{Z}$ and $\mathbf{Z}^{\prime },$ which are
continuously decreasing functions of $\omega $ at any fixed $k_{x}.$
Therefore, each equation may have at most three solutions within a stopband
at a fixed $k_{x}$. Thus, \textit{any transversely periodic half-space
admits not more than three surface waves per stopband with different
frequencies at a fixed }$k_{x}$\textit{. }[Note aside that counting
dispersion branches $\omega \left( k_{x}\right) $ per stopband at varying $%
k_{x}$ would be ambiguous since they may terminate and start again via
meeting the band edge and breaking away from it, see \S \ref{SSSec3.2.2}]

The reason for considering the surface wave problem in a given half-space
and, side by side, in an "inverted" one lies in their non-trivial intrinsic
relation. It becomes clear from examining the matrix $\mathbf{\Upsilon }_{3}$
introduced in (\ref{55.4})$_{2}$. By precise analogy with the argumentation
of \S \ref{SSec6.2}, the eigenvalues of $\mathbf{\Upsilon }_{3}$ at a fixed $%
k_{x}$ are real monotonically decreasing functions of $\omega $ of arbitrary
sign at the band edges, therefore the equation%
\begin{equation}
\det \mathbf{\Upsilon }_{3}\left[ \omega ,k_{x}\right] =0  \label{56.3}
\end{equation}%
considered at a fixed $k_{x}$ may have up to three solutions per stopband.
Since $\mathbf{\Upsilon }_{3}$ is a product of traction matrices related to
the mutually "inverse" half-spaces, this is the maximum number of solutions
Eqs. (\ref{56.1}) and (\ref{56.2}) in total. Thus, \textit{three is the
maximum total number of surface waves per stopband at a fixed }$k_{x}$%
\textit{\ in a pair of mutually "inverse" transversely periodic half-spaces }%
(see their definition in \S \ref{SSec7.1}). In other words, if, at some
fixed $k_{x}$, one of these half-spaces admits three or two or one waves in
a given stopband, the other cannot support a surface wave or admits at most
one or two, respectively. More can be said in the case of the lowest
stopband due to the additional property of positive definiteness of $\mathbf{%
\Upsilon }_{3}$ at $\omega =0$. Since $\mathbf{\Upsilon }_{3}$ normally
diverges at the stopband edges $\hat{\omega}$ unless $\det \mathbf{W}_{3}%
\left[ \hat{\omega},k_{x}\right] =0$ (see \S \ref{SSec7.2}), at least one of
its eigenvalues at any given $k_{x}$ varies monotonically from the positive
value at $\omega =0$ to $-\infty $ at $\omega =\hat{\omega}$ and hence turns
to zero in between. Thus, \textit{at least one surface wave is guaranteed to
exist for one of the mutually "inverse" half-spaces in the lowest stopband }$%
0<\omega <\hat{\omega}$ \textit{unless accidentally }$\mathbf{W}_{3}$ 
\textit{is singular at }$\hat{\omega}\left( k_{x}\right) .$

Essential simplification occurs if the periodicity profile is symmetric and
thus a given ("direct") and "inverted" half-spaces are identical. According
to \S \ref{SSSec3.2.4}, the monodromy matrix $\mathbf{M}\left( T,0\right) $
and the eigenvector matrix $\mathbf{W}$ in this case satisfy the same
identities as their counterparts in the homogeneous medium, and so
Barnett-Lothe's proof of the Rayleigh wave uniqueness theorem is directly
applicable. Thus, \textit{a transversely periodic half-space with a
symmetric profile does not admit more than one surface wave in a stopband at
any }$k_{x}$\textit{; it always exists in the lowest stopband unless its
upper edge renders }$\mathbf{W}_{3}$ \textit{singular. }

Similar reasoning shows that the clamped boundary condition also admits the
existence of up to three surface waves in any upper stopband, but precludes
any wave in the lowest stopband (the latter in contrast to the traction-free
case).

The above results concerning the surface waves in (full) stopbands were
obtained in \cite{DSh18}. Stringent but feasible conditions for the surface
wave existence beyond the stopbands were established in \cite{DSh19*}. The
analysis of the localized waves at the interface of two transversely
periodic half-spaces was carried out in \cite{DSh21}.

In conclusion, two comments are in order. First, as was mentioned above, the
dispersion branches may be segmented within stopbands by starting and
terminating at the band edges. Generally, such broken dispersion branches
are distributed irregularly on the $\left( \omega ,k_{x}\right) $-plane; all
the more remarkable is that the case of the SH waves allows deriving simple
conditions on the (periodic) material properties for achieving perfectly
regular spectral pattern; for instance, the periodicity profile may be
chosen so that all dispersion branches are confined in between certain
constant values of the ratio $s=k_{x}/\omega $ \cite{ShPG}. Also
interestingly, any given stopband admits at most one SH surface wave at a
fixed $k_{x}$ if the surface of the periodic half-space is traction-free,
but this restriction is lifted if this surface is loaded by a foreign layer
of finite width, which thereby breaks the periodicity (the latter setting
may be viewed as the Love wave problem for a periodic substrate, see \S \ref%
{SSSec9.3.1}) \cite{ShKKP}.

Secondly, let us briefly mention some studies of surface waves in a
half-space with aperiodic transverse inhomogeneity. Most of them have dealt
with the case of continuously inhomogeneous (functionally graded) media,
avoiding the case of piecewise continuous inhomogeneity (layered media).
Under the assumptions of exponential depth-dependence or within the WKB
approximate approach and appropriate profile models, the dispersion of the
Rayleigh and SH surface waves was derived in \cite{D1,Sh11} and \cite{AB}-%
\cite{T1}, respectively. The conditions on the depth-dependence profile
required for the SH surface wave to exist and possible peculiarities in its
dispersion spectrum were analyzed in \cite{SShS}. A general treatment of the
surface wave problem in arbitrarily inhomogeneous half-space may be found in 
\cite{BK}.

\section{Impedance of a laterally periodic vertically homogeneous half-space 
\label{Sec8}}

\subsection{Definition\label{SSec8.1}}

Consider an anisotropic half-space $y\geq 0,$ which is periodic along the
lateral axis $X$ and homogeneous along the depth axis $Y$, so that its
density and stiffness coefficients are described by $T$-periodic functions $%
\rho \left( x\right) =\rho \left( x+T\right) $ and $c_{ijkl}\left( x\right)
=c_{ijkl}\left( x+T\right) .$ According to \S \ref{SSSec3.2.3}, the
PWE-processed ODS (\ref{17})-(\ref{20}) with a constant system matrix $%
\widetilde{\mathbf{Q}}_{0}=i\widetilde{\mathbf{N}}_{0}\left[ \omega ,K_{x}%
\right] $ has partial solutions of the form $\mathbf{\tilde{\eta}}_{\alpha
}\left( y\right) =\mathbf{\tilde{\xi}}_{\alpha }e^{ik_{y\alpha }y}$ (\ref%
{20.1}). In what follows, the attention is restricted to the subsonic
frequency range $\omega <\hat{\omega}\left( K_{x}\right) ,$ in which all the
eigenvalues $k_{y1,}...,k_{y,6M}$ of $\widetilde{\mathbf{N}}_{0}$ are
complex and hence complex conjugated. We set their numbering so that%
\begin{equation}
k_{y\alpha }=k_{y,3M+\alpha }^{\ast },\ \ \func{Im}k_{y\alpha }>0,\ \ \alpha
=1,...,3M.  \label{57.-1}
\end{equation}%
Then, as mentioned in \S \ref{SSSec3.2.3}, the matrix of eigenvectors $%
\mathbf{\tilde{\xi}}_{\alpha }=(\mathbf{\tilde{A}}_{\alpha }$ $\mathbf{%
\tilde{B}}_{\alpha })^{T}$ of $\widetilde{\mathbf{N}}_{0},$%
\begin{equation}
\widetilde{\mathbf{\Xi }}=\left\Vert \mathbf{\tilde{\xi}}_{1}...\mathbf{%
\tilde{\xi}}_{6M}\right\Vert =\left( 
\begin{array}{cc}
\left\Vert \mathbf{\tilde{A}}_{1}...\mathbf{\tilde{A}}_{3M}\right\Vert & 
\left\Vert \mathbf{\tilde{A}}_{3M+1}...\mathbf{\tilde{A}}_{6M}\right\Vert \\ 
\left\Vert \mathbf{\tilde{B}}_{1}...\mathbf{\tilde{B}}_{3M}\right\Vert & 
\left\Vert \mathbf{\tilde{B}}_{3M+1}...\mathbf{\tilde{B}}_{6M}\right\Vert%
\end{array}%
\right) \equiv \left( 
\begin{array}{cc}
\widetilde{\mathbf{\Xi }}_{1} & \widetilde{\mathbf{\Xi }}_{2} \\ 
\widetilde{\mathbf{\Xi }}_{3} & \widetilde{\mathbf{\Xi }}_{4}%
\end{array}%
\right) ,  \label{57.0}
\end{equation}%
satisfies the orthonormality relation%
\begin{equation}
\widetilde{\mathbf{\Xi }}^{+}\widetilde{\mathbf{\mathbb{T}}}\widetilde{%
\mathbf{\Xi }}=\widetilde{\mathbf{\mathbb{T}}}\ \Leftrightarrow \ \widetilde{%
\mathbf{\Xi }}^{-1}=\widetilde{\mathbf{\mathbb{T}}}\widetilde{\mathbf{\Xi }}%
^{+}\widetilde{\mathbf{\mathbb{T}}}~\Leftrightarrow \ \widetilde{\mathbf{\Xi 
}}\widetilde{\mathbf{\mathbb{T}}}\widetilde{\mathbf{\Xi }}^{+}=\widetilde{%
\mathbf{\mathbb{T}}}.  \label{57}
\end{equation}%
The transonic frequency $\omega =\hat{\omega}\left( K_{x}\right) $ indicates
the eigenvalue degeneracy $k_{y\alpha }=k_{y,3M+\alpha }\equiv k_{y,\deg },\
\alpha \in \left\{ 1,...,3M\right\} ,$ at which the matrix $\widetilde{%
\mathbf{N}}_{0}$ is non-semisimple. A particular case is when the
periodicity profile is symmetric, i.e. may be described by the even
functions $\rho \left( x\right) =\rho \left( -x\right) $ and $c_{ijkl}\left(
x\right) =c_{ijkl}\left( -x\right) $, where $x=0$ is taken at the period
edge or midpoint. Then the matrix $\widetilde{\mathbf{N}}_{0}$ is real,
hence $\widetilde{\mathbf{\Xi }}\widetilde{\mathbf{\mathbb{T}}}=\widetilde{%
\mathbf{\Xi }}^{\ast }$ and identity (\ref{57}) reduces to the form 
\begin{equation}
\widetilde{\mathbf{\Xi }}^{T}\widetilde{\mathbf{\mathbb{T}}}\widetilde{%
\mathbf{\Xi }}=\widetilde{\mathbf{I}}\ \Leftrightarrow \ \widetilde{\mathbf{%
\Xi }}^{-1}=\widetilde{\mathbf{\Xi }}^{T}\widetilde{\mathbf{\mathbb{T}}}%
\Leftrightarrow \ \widetilde{\mathbf{\Xi }}\widetilde{\mathbf{\Xi }}^{T}=%
\widetilde{\mathbf{\mathbb{T}}}.  \label{57.2}
\end{equation}

By (\ref{15.1}) and (\ref{15.2}), the surface-localized wave field vanishing
at $y\rightarrow \infty $ may be written as%
\begin{equation}
\left( 
\begin{array}{c}
\mathbf{u}\left( x,y\right) \\ 
i\mathbf{t}_{2}\left( x,y\right)%
\end{array}%
\right) =e^{iK_{x}x}\sum\limits_{\alpha =1}^{3M}\tilde{c}_{\alpha }\left( 
\begin{array}{c}
\mathbf{A}_{\alpha }\left( x\right) \\ 
\mathbf{B}_{\alpha }\left( x\right)%
\end{array}%
\right) e^{ik_{y\alpha }y}=\sum\limits_{n=-N}^{N}\left\Vert \mathbf{\hat{\xi}%
}_{1}^{\left( n\right) }...\mathbf{\hat{\xi}}_{3M}^{\left( n\right)
}\right\Vert \mathrm{diag}\left( e^{ik_{y\alpha }y}\right) e^{ik_{n}x}%
\mathbf{\tilde{c},}  \label{57.1}
\end{equation}%
where $\left\Vert ...\right\Vert $ is a $6\times 3M$ matrix formed of
enclosed column vectors $\mathbf{\hat{\xi}}_{\alpha }^{\left( n\right) }=(%
\mathbf{\hat{A}}_{\alpha }^{\left( n\right) }\mathbf{\ \hat{B}}_{\alpha
}^{\left( n\right) })^{T}$ normalized via (\ref{57}), and $\mathbf{\tilde{c}}%
=\left( \tilde{c}_{1}...\tilde{c}_{3M}\right) ^{T}$ is the vector of
disposable constants. Following \cite{ShWM}, introduce the $3\times 3$
impedance $\mathbf{Z}\left( x\right) $ and admittance $\mathbf{Z}^{-1}\left(
x\right) \equiv \mathbf{Y}\left( x\right) $ as $3\times 3$ matrices
connecting the displacement and traction vectors in (\ref{57.1}), namely, $%
\mathbf{t}_{2}\left( x,y\right) =-\mathbf{Z}\left( x\right) \mathbf{u}\left(
x,y\right) $ and so%
\begin{equation}
\mathbf{B}_{\alpha }\left( x\right) =-i\mathbf{Z}\left( x\right) \mathbf{A}%
_{\alpha }\left( x\right) ,\ \mathbf{A}_{\alpha }\left( x\right) =i\mathbf{Y}%
\left( x\right) \mathbf{B}_{\alpha }\left( x\right) ,\ \alpha =1,...,3M,
\label{58}
\end{equation}%
where the dependence on the parameters $\omega $ and $K_{x}$ is understood
and suppressed. As $T$-periodic matrix functions, $\mathbf{Z}\left( x\right) 
$ and $\mathbf{Y}\left( x\right) $ expand in Fourier series 
\begin{equation}
\mathbf{Z}\left( x\right) =\dsum\nolimits_{n=-N}^{N}\widehat{\mathbf{Z}}%
^{\left( n\right) }e^{ignx},\ \mathbf{Y}\left( x\right)
=\dsum\nolimits_{n=-N}^{N}\widehat{\mathbf{Y}}^{\left( n\right) }e^{ignx},
\label{59}
\end{equation}%
truncated similarly to (\ref{57.1}). Substituting both expansions and (\ref%
{59}) into (\ref{58}) gives the relations 
\begin{equation}
\mathbf{\hat{B}}_{\alpha }^{\left( n\right) }=-i\widehat{\mathbf{Z}}^{\left(
n-m\right) }\mathbf{\hat{A}}_{\alpha }^{\left( m\right) },\ \ \ \mathbf{\hat{%
A}}_{\alpha }^{\left( n\right) }=i\widehat{\mathbf{Y}}^{\left( n-m\right) }%
\mathbf{\hat{B}}_{\alpha }^{\left( m\right) },\ n,m=-N,...,N;\ \alpha
=1,...,3M.  \label{61}
\end{equation}%
They may be collected into the form 
\begin{equation}
\mathbf{\tilde{B}}_{\alpha }=-i\widetilde{\mathbf{Z}}\mathbf{\tilde{A}}%
_{\alpha },\ \mathbf{\tilde{A}}_{\alpha }=i\widetilde{\mathbf{Y}}\mathbf{%
\tilde{B}}_{\alpha },\ \alpha =1,...,3M,  \label{62}
\end{equation}%
with $3M\times 3M$ To\"{e}plitz matrices 
\begin{equation}
\widetilde{\mathbf{Z}}={\Large \{}\widehat{\mathbf{Z}}^{\left( n-m\right) }%
{\Large \}},{\Large \ }\widetilde{\mathbf{Y}}={\Large \{}\widehat{\mathbf{Y}}%
^{\left( n-m\right) }{\Large \}},\ n,m=-N,...,N,  \label{62.1}
\end{equation}%
whose $\left( nm\right) $th block is the $\left( n-m\right) $th coefficient
in the matrix Fourier series (\ref{59}). Rewriting $3M$ vector equations (%
\ref{62}) in the matrix form and invoking notations (\ref{57.0}) yields $%
\widetilde{\mathbf{\Xi }}_{3}=-i\widetilde{\mathbf{Z}}\widetilde{\mathbf{\Xi 
}}_{1}$ and $\widetilde{\mathbf{\Xi }}_{1}=i\widetilde{\mathbf{Y}}\widetilde{%
\mathbf{\Xi }}_{3}$, i.e.%
\begin{equation}
\widetilde{\mathbf{Z}}\left[ \omega ,K_{x}\right] =i\widetilde{\mathbf{\Xi }}%
_{3}\widetilde{\mathbf{\Xi }}_{1}^{-1},\ \ \widetilde{\mathbf{Y}}\left[
\omega ,K_{x}\right] =-i\widetilde{\mathbf{\Xi }}_{1}\widetilde{\mathbf{\Xi }%
}_{3}^{-1}.  \label{63}
\end{equation}%
Equation (\ref{63}) expresses the Fourier coefficients of $\mathbf{Z}\left(
x\right) $ and $\mathbf{Y}\left( x\right) $ in terms of the eigenvectors of $%
\widetilde{\mathbf{N}}_{0}.$

Now let the given superlattice $y\geq 0$ be rotated 180$%
{{}^\circ}%
$ about the axis $Y$. The density $\rho ^{\prime }\left( x\right) $ and
stiffness coefficients $c_{ijkl}^{\prime }\left( x\right) $ of the
"inverted" laterally periodic half-space are of the form 
\begin{equation}
\rho ^{\prime }\left( x\right) =\rho \left( -x\right) ,\ c_{ijkl}^{\prime
}\left( x\right) =c_{ijkl}\left( -x\right)  \label{63.1}
\end{equation}%
(the prime is not a derivative). This leads to the system matrix $\widetilde{%
\mathbf{Q}}_{0}^{\prime }=i\widetilde{\mathbf{N}}_{0}^{\prime }\left[ \omega
,K_{x}\right] $ with $\widetilde{\mathbf{N}}_{0}^{\prime }=\widetilde{%
\mathbf{N}}_{0}^{\ast },$ i.e. the eigenvalue and eigenvector sets of $%
\widetilde{\mathbf{N}}_{0}$ and $\widetilde{\mathbf{N}}_{0}^{\prime }$ are
complex conjugates of each other. With due regard for the ordering (\ref%
{57.-1}), the eigenvalues $k_{y\alpha }^{\prime }$ of $\widetilde{\mathbf{N}}%
_{0}^{\prime }$ satisfying $\func{Im}k_{y\alpha }^{\prime }>0$ are equal to $%
k_{y,3M+\alpha }^{\ast }$ and hence the corresponding eigenvectors $\mathbf{%
\tilde{\xi}}_{\alpha }^{\prime }$ coincide with $\mathbf{\tilde{\xi}}%
_{3M+\alpha }^{\ast }=(\mathbf{\tilde{A}}_{3M+\alpha }^{\ast }~\mathbf{%
\tilde{B}}_{3M+\alpha }^{\ast })^{T}$. Note that $\widetilde{\mathbf{N}}_{0}$
and $\widetilde{\mathbf{N}}_{0}^{\prime }$ imply the same transonic
frequency $\hat{\omega}\left( K_{x}\right) ,$ at which they become
non-semisimple and acquire the same eigenvector $\mathbf{\tilde{\xi}}_{\deg
} $ such that corresponds to the (real) degenerate eigenvalue $k_{y,\deg }$
and satisfies the self-orthogonality relation $\mathbf{\tilde{\xi}}_{\deg
}^{+}\mathbf{\mathbb{T}\hat{\xi}}_{\deg }=0.$

It follows that the surface-localized wave field vanishing in the infinite
depth $y\rightarrow \infty $ of the "inverted" half-space may be written as 
\begin{equation}
\left( 
\begin{array}{c}
\mathbf{u}\left( x,y\right) \\ 
i\mathbf{t}_{2}\left( x,y\right)%
\end{array}%
\right) =e^{iK_{x}x}\sum\limits_{\alpha =1}^{3M}\tilde{c}_{\alpha }^{\prime
}\left( 
\begin{array}{c}
\mathbf{A}_{\alpha }^{\prime }\left( x\right) \\ 
\mathbf{B}_{\alpha }^{\prime }\left( x\right)%
\end{array}%
\right) e^{ik_{y\alpha }^{\prime }y}=\sum\limits_{n=-N}^{N}\left\Vert 
\mathbf{\hat{\xi}}_{3M+1}^{\left( n\right) \ast }...\mathbf{\hat{\xi}}%
_{6M}^{\left( n\right) \ast }\right\Vert \mathrm{diag}\left(
e^{ik_{y,3M+\alpha }^{\ast }y}\right) e^{ik_{n}x}\mathbf{\tilde{c}}^{\prime
}.  \label{66}
\end{equation}%
Introduce the impedance and admittance matrices relating the displacement
and traction amplitudes of the wave field (\ref{66}), namely, 
\begin{equation}
\mathbf{B}_{\alpha }^{\prime }\left( x\right) =-i\mathbf{Z}^{\prime \ast
}\left( -x\right) \mathbf{A}_{\alpha }^{\prime }\left( x\right) ,\ \mathbf{A}%
_{\alpha }^{\prime }\left( x\right) =i\mathbf{Y}^{\prime \ast }\left(
-x\right) \mathbf{B}_{\alpha }\left( x\right) ,\ \alpha =1,...,3M,
\label{61'}
\end{equation}%
where 
\begin{equation}
\mathbf{Z}^{\prime \ast }\left( -x\right) =\sum\nolimits_{n=-N}^{N}\widehat{%
\mathbf{Z}}^{\prime \left( n\right) \ast }e^{ignx},\ \mathbf{Y}^{\prime \ast
}\left( -x\right) =\sum\nolimits_{n=-N}^{N}\widehat{\mathbf{Y}}^{\prime
\left( n\right) \ast }e^{ignx}  \label{65}
\end{equation}%
with $\widehat{\mathbf{Z}}^{\prime \left( n\right) }$ and $\widehat{\mathbf{Y%
}}^{\prime \left( n\right) }$ being the Fourier coefficients of $\mathbf{Z}%
^{\prime }\left( x\right) $ and $\mathbf{Y}^{\prime }\left( x\right) $%
\footnote{{\footnotesize Explicit form of definition (\ref{61'}) is
motivated by compatibility of the ensuing Eq. (\ref{68}) with the notations
adopted for impedances and admittances in \S 6.1. }}. Inserting (\ref{65})
and (\ref{66}) in (\ref{61'}) and taking the complex conjugate yields%
\begin{equation}
\mathbf{\hat{B}}_{3M+\alpha }^{\left( n\right) }=i\widehat{\mathbf{Z}}%
^{\prime \left( n-m\right) }\mathbf{\hat{A}}_{3M+\alpha }^{\left( m\right)
},\ \ \ \mathbf{\hat{A}}_{3M+\alpha }^{\left( n\right) }=-i\widehat{\mathbf{Y%
}}^{\prime \left( n-m\right) }\mathbf{\hat{B}}_{3M+\alpha }^{\left( m\right)
},\ n,m=-N,...,N;\ \alpha =1,...,3M.  \label{67}
\end{equation}%
This may be written in the $3M\times 3M$ form as $\widetilde{\mathbf{\Xi }}%
_{4}=-i\widetilde{\mathbf{Z}}^{\prime }\widetilde{\mathbf{\Xi }}_{2}^{-1}$
and $\widetilde{\mathbf{\Xi }}_{2}=i\widetilde{\mathbf{Y}}^{\prime }%
\widetilde{\mathbf{\Xi }}_{2}^{-1}$, so that the block To\"{e}plitz matrices 
$\widetilde{\mathbf{Z}}^{\prime }={\Large \{}\widehat{\mathbf{Z}}^{\prime
\left( n-m\right) }{\Large \}}$ and $\widetilde{\mathbf{Y}}^{\prime }=%
{\Large \{}\widehat{\mathbf{Y}}^{\prime \left( n-m\right) }{\Large \}}$ are
given by 
\begin{equation}
\widetilde{\mathbf{Z}}^{\prime }\left[ \omega ,K_{x}\right] =-i\widetilde{%
\mathbf{\Xi }}_{4}\widetilde{\mathbf{\Xi }}_{2}^{-1},\ \ \widetilde{\mathbf{Y%
}}^{\prime }\left[ \omega ,K_{x}\right] =i\widetilde{\mathbf{\Xi }}_{2}%
\widetilde{\mathbf{\Xi }}_{4}^{-1},  \label{68}
\end{equation}%
cf. (\ref{63}).

It is clear that if the periodicity profile is symmetric, then the "direct"
and "inverted" superlattices are identical; accordingly, by (\ref{57.2})$%
_{1} $ and (\ref{63}), (\ref{68}), $\widetilde{\mathbf{\Xi }}_{1}=\widetilde{%
\mathbf{\Xi }}_{2}^{\ast },$ $\widetilde{\mathbf{\Xi }}_{3}=\widetilde{%
\mathbf{\Xi }}_{4}^{\ast }$ and hence $\widetilde{\mathbf{Z}}=$ $\widetilde{%
\mathbf{Z}}^{\prime \ast },$ which implies $\mathbf{Z}\left( x\right) =%
\mathbf{Z}^{\prime \ast }\left( x\right) $, as it must.

\subsection{Properties\label{SSec8.2}}

Equating the corresponding blocks of the orthonormality relation (\ref{57})
proves that the $3M\times 3M$ matrices$~\widetilde{\mathbf{Z}}$ and $%
\widetilde{\mathbf{Z}}^{\prime }$ and hence their inverses, the matrices $%
\widetilde{\mathbf{Y}}$ and $\widetilde{\mathbf{Y}}^{\prime }$, are
Hermitian, while their sums satisfy the equalities%
\begin{equation}
\widetilde{\mathbf{Z}}+\mathbf{\widetilde{\mathbf{Z}}}^{\prime }=-2%
\widetilde{\mathbf{\Upsilon }}_{2}^{-1},\ \widetilde{\mathbf{Y}}+\mathbf{%
\widetilde{\mathbf{Y}}}^{\prime }=2\widetilde{\mathbf{\Upsilon }}_{3}^{-1},
\label{71}
\end{equation}%
where $\widetilde{\mathbf{\Upsilon }}_{2}$ and $\widetilde{\mathbf{\Upsilon }%
}_{3}$ are the off-diagonal blocks of the $6M\times 6M$ matrix 
\begin{equation}
\widetilde{\mathbf{\Upsilon }}=i\widetilde{\mathbf{\Xi }}\mathrm{diag}\left( 
\mathbf{I,}-\mathbf{I}\right) \widetilde{\mathbf{\Xi }}^{-1}=i\left( 
\begin{array}{cc}
\widetilde{\mathbf{I}}-2\widetilde{\mathbf{\Xi }}_{2}\widetilde{\mathbf{\Xi }%
}_{3}^{+} & 2\widetilde{\mathbf{\Xi }}_{1}\widetilde{\mathbf{\Xi }}_{2}^{+}
\\ 
2\widetilde{\mathbf{\Xi }}_{3}\widetilde{\mathbf{\Xi }}_{4}^{+} & \widetilde{%
\mathbf{I}}-2\widetilde{\mathbf{\Xi }}_{4}\widetilde{\mathbf{\Xi }}_{1}^{+}%
\end{array}%
\right) \equiv \left( 
\begin{array}{cc}
\widetilde{\mathbf{\Upsilon }}_{1} & \widetilde{\mathbf{\Upsilon }}_{3} \\ 
\widetilde{\mathbf{\Upsilon }}_{2} & \widetilde{\mathbf{\Upsilon }}_{1}^{+}%
\end{array}%
\right) ,\   \label{50.1}
\end{equation}%
cf. (\ref{50}) and (\ref{55.4}).\ 

The above matrices can be shown to possess similar sign-definiteness
properties as their counterparts in homogeneous and transversely periodic
half-spaces. That is, the matrices $\widetilde{\mathbf{Z}},\ \mathbf{%
\widetilde{\mathbf{Z}}}^{\prime }$ and hence their inverses $\widetilde{%
\mathbf{Y}},$ $\widetilde{\mathbf{Y}}^{\prime }$ are positive definite in
the limit $\omega \rightarrow 0;$ the frequency derivatives of $\widetilde{%
\mathbf{Z}},~\mathbf{\widetilde{\mathbf{Z}}}^{\prime }$ and $\widetilde{%
\mathbf{Y}},$ $\widetilde{\mathbf{Y}}^{\prime }$ at $\omega \leq \hat{\omega}
$ are, respectively, negative-definite and positive-definite. The matrices $%
\widetilde{\mathbf{\Upsilon }}_{2}$ and $\widetilde{\mathbf{\Upsilon }}_{3}$
are finite at $\omega <\hat{\omega}$; by (\ref{71}), $\widetilde{\mathbf{%
\Upsilon }}_{2}$ is negative definite and $\widetilde{\mathbf{\Upsilon }}%
_{3} $ is positive definite at $\omega \rightarrow 0$ and they both have
negative-definite derivatives in $\omega $. In consequence, $\widetilde{%
\mathbf{\Upsilon }}_{2}$ and therefore $\widetilde{\mathbf{\Xi }}_{1},$ $%
\widetilde{\mathbf{\Xi }}_{2}$ are non-singular; hence, by definition (\ref%
{63})$_{2}$ and (\ref{68})$_{2}$, so are $\widetilde{\mathbf{Y}}\ $\ and $%
\widetilde{\mathbf{Y}}^{\prime }$. Thus, the eigenvalues of $\widetilde{%
\mathbf{Z}}$ and$\ \widetilde{\mathbf{Z}}^{\prime }$ are finite and hence 
\textit{continuously }decreasing at $\omega \leq \hat{\omega}.$

A block To\"{e}plitz matrix is known to take over the properties of its
generating matrix, i.e. the periodic matrix function of $x$ whose Fourier
coefficients it consists of. In particular, the matrix function generating a
Hermitian and sign-definite (or singular) To\"{e}plitz matrix is itself
Hermitian and likewise sign-definite (or singular) for $x\in \left[ 0,T%
\right] $, see e.g. \cite{S}. Hence, the $3\times 3$ impedances $\mathbf{Z}%
\left( x\right) $ and $\mathbf{Z}^{\prime }\left( x\right) $ considered in
the subsonic range $\omega \leq \hat{\omega}$ are finite Hermitian matrix
functions that are positive definite at $\omega =0$ and have
negative-definite frequency derivatives for $\omega \leq \hat{\omega}.$ The
properties of their inverse, the admittances $\mathbf{Y}\left( x\right) $ and%
$\ \mathbf{Y}^{\prime }\left( x\right) $, follow as a consequence.

The explicit form of the matrices $\mathbf{\Upsilon }_{2}\left( x\right) $
and $\mathbf{\Upsilon }_{3}\left( x\right) $ generating $\widetilde{\mathbf{%
\Upsilon }}_{2}=2i\widetilde{\mathbf{\Xi }}_{1}\widetilde{\mathbf{\Xi }}%
_{2}^{+}$ and $\widetilde{\mathbf{\Upsilon }}_{3}=2i\widetilde{\mathbf{\Xi }}%
_{3}\widetilde{\mathbf{\Xi }}_{4}^{+}$ is 
\begin{equation}
\mathbf{\Upsilon }_{2}\left( x\right) =2i\left\{ \mathbf{A}_{\alpha }\left(
x\right) \right\} \left\{ \mathbf{A}_{\alpha }^{\prime }\left( -x\right)
\right\} ^{T},\ \mathbf{\Upsilon }_{3}\left( x\right) =2i\left\{ \mathbf{B}%
_{\alpha }\left( x\right) \right\} \left\{ \mathbf{B}_{\alpha }^{\prime
}\left( -x\right) \right\} ^{T},  \label{80}
\end{equation}%
where $\left\{ \left( \cdot \right) _{\alpha }\right\} $ are the $3\times 3M$
matrices with the 3-component columns $\left( \cdot \right) _{\alpha },\
\alpha =1,...,3M$. Combining (\ref{80}) with (\ref{58})$_{1}$ and (\ref{61'})%
$_{1}$ yields 
\begin{equation}
\mathbf{\Upsilon }_{3}\left( x\right) =-2i\mathbf{Z}\left( x\right) \mathbf{%
\Upsilon }_{2}\left( x\right) \mathbf{Z}^{\prime }\left( x\right) ,
\label{80.2}
\end{equation}%
where the Hermiticity of $\mathbf{Z}^{\prime }\left( x\right) $ was used.
According to the properties of $\widetilde{\mathbf{\Upsilon }}_{2}$ and $%
\widetilde{\mathbf{\Upsilon }}_{3}$ mentioned above, $\mathbf{\Upsilon }%
_{2}\left( x\right) $ and $\mathbf{\Upsilon }_{3}\left( x\right) $ are
finite Hermitian matrices, which are, respectively, negative and positive
definite at $\omega =0$ and have negative definite frequency derivatives
within $\omega <\hat{\omega}.$ As a result, $\mathbf{\Upsilon }_{2}\left(
x\right) $ is non-singular and hence so are $\mathbf{Y}\left( x\right) $ and$%
\ \mathbf{Y}^{\prime }\left( x\right) $. The matrices $\mathbf{\Upsilon }%
_{2}\left( x\right) \ $and $\mathbf{\Upsilon }_{3}\left( x\right) $ diverge
at the transonic state $\omega =\hat{\omega}$ unless, respectively, $\left\{ 
\mathbf{A}_{J}\left( x\right) \right\} $ or $\left\{ \mathbf{B}_{J}\left(
x\right) \right\} $ is accidentally singular at $\hat{\omega}$.

Note that the above non-singularity of $\widetilde{\mathbf{\Xi }}_{1}$ and%
\emph{\ }$\widetilde{\mathbf{\Xi }}_{2},$ or, equally, of $\mathbf{\Upsilon }%
_{2}\left( x\right) ,$ rules out the possibility of surface waves on a
clamped boundary, which is the same feature as that for the homogeneous
half-spaces and for the transversely periodic half-spaces within the lowest
stopband.

\subsection{Direct evaluation of the impedance\label{SSec8.3}}

The PWE version of the Barnett-Lothe integral formalism was outlined in \cite%
{NSK} by invoking the sign function of the matrix $i\widetilde{\mathbf{N}}%
_{0}\left[ \omega ,K_{x}\right] .$ It was shown to satisfy the equalities 
\begin{equation}
\mathrm{sign}(i\widetilde{\mathbf{N}}_{0})=i\langle \widetilde{\mathbf{N}}%
_{\varphi }\rangle =i\widetilde{\mathbf{\Upsilon }},  \label{49.1}
\end{equation}%
where $\langle \widetilde{\mathbf{N}}_{\varphi }\rangle =\frac{1}{\pi }%
\int_{0}^{\pi }\widetilde{\mathbf{N}}_{\varphi }\mathrm{d}\varphi $ and the
matrix $\widetilde{\mathbf{N}}_{\varphi }(=\mathbf{\mathbb{T}\widetilde{%
\mathbf{N}}}_{\varphi }^{+}\mathbf{\mathbb{T)}}\ $is constructed according
to (\ref{19}) and (\ref{20}) up to replacing a fixed frame $\left( \mathbf{e}%
_{1},\mathbf{e}_{2}\right) $ with the rotating one in the same plane. The
matrix $\widetilde{\mathbf{\Upsilon }}$ is defined in (\ref{50.1}).
Combining the latter with (\ref{63}) and (\ref{68}) yields 
\begin{equation}
\begin{array}{c}
\widetilde{\mathbf{Z}}=-\widetilde{\mathbf{\Upsilon }}_{2}^{-1}{\large (}%
\mathbf{I}+i\widetilde{\mathbf{\Upsilon }}_{1}{\large )},\ \widetilde{%
\mathbf{Y}}=\widetilde{\mathbf{\Upsilon }}_{3}^{-1}{\large (}\mathbf{I}+i%
\widetilde{\mathbf{\Upsilon }}_{1}^{T}{\large )}, \\ 
\widetilde{\mathbf{Z}}^{\prime }=-\widetilde{\mathbf{\Upsilon }}_{2}^{-1}%
{\large (}\mathbf{I}-i\widetilde{\mathbf{\Upsilon }}_{1}{\large )},\ 
\widetilde{\mathbf{Y}}^{\prime }=\widetilde{\mathbf{\Upsilon }}_{3}^{-1}%
{\large (}\mathbf{I}-i\widetilde{\mathbf{\Upsilon }}_{1}^{T}{\large )},%
\end{array}
\label{53.1}
\end{equation}%
which is similar to (\ref{53}).

Equations (\ref{49.1}) and (\ref{53.1}) allow direct evaluation of the
matrices $\widetilde{\mathbf{Z}}$ and $\widetilde{\mathbf{Z}}^{\prime },$
bypassing the eigenvalue problem. Regarding numerical implementation, the
link of the sign function to the projectors, which in turn are expressed as
contour integrals of the resolvent of the system matrix, proves suitable for
handling the large-size PWE matrices \cite{NSK}.

\subsection{Surface waves\label{SSec8.4}}

Let the surface of the half-space $y\geq 0$ with a laterally periodic
profile of material properties $\rho \left( x\right) $ and $c_{ijkl}\left(
x\right) $ and its "inverted" version with $\rho ^{\prime }\left( x\right)
=\rho \left( -x\right) $ and $c_{ijkl}^{\prime }\left( x\right)
=c_{ijkl}\left( -x\right) $ maintain zero traction $\mathbf{t}_{2}\left(
x,0\right) =\mathbf{0}$ at any $x$. By (\ref{57.1}) and (\ref{66}), this
means the vanishing of, respectively, the following linear combinations:%
\begin{equation}
\begin{array}{l}
\ \widetilde{\mathbf{\Xi }}_{3}\mathbf{\tilde{c}}=\mathbf{0}\
\Leftrightarrow \left\Vert \mathbf{B}_{1}\left( x\right) ...\mathbf{B}%
_{3M}\left( x\right) \right\Vert \mathbf{\tilde{c}}=\mathbf{0\ }\forall x,
\\ 
\widetilde{\mathbf{\Xi }}_{4}^{\ast }\mathbf{\tilde{c}}^{\prime }=\mathbf{0}%
\ \Leftrightarrow \left\Vert \mathbf{B}_{1}^{\prime }\left( x\right) ...%
\mathbf{B}_{3M}^{\prime }\left( x\right) \right\Vert \mathbf{\tilde{c}}%
^{\prime }=\mathbf{0\ }\forall x\mathbf{.}%
\end{array}
\label{81.0}
\end{equation}%
Therefore, according to Eqs. (\ref{63}) and (\ref{68}), the dispersion
branches $\omega \left( K_{x}\right) $ of the subsonic ($\omega <\hat{\omega}
$) surface waves satisfying the traction-free boundary condition in the two
above cases may be defined, respectively, from the equations%
\begin{equation}
\begin{array}{c}
\det \widetilde{\mathbf{Z}}\left[ \omega ,K_{x}\right] =0\Leftrightarrow
\det \mathbf{Z}\left( x;\omega ,K_{x}\right) =0\ \forall x, \\ 
\det \widetilde{\mathbf{Z}}^{\prime }\left[ \omega ,K_{x}\right] =0\
\Leftrightarrow \det \mathbf{Z}^{\prime }\left( x;\omega ,K_{x}\right) =0\
\forall x,%
\end{array}
\label{81}
\end{equation}%
where a semicolon is used to set apart dependence on the variable $x$ and on
the parameters $\omega ,~K_{x}.$

The left-hand algebraic equations (\ref{81})$_{1},$ which involve
numerically accessible PWE matrices (see \S \ref{SSec8.3}), are suitable for
computing the surface-wave branches $\omega \left( K_{x}\right) $. On the
other hand, the right-hand equations (\ref{81})$_{2},$ involving $3\times 3$
matrices, provide an insight into a possible number of branches. According
to \S \ref{SSec8.2}, the three eigenvalues of $\mathbf{Z}\left( x\right) $
and$\ \mathbf{Z}^{\prime }\left( x\right) $ taken at any fixed $x$ and $%
K_{x} $ are continuously decreasing functions of $\omega \leq $ $\hat{\omega}
$ positive at $\omega =0.$ It follows that each of Eqs. (\ref{81})$_{2}$
considered at fixed $x$ and $K_{x}$ may have at most three solutions for $%
\omega =\omega \left( x,K_{x}\right) $. Hence, they cannot have more than
three solutions $\omega =\omega \left( K_{x}\right) $ such that are the same
for any $x\in \left[ 0,T\right] .$ In other words, \textit{a half-space with
a generic profile of lateral periodicity admits at most three subsonic
surface waves at a fixed }$K_{x}$\textit{. }[Note that counting branches $%
\omega =\omega \left( K_{x}\right) $ for varying $K_{x}$ instead of counting
solutions at a fixed $K_{x}$ could be misleading since the subsonic branches
may terminate and start at the transonic frequency $\hat{\omega}\left(
K_{x}\right) ,$ see a similar remark in \S \ref{SSec7.3}.]

The above holds true for either of the two half-spaces with mutually
"inverse" profiles of periodicity. What is more, their joint consideration
reveals an additional constraint on the number of surface wave solutions,
which follows from the equation 
\begin{equation}
\det \mathbf{\Upsilon }_{3}\left( x;\omega ,K_{x}\right) =0.  \label{81.1}
\end{equation}%
By virtue of definition (\ref{80})$_{2}$ of $\mathbf{\Upsilon }_{3}\left(
x\right) $ and its properties mentioned below (\ref{80.2}), Eq. (\ref{81.1})
may have a maximum of three solutions $\omega =\omega \left( K_{x}\right) $
that are split, one way or another, between the solutions of Eqs. (\ref{81})$%
_{1}$ and (\ref{81})$_{2}.$ Thus, \textit{for any fixed }$K_{x},$\textit{\
the maximum total number of subsonic surface waves admissible in two
half-spaces with mutually "inverse" profiles of lateral periodicity is three 
}(see \cite{ShWM} for an explicit analytical example). Furthermore, due to
the divergence of $\mathbf{\Upsilon }_{3}\left( x\right) $ at the normal
(not exceptional) transonic state $\omega =\hat{\omega}$ such that keeps $%
\left\{ \mathbf{B}_{J}\left( x\right) \right\} $ non-singular, at least one
of the eigenvalues of $\mathbf{\Upsilon }_{3}\left( x\right) $ varies
monotonically from the positive value at $\omega =0$ to $-\infty $ at $%
\omega =\hat{\omega}$ and hence turns to zero in between. Thus, with
reference to (\ref{80}), \textit{at least one surface wave is guaranteed to
exist for one of the mutually inverse superlattices, provided that the
transonic state }$\hat{\omega}\left( K_{x}\right) $ \textit{at a given }$%
K_{x}$\textit{\ is normal. }

Consider briefly the case of a half-space with a symmetric periodicity
profile. Then $\widetilde{\mathbf{Z}}^{\prime }$ and $\mathbf{Z}^{\prime
}\left( x\right) $ are complex conjugates of $\widetilde{\mathbf{Z}}$ and $%
\mathbf{Z}\left( x\right) $ (see below (\ref{68})), so the positive
definiteness of $\widetilde{\mathbf{\Upsilon }}_{2}$ implies the same for $%
\func{Re}\widetilde{\mathbf{Z}}$ (see (\ref{71})$_{1}$). It also follows
that the zero eigenvalue of the matrix $\mathbf{\Upsilon }_{3}\left(
x\right) $ if it exists must be a double one (see (\ref{80.2})). Either of
these arguments suffices to prove that such a half-space admits only one
surface wave, which is likewise the case of the unique Rayleigh wave in a
homogeneous half-space.

Thus, we observe that, despite technical dissimilarities, the surface wave
problems in the transversely and laterally periodic half-spaces are similar
in that either of them admits at most three solutions with different
frequencies at a fixed wavenumber $k_{x}$ or $K_{x},$ which may exist in a
stopband or subsonic interval in aggregate in a pair of half-spaces with
mutually "inverse" profiles obtained by inversion $y\rightarrow -y$ or $%
x\rightarrow -x$ (see \S \S \ref{SSec7.1} and \ref{SSec8.1}). Such a
conjunction of the results suggests the possibility of a unified proof due
to some "rabbit hole" between the two above surface wave problems.

\section{Impedance of a transversely inhomogeneous{\protect\Large \ }plate 
\label{Sec9}}

\subsection{Definition and properties\label{SSec9.1}}

Consider the wave field $\mathbf{u}\left( x,y\right) $ (\ref{2}) propagating
in an infinite transversely inhomogeneous (multilayered and/or functionally
graded) plate with planar faces orthogonal to the axis $Y.$ Recall the
definition of the matricant $\mathbf{\eta }\left( y_{2}\right) =\mathbf{M}%
\left( y_{2},y_{1}\right) \mathbf{\eta }\left( y_{1}\right) $ expanded in
blockwise notations as 
\begin{equation}
\left( 
\begin{array}{c}
\mathbf{a}\left( y_{2}\right) \\ 
\mathbf{b}\left( y_{2}\right)%
\end{array}%
\right) =\left( 
\begin{array}{cc}
\mathbf{M}_{1} & \mathbf{M}_{2} \\ 
\mathbf{M}_{3} & \mathbf{M}_{4}%
\end{array}%
\right) \left( 
\begin{array}{c}
\mathbf{a}\left( y_{1}\right) \\ 
\mathbf{b}\left( y1\right)%
\end{array}%
\right) .  \label{84}
\end{equation}%
Rearranging (\ref{84}) provides two types of plate impedance.

One type is the $6\times 6$ impedance matrix $\mathbf{Z}$ and its inverse,
admittance $\mathbf{Y}=\mathbf{Z}^{-1},$ which link the displacement $%
\mathbf{a}\left( y\right) $ taken at a pair of points $y_{1},$ $y_{2}$ and
the traction $\mathbf{b}\left( y\right) $ taken at the same points $y_{1},$ $%
y_{2},$ namely,%
\begin{equation}
\left( 
\begin{array}{c}
\mathbf{b}\left( y_{1}\right) \\ 
-\mathbf{b}\left( y_{2}\right)%
\end{array}%
\right) =-i\mathbf{Z}\left( y_{2},y_{1}\right) \left( 
\begin{array}{c}
\mathbf{a}\left( y_{1}\right) \\ 
\mathbf{a}\left( y_{2}\right)%
\end{array}%
\right) ,  \label{82}
\end{equation}%
where taking the tractions with inverse signs observes their definition as
internal forces (this secures explicit Hermiticity of $\mathbf{Z}$, see
below). From (\ref{84}) and (\ref{82}), 
\begin{equation}
\begin{array}{c}
\mathbf{Z}\left( y_{2},y_{1}\right) =i\left( 
\begin{array}{cc}
-\mathbf{M}_{2}^{-1}\mathbf{M}_{1} & \mathbf{M}_{2}^{-1} \\ 
\mathbf{M}_{4}\mathbf{M}_{2}^{-1}\mathbf{M}_{1}-\mathbf{M}_{3} & -\mathbf{M}%
_{4}\mathbf{M}_{2}^{-1}%
\end{array}%
\right) \ (\det \mathbf{Z}=-\dfrac{\det \mathbf{M}_{3}}{\det \mathbf{M}_{2}}%
), \\ 
\mathbf{Y}\left( y_{2},y_{1}\right) =\mathbf{Z}^{-1}\left(
y_{2},y_{1}\right) =i\left( 
\begin{array}{cc}
\mathbf{M}_{3}^{-1}\mathbf{M}_{4} & \mathbf{M}_{3}^{-1} \\ 
\mathbf{M}_{1}\mathbf{M}_{3}^{-1}\mathbf{M}_{4}-\mathbf{M}_{2} & \mathbf{M}%
_{1}\mathbf{M}_{3}^{-1}%
\end{array}%
\right) .\ 
\end{array}
\label{83}
\end{equation}%
where $\mathbf{M}_{1...4}$ are the $3\times 3\mathbf{\ }$blocks of the
matricant $\mathbf{M}\left( y_{2},y_{1}\right) $ \cite{ShPD,Sh2}\footnote{%
{\footnotesize Note a misprint in \cite{Sh2} in that the sign of the left
off-diagonal block of the admittance }$\mathbf{Y}${\footnotesize \ must be
inverted.}}. If the variation of material properties is symmetric about the
plate midplane (in particular, if the plate is homogeneous), then Eq. (\ref%
{28.2}) implies $\mathbf{M}_{1}=\mathbf{M}_{4}^{T}$ and hence $\mathbf{Z=%
\mathbb{T}Z}^{T}\mathbf{\mathbb{T}}$. If the plate material has a symmetry
plane parallel to the faces or orthogonal to the propagation direction $X$,
then $\mathbf{Z}=-\mathbf{\mathbb{G}Z}^{\ast }\mathbf{\mathbb{G}}$ by (\ref%
{32})$_{2}$. If both conditions apply, then $\mathbf{Z}=\mathbf{\mathbb{Y}Z}%
^{+}\mathbf{\mathbb{Y}}$ where $\mathbf{\mathbb{Y=TG}}$.

Other versions of the $6\times 6$ two-point matrices available in the
literature either adhere to the standard stiffness-matrix pattern, where the
displacement and traction are kept on the opposite sides of the two-point
relation, or adopt a mixed compliance-stiffness pattern, where the
displacement and traction referred to the opposite edge points are kept on
the same side, see \cite{RW} and \cite{Tan,G...W}, respectively. These two
patterns exhibit different trends at $y_{1}\rightarrow y_{2}:$ the blocks of
the former, by (\ref{83}), diverge as $\mathbf{M}_{2}^{-1}\left(
y_{2},y_{1}\right) \sim $ $\left[ \int_{y_{1}}^{y_{2}}\mathbf{N}_{2}\left(
y\right) dy\right] ^{-1}\sim \left( y_{1}-y_{2}\right) ^{-1},$ while the
latter, merely by construction, approaches the identity matrix $\mathbf{I}$
or its block permutation $\mathbf{\mathbb{T}}$.

A different type of plate impedance is the $3\times 3$ matrix $\mathbf{z}%
\left( y\right) $, which relates the displacement and traction at an
arbitrary point $y_{2},$ 
\begin{equation}
\mathbf{b}\left( y_{2}\right) =-i\mathbf{z}\left( y_{2}\right) \mathbf{a}%
\left( y_{2}\right) ,  \label{85}
\end{equation}%
given its value $\mathbf{z}\left( y_{1}\right) $ at the reference point $%
y=y_{1}$. Accordingly, it was called the conditional impedance in \cite%
{ShPD,Sh2} and its equivalent was referred to as the surface one in \cite{HC}
(we shall use the former). From (\ref{84}) and (\ref{85}), \ \ \ \ 
\begin{equation}
\mathbf{z}\left( y_{2}\right) |_{\mathbf{z}\left( y_{1}\right) }=i\left( 
\mathbf{M}_{3}-i\mathbf{M}_{4}\mathbf{z}\left( y_{1}\right) \right) \left( 
\mathbf{M}_{1}-i\mathbf{M}_{2}\mathbf{z}\left( y_{1}\right) \right) ^{-1}=-%
\mathbf{Z}_{4}+\mathbf{Z}_{3}\left( \mathbf{z}\left( y_{1}\right) -\mathbf{Z}%
_{1}\right) ^{-1}\mathbf{Z}_{2}  \label{86}
\end{equation}%
with $\mathbf{M}_{1...4}$ and $\mathbf{Z}_{1...4}$ being the blocks of $%
\mathbf{M}\left( y_{2},y_{1}\right) $ and $\mathbf{Z}\left(
y_{2},y_{1}\right) .$ In particular, if the traction-free or clamped
condition $\mathbf{b}\left( y_{1}\right) =\mathbf{0}$ or $\mathbf{a}\left(
y_{1}\right) =\mathbf{0}$ is imposed at $y=y_{1}$ (which does not in itself
restrict the parameters $\omega ,$ $k_{x}$), then Eq. (\ref{86}) yields%
\begin{equation}
\begin{array}{c}
\mathbf{z}\left( y_{2}\right) |_{\mathbf{z}\left( y_{1}\right) =\mathbf{0}}=i%
\mathbf{M}_{3}\mathbf{M}_{1}^{-1},\ \mathbf{z}\left( y_{2}\right) |_{\mathbf{%
y}\left( y_{1}\right) =\mathbf{0}}=i\mathbf{M}_{4}\mathbf{M}_{2}^{-1}\
\Rightarrow \\ 
\ \mathbf{y}\left( y_{2}\right) |_{\mathbf{z}\left( y_{1}\right) =\mathbf{0}%
}=-i\mathbf{M}_{1}\mathbf{M}_{3}^{-1},\ \mathbf{y}\left( y_{2}\right) |_{%
\mathbf{y}\left( y_{1}\right) =\mathbf{0}}=-i\mathbf{M}_{2}\mathbf{M}%
_{4}^{-1},%
\end{array}
\label{87}
\end{equation}%
where $\mathbf{y}=\mathbf{z}^{-1}$ is the conditional admittance.

The impedances "reciprocal" to (\ref{82}) and (\ref{87})$_{1,2}$ under the
inversion $y_{1}\rightleftarrows y_{2}$ follow in the form $\mathbf{Z}\left(
y_{1},y_{2}\right) =-\mathbf{\mathbb{T}Z}\left( y_{2},y_{1}\right) \mathbf{%
\mathbb{T}}$ and 
\begin{equation}
\mathbf{z}\left( y_{1}\right) |_{\mathbf{z}\left( y_{2}\right) }=-i\left( 
\mathbf{M}_{4}+i\mathbf{z}\left( y_{2}\right) \mathbf{M}_{4}\right)
^{-1}\left( \mathbf{M}_{3}+i\mathbf{z}\left( y_{2}\right) \mathbf{M}%
_{1}\right) =\mathbf{Z}_{1}-\mathbf{Z}_{2}\left( \mathbf{z}\left(
y_{2}\right) +\mathbf{Z}_{4}\right) ^{-1}\mathbf{Z}_{3},  \label{88}
\end{equation}%
where $\mathbf{z}\left( y_{1}\right) |_{\mathbf{z}\left( y_{2}\right) }$ is
expressed via the blocks of $\mathbf{M}\left( y_{2},y_{1}\right) $ and $%
\mathbf{Z}\left( y_{2},y_{1}\right) $ like in (\ref{86}). Given a $T$%
-periodic structure $\left[ y_{1},y_{1}+nT\right] ,$ the above formulas can
be specialized by replacing the blocks of $\mathbf{M}\left(
y_{2},y_{1}\right) $ with those of $\mathbf{M}\left( nT,0\right) =\mathbf{M}%
^{n}\left( T,0\right) $.

By virtue of (\ref{26}), the plate impedances $\mathbf{Z}$ and $\mathbf{z}$
of the above form are Hermitian matrices. Furthermore, when regarded at
fixed points $y_{1},\ y_{2}$ and fixed $k_{x}$ as functions of $\omega ,$
they manifest specific sign-definiteness stemming from energy
considerations, see Appendix 2. Given Stroh's ODS formulation with (\ref{8})$%
_{1}$ or (\ref{8})$_{3}$, the matrix $\mathbf{Z}\left( y_{2},y_{1}\right) $
is positive definite at $\omega =0$ and its derivative in $\omega $ is
piecewise continuous and negative definite between the poles, whereas the
matrices $\mathbf{z}\left( y_{2}\right) |_{\mathbf{z}\left( y_{1}\right) =%
\mathbf{0}}$ and $\mathbf{z}\left( y_{2}\right) |_{\mathbf{y}\left(
y_{1}\right) =\mathbf{0}}$ are negative definite at $\omega =0$ and their
derivatives are positive definite between the poles (all aforementioned
signs must be inverted if (\ref{8})$_{2}$ is chosen or $y_{1}$ and $y_{2}$
are swapped).

Expressing the plate impedances through the matricant, while analytically
straightforward, retains the issue of numerical instabilities at large
frequency-thickness values. However, the impedance admits other
computational schemes that are stable. One of them is the recursive
approach, which was seemingly first proposed in \cite{HBBH}. It is
especially transparent with respect to the $3\times 3$ conditional
impedance, whose definition implies a recursive identity $\mathbf{z}\left(
y_{2}\right) |_{\mathbf{z}\left( y_{1}\right) }=\mathbf{z}\left(
y_{2}\right) |_{\mathbf{z}\left( \widetilde{y}\right) |_{\mathbf{z}\left(
y_{1}\right) }}\ \forall \widetilde{y}\in \left[ y_{1},y_{2}\right] $ (the
latter restriction on $\widetilde{y}$ is actually optional). Hence, the
impedance $\mathbf{z}$ for a given layer can be obtained by means of
successive calculations, that involve fictitious sublayers of sufficiently
small thickness to ensure stable evaluation of the matricant $\mathbf{M}$
through each sublayer. The recursive formulas for the $6\times 6$ stiffness
matrix, equivalent to $\mathbf{Z}$, and a similar one for the
compliance-stiffness matrix were elaborated and implemented by various
authors, see \cite{RW}-\cite{HC}, \cite{PPV}. Another option for stable
computing is due to the fact that both $\mathbf{Z}\left( y,y_{1}\right) \ $%
and $\mathbf{z}\left( y\right) $ considered as functions of $y$ satisfy the
matrix differential Riccati equation (its explicit $6\times 6$\ and $3\times
3$\ forms adjusted to the present notations may be found in \cite{NSh}).
Numerical integration of this equation proves efficient for evaluating the
impedance in various types of structures, possibly in conjunction with the
recursive scheme, see \cite{Z}-\cite{BDG}. An additional numerically
advantageous feature of using the impedance matrix is the above piecewise
monotonicity of its impedance eigenvalues. It implies a strict alternation
of their zeros and poles on the $\omega $ axis, which enables the tracing
and counting of zeros through much easier counting of poles, see
Wittrick-Williams algorithm \cite{G...W,WW}.

\subsection{Lamb wave spectrum\label{SSec9.2}}

\subsubsection{Overview\label{SSSec9.2.1}}

By definition (\ref{84}), the vanishing of the determinant of the blocks $%
\mathbf{M}_{1},$ or $\mathbf{M}_{2},$ or $\mathbf{M}_{3},$ or $\mathbf{M}%
_{4} $ is the equation for the guided wave spectrum $\omega \left(
k_{x}\right) $ in a plate with, respectively, the face $y_{2}$ clamped and
the face $y_{1}$ free (c/f), or with both faces clamped (c/c), or with both
faces free (f/f), or with the face $y_{2}$ free and the face $y_{1}$ clamped
(f/c). The equivalent real-valued formulation is available via the
determinant of the appropriate $6\times 6$ or $3\times 3$ admittance or
impedance Hermitian matrices (\ref{83}) and (\ref{87}) considered as
functions of the parameters $\omega $ and $k_{x}$ at fixed $y_{1}$ and $%
y_{2} $. The above sign-definiteness properties of these matrices entail the
following hierarchy among the lower bounds of the frequency spectra in an
arbitrary given plate under different boundary conditions (these are
indicated by the subscript):%
\begin{equation}
\min \omega _{\mathrm{f/f}}\leq \min \left( \omega _{\mathrm{c/f}},\omega _{%
\mathrm{f/c}}\right) \leq \min \omega _{\mathrm{c/c}}\ \ (\mathrm{fixed}\
k_{x}).  \label{90}
\end{equation}%
In the case of SH waves, when $M_{1...4}$ are scalars and hence the matrices
(\ref{87}) reduce to scalar functions of a tangent- or cotangent-type shape,
a similar inequality extends to the entire infinite set of the dispersion
branches, i.e. $\omega _{\mathrm{f/f}}\leq \omega _{\mathrm{c/f}},\omega _{%
\mathrm{f/c}}\leq \omega _{\mathrm{c/c}}$ at any fixed $k_{x}$, see \cite%
{ShPK}.

The following exposition will be confined to the case of guided waves in a
plate $\left[ 0,H\right] $ with both faces $y_{1}=0$ and$~y_{2}=H$ free of
traction (the Lamb waves). According to the above background, the
corresponding dispersion equation may be expressed in either form 
\begin{equation}
\det \mathbf{M}_{3}\left[ \omega ,k_{x}\right] =0\Leftrightarrow \det 
\mathbf{Z}\left[ \omega ,k_{x}\right] =0\Leftrightarrow \det \mathbf{z}\left[
\omega ,k_{x}\right] =0,  \label{91}
\end{equation}%
where $\mathbf{z}\left[ \omega ,k_{x}\right] $ may be specified as $\mathbf{z%
}\left( H\right) |_{\mathbf{z}\left( 0\right) =\mathbf{0}}$ or $\mathbf{z}%
\left( 0\right) |_{\mathbf{z}\left( H\right) =\mathbf{0}}$ given by (\ref{87}%
)$_{1}$ or (\ref{88})$_{2}$. The spectrum $\omega _{J}\left( k_{x}\right) ,$ 
$J=1,2,...$, defined by (\ref{91}) contains three fundamental branches that
originate at $\omega =0,\ k_{x}=0$ (such branches are absent in the spectrum
of a plate if at least one of its faces is clamped) and a countably infinite
set of the upper branches with cutoffs at the vertical resonance frequencies 
$\omega _{J}\left( 0\right) \neq 0$. Analytical estimates of the branches
(see below) hinge on the dispersion equation in the form (\ref{91})$_{1}$,
whereas the impedance-related formulations (\ref{91})$_{2,3}$ facilitate
numerical implementation due to the aforementioned methods of stable
computation of the plate impedance.

Invoking impedance also appears helpful for analyzing some general
properties of the spectrum. As an example of this point, assume an $N$%
-layered plate $\left[ 0,H\right] $ and consider the block-diagonal
("global-matrix type") form of the dispersion equation: 
\begin{equation}
\begin{array}{c}
\det \left[ \mathrm{diag}\left( \mathbf{Z}_{1},\mathbf{Z}_{3},...,\mathbf{Z}%
_{N-1},\mathbf{\hat{0}}\right) +\mathrm{diag}\left( \mathbf{\hat{0},Z}_{2},%
\mathbf{Z}_{4},...,\mathbf{Z}_{N}\right) \right] =0\ \mathrm{for\ even}\ N,
\\ 
\det \left[ \mathrm{diag}\left( \mathbf{Z}_{1},\mathbf{Z}_{3},...,\mathbf{Z}%
_{N}\right) +\mathrm{diag}\left( \mathbf{\hat{0},Z}_{2},...,\mathbf{Z}_{N-1},%
\mathbf{\hat{0}}\right) \right] =0\ \mathrm{for\ odd}\ N,%
\end{array}
\label{99}
\end{equation}%
where $\mathbf{\hat{0}}$ is the $3\times 3$ zero matrix and $\mathbf{Z}%
_{j}\equiv \mathbf{Z}(y_{2}^{\left( j\right) },y_{1}^{\left( j\right) })$ is
the $6\times 6$ impedance (\ref{82}) of the $j$th layer $[y_{2}^{\left(
j\right) },y_{1}^{\left( j\right) }]$ (with $y_{1}^{\left( 1\right) }=0,\
y_{2}^{\left( N\right) }=H$). Due to the above sign properties of $\mathbf{Z}%
_{j}$ and the fact that the global matrix in (\ref{99}) is positive definite
when so are all diagonal blocks $\mathbf{Z}_{j}$, it follows that at any
fixed $k_{x},$ the least guided wave frequency in the layered plate is
always greater than the least value among the frequencies $\omega _{n,j}$ in
the constituent layers $j=1,...,N$ assumed traction-free (solutions of $\det 
\mathbf{Z}_{j}=0$), i.e., the latter is the lower bound of the layered plate
spectrum.

Next, we will present some explicit results on the Lamb wave spectrum. In
the rest of this section, the previously used notations of the unit vectors $%
\mathbf{e}_{1}\Vert X$ and $\mathbf{e}_{2}\Vert Y$ are replaced by $\mathbf{m%
}$\ and $\mathbf{n},$\ which are more conventional in the context of plate
waves.

\subsubsection{Longwave approximation for the fundamental branches\label%
{SSSec9.2.2}}

\paragraph{Velocities at $\protect\omega \rightarrow 0,\ k_{x}\rightarrow 0$%
\label{SSSSec9.2.2.1}}

Consider the origin of the fundamental dispersion branches characterized by
finite velocity $v=\omega /k_{x}$ at $k_{x}\rightarrow 0.$ The dispersion
equation (\ref{91})$_{1}$ with $\mathbf{M}_{3}\left( H,0\right) $ truncated
by the first-order term of (\ref{10}) reduces to the form 
\begin{equation}
\det \left( \left\langle \mathbf{N}_{3}\right\rangle -\left\langle \rho
\right\rangle v^{2}\right) =0,  \label{92.1}
\end{equation}%
where%
\begin{equation}
\left\langle \mathbf{...}\right\rangle =\frac{1}{H}\int_{0}^{H}...\left(
y\right) dy  \label{92}
\end{equation}%
indicates an average through the plate. Hence the longwave limit of the
fundamental wave velocities $v_{J}\left( k_{x}\right) |_{k_{x}\rightarrow
0}\equiv v_{0J}$ $(J=1,2,3)$ is set by the eigenvalues $0\leq $ $\overline{%
\lambda }_{2}\leq \overline{\lambda }_{3}$ of the (positive semi-definite)
matrix $\left\langle \mathbf{N}_{3}\right\rangle $, i.e.%
\begin{equation}
v_{01}^{2}=0,\ v_{02}^{2}=\frac{\overline{\lambda }_{2}}{\left\langle \rho
\right\rangle },\ v_{03}^{2}=\frac{\overline{\lambda }_{3}}{\left\langle
\rho \right\rangle },  \label{93}
\end{equation}%
while the corresponding mutually orthogonal eigenvectors $\mathbf{n,\ }%
\overline{\mathbf{p}}_{2},\ \overline{\mathbf{p}}_{3}$ of $\left\langle 
\mathbf{N}_{3}\right\rangle $ define the wave polarizations. Note that the
eigenvalues and eigenvectors $\overline{\lambda }_{2,3}$ and $\overline{%
\mathbf{p}}_{2,3}$ of the averaged $\left\langle \mathbf{N}_{3}\right\rangle 
$ are generally not equal to the averaged eigenvalues and eigenvectors $%
\left\langle \lambda _{2,3}\right\rangle $ and $\left\langle \mathbf{p}%
_{2,3}\right\rangle $ of $\mathbf{N}_{3}\left( y\right) $. \ 

Applying Weyl's inequality to the definition (\ref{7}) of $\mathbf{N}_{3}$
shows that $v_{02}\leq c_{2}$ and $v_{03}\leq c_{3}$, where $\left(
0<\right) c_{1}^{2}\leq c_{2}^{2}\leq c_{3}^{2}$ $\ $are the eigenvalues of
the matrix $\left\langle \rho \right\rangle ^{-1}\left\langle \left( \mathbf{%
mm}\right) \right\rangle $. One may also determine the azimuthal
orientations $\theta $ of the propagation direction $\mathbf{m=m}\left(
\theta \right) $ that provide extreme values of the velocities $v_{0J}$ in a
given plate with a fixed normal $\mathbf{n}.$ Differentiating Eq. (\ref{92.1}%
) yields the equation for the sought value of $\theta $, 
\begin{equation}
\mathbf{m}\times {\large [}\left\langle \left( \overline{\mathbf{p}}_{J}%
\overline{\mathbf{p}}_{J}\right) \right\rangle -\langle \left( \overline{%
\mathbf{p}}_{J}\mathbf{n}\right) \left( \mathbf{nn}\right) ^{-1}\left( 
\mathbf{n}\overline{\mathbf{p}}_{J}\right) \rangle {\large ]}\mathbf{m=0,\ }%
J=2,3,  \label{93.1}
\end{equation}%
where $\times $ means vector product, $\overline{\mathbf{p}}_{J}=\overline{%
\mathbf{p}}_{J}\left( \theta \right) $, and the notation (\ref{4.1}) is
used. In particular, it is seen that if some $\theta _{l}$ renders the
polarization $\overline{\mathbf{p}}_{J}\left( \theta _{l}\right) $
longitudinal ($\parallel \mathbf{m}$), then $v_{0J}\left( \theta _{l}\right) 
$ is an extremum.

The above generalizes the results of \cite{Sh2} to the case of transversely
inhomogeneous plates. For the homogeneous plate, it can also be shown that
the velocity $v_{03}$ is always greater than the transonic velocity $\hat{v}$
and the Rayleigh-wave velocity $v_{\mathrm{R}}$ (recall that the inequality $%
v_{\mathrm{R}}>\hat{v}$ is extraordinary but possible).

\paragraph{Leading-order dispersion dependence\label{SSSSec9.2.2.2}}

The longwave onset of the fundamental dispersion branches can be found from
Eq. (\ref{91})$_{1}$ with $\mathbf{M}_{3}\left( H,0\right) $ approximated by
several terms of expansion in powers of $k_{x}H\ll 1$. This task is
relatively tractable in the case of a homogeneous plate (see (\ref{11.0})),
for which the leading-order terms of the dependence $v_{J}\left(
k_{x}\right) $ or $v_{J}\left( \omega \right) $ are detailed in \cite{Sh2}
and the next-order terms are provided in \cite{PShK}. Unfortunately, the
accuracy of the power series approximation is known to deteriorate rapidly
as the variable approaches the convergence radius. On the other hand, the
Taylor coefficients are the ingredients of the Pad\'{e} approximation, whose
application extends the fitting range. Moreover, it allows approximating the
entire lowest (flexural) velocity branch provided its shortwave limit (the
Rayleigh velocity) is known \cite{PShK}.

In the case of an arbitrary transversely inhomogeneous anisotropic plate, a
compact explicit expression is obtainable only for the leading-order
dispersion coefficient describing the slope at the onset of the flexural
branch $v_{1}\left( k_{x}\right) =\overline{\kappa }k_{x}H+...$ This
coefficient $\overline{\kappa }$ admits several equivalent representations
elaborated in \cite{AM,ShPDB}; e.g., one of them reads 
\begin{equation}
\left\langle \rho \right\rangle \overline{\kappa }^{2}=\sum\nolimits_{\alpha
=2,3}\frac{1}{16\overline{\lambda }_{\alpha }}\left[ \int\nolimits_{0}^{1}%
\int\nolimits_{0}^{\varsigma }\left( \varsigma -\varsigma _{1}\right)
^{2}f_{\alpha }\left( \varsigma \right) f_{\alpha }\left( \varsigma
_{1}\right) \mathrm{d}\varsigma \mathrm{d}\varsigma _{1}\right] ,
\label{95.0}
\end{equation}%
where $f_{\alpha }\left( y\right) =\mathbf{m}^{T}\mathbf{N}_{3}\left(
y\right) \overline{\mathbf{p}}_{\alpha }$.

Note the inequality%
\begin{equation}
\left\langle \rho \right\rangle \overline{\kappa }^{2}<\tfrac{1}{4}\mathbf{m}%
^{T}\left\langle \mathbf{N}_{3}\right\rangle \mathbf{m}\equiv 3\left\langle
\rho \kappa ^{2}\right\rangle ,  \label{95}
\end{equation}%
which bounds the difference between the exact result (\ref{95.0}) and an
"intuitively suggestive" evaluation $\left\langle \rho \kappa
^{2}\right\rangle $ obtained by averaging the same coefficient $\rho \kappa
^{2}=\tfrac{1}{12}\mathbf{m}^{T}\mathbf{N}_{3}\mathbf{m}$ as in a
homogeneous plate but with varying $\mathbf{N}_{3}=\mathbf{N}_{3}\left(
y\right) $ (here $\kappa $ is an anisotropic analogue of Kirhhoff's
coefficient). If the plate consists of $n$ homogeneous layers with
coefficients $\kappa _{i}$, then $\overline{\kappa }>\min \left( \kappa
_{1},...,\kappa _{n}\right) .$ For a periodic plate consisting of $N$
periods, the discrepancy between $\left\langle \rho \right\rangle \overline{%
\kappa }^{2}$ and $\left\langle \rho \kappa ^{2}\right\rangle $ is of the
order of $N^{-2}$ and hence is small at large $N.$ The same is valid for the
coefficient of the leading-order dispersion ($\sim \left( kH\right) ^{2}$)
at the onset of the two higher fundamental velocity branches $v_{2,3}\left(
k_{x}\right) .$ For more details, see \cite{ShPDB}.

\subsubsection{Vicinity of the cutoffs\label{SSSec9.2.3}}

Guided wave propagation near the vertical resonances (cutoffs) may lead to
interesting phenomena of "negative" and zero group velocity, which have
attracted much interest due to their advantageous applications in NDT
imaging and, more recently, in some other modern application areas \cite{PCR}%
-\cite{LGBLLPA}. The study of group velocity in anisotropic plates makes
relevant the dependence on the azimuthal orientation of the sagittal plane $%
\left( \mathbf{m}\left( \theta \right) \mathbf{,n}\right) ,$ rotating by the
angle $\theta $ about the fixed normal to the plate $\mathbf{n}$. Assume $%
k_{x}H\ll 1$ and consider the near-cutoff asymptotics of the frequency and
group velocity along the $J$th dispersion branch ($J>3$): 
\begin{equation}
\begin{array}{c}
\omega _{J}\left( k_{x},\theta \right) =\Omega _{J}+b_{J}\left( \theta
\right) \left( k_{x}H\right) ^{2}+O(\left( kH\right) ^{4}), \\ 
\mathbf{g}_{J}\left( k_{x},\theta \right) =\dfrac{\partial \omega _{J}}{%
\partial k_{x}}\mathbf{m}+\dfrac{1}{k_{x}}\dfrac{\partial \omega _{J}}{%
\partial \theta }\mathbf{t}= \\ 
=2k_{x}H^{2}\left[ b_{J}\left( \theta \right) \mathbf{m}+\frac{1}{2}%
b_{J}^{\prime }\left( \theta \right) \mathbf{t}\right] +O(\left( kH\right)
^{3})\equiv g_{J}^{\left( \mathbf{m}\right) }\mathbf{m}+g_{J}^{\left( 
\mathbf{t}\right) }\mathbf{t},%
\end{array}
\label{96}
\end{equation}%
where $\omega _{J}|_{k_{x}=0}\equiv \Omega _{J}$ is the cutoff frequency and 
$\mathbf{t}\left( \theta \right) =\mathbf{m}\left( \theta \right) \times 
\mathbf{n}$. It is seen that the signs of $b_{J}\left( \theta \right) $ and $%
b_{J}^{\prime }\left( \theta \right) $ determine those of the in-plane and
out-of-plane group velocity components within some vicinity of the cutoff.
Moreover, since $\partial \omega _{J}/\partial k_{x}$ must become positive
as $k_{x}$ grows, a negative value of $b_{J}\left( \theta \right) $
guarantees the vanishing of the in-plane group velocity $g_{J}^{\left( 
\mathbf{m}\right) }\left( k_{x},\theta \right) $ at some $k_{x}$. When $%
\left( \mathbf{m}\left( \theta \right) \mathbf{,n}\right) $ is a symmetry
plane, i.e. $\mathbf{g}_{J}=g_{J}^{\left( \mathbf{m}\right) }\mathbf{m}$,
the inequality $b_{J}\left( \theta \right) <0$ implies backward propagation
without steering and ensures the existence of the zero group velocity (ZGV)
point $\mathbf{g}_{J}\left( k_{x},\theta \right) =\mathbf{0}$ on the $J$th
branch. Note that this prediction does not require computing the dispersion
branch in full. At the same time, it is understood that the above criterion
is sufficient but not necessary, and also that there may be more than one
point with $g_{J}^{\left( \mathbf{m}\right) }=0$ on a given branch, see
examples in \cite{KPGP}.

Deriving explicit expressions of $\Omega _{J}$ and $b_{J}\left( \theta
\right) $ for an arbitrary inhomogeneous plate is a hardly amenable task
(unless within the WKB approximation, see e.g. \cite{ShPK} for the SH
waves). Let us confine ourselves to the case of homogeneous plates. Denote
the phase velocities and the unit polarizations of the three bulk modes
propagating along the normal $\mathbf{n}$ by $c_{\alpha }$ and $\mathbf{a}%
_{\alpha },$ $\alpha =1,2,3.$ The cutoff frequencies are given by the
well-known formula $\Omega _{n,\alpha }=\pi nc_{\alpha }/H,\ $where $%
n=1,2,...$ (the pair $n,\alpha \ $thus plays the role of the branch index $J$%
). Expanding the matricant (\ref{11.0}), defined through the system matrix $%
\mathbf{Q}_{0}$ in either of the forms (\ref{8})$_{2,3},$ near $\Omega
_{n,\alpha }$ and plugging it in Eq. (\ref{91})$_{1}$ yields

\begin{equation}
\begin{array}{c}
b_{n,\alpha }\left( \theta \right) \equiv \dfrac{c_{\alpha }}{2\pi nH}\left[
W_{\alpha }^{\left( 1\right) }\left( \theta \right) +W_{n,\alpha }^{\left(
2\right) }\left( \theta \right) \right] ,\  \\ 
\ W_{\alpha }^{\left( 1\right) }\left( \theta \right) =\dfrac{1}{c_{\alpha
}^{2}}\left[ \left( mm\right) _{\alpha \alpha }-\dfrac{\left( mn\right)
_{\alpha \alpha }^{2}}{c_{\alpha }^{2}}+\sum\limits_{\beta =1,\,\beta \neq
\alpha }^{3}\dfrac{\left( \left( mn\right) _{\alpha \beta }+\left( mn\right)
_{\beta \alpha }\right) ^{2}}{c_{\alpha }^{2}-c_{\beta }^{2}}\right] , \\ 
W_{n,\alpha }^{\left( 2\right) }\left( \theta \right) =\dfrac{4}{\pi n}%
\sum\limits_{\beta =1,\,\beta \neq \alpha }^{3}\dfrac{\left( c_{\alpha
}^{2}\left( mn\right) _{\alpha \beta }+c_{\beta }^{2}\left( mn\right)
_{\beta \alpha }\right) ^{2}}{c_{\alpha }^{3}c_{\beta }\left( c_{\alpha
}^{2}-c_{\beta }^{2}\right) ^{2}}\tan \left[ \dfrac{\pi n}{2}\left( 1-\dfrac{%
c_{\alpha }}{c_{\beta }}\right) \right] ,%
\end{array}
\label{96.1}
\end{equation}%
where $\left( mm\right) _{\alpha \alpha }=\frac{1}{\rho }a_{\alpha
i}m_{j}c_{ijkl}m_{k}a_{\alpha l}$ and $\left( mn\right) _{\alpha \beta }=%
\frac{1}{\rho }a_{\alpha i}m_{j}c_{ijkl}n_{k}a_{\beta l}$ are the elements
of the matrices $\left( mm\right) $ and $\left( mn\right) $ (\ref{4.1}) in
the basis of vectors $\left\{ \mathbf{a}_{1},\mathbf{a}_{2},\mathbf{a}%
_{3}\right\} $ \cite{ShP}. The expression for $b_{n,\alpha }^{\prime }\left(
\theta \right) $ follows through differentiating the vector $\mathbf{m}%
\left( \theta \right) $ in (\ref{96.1}). If the sagittal plane $\left( 
\mathbf{m}\left( \theta \right) \mathbf{,n}\right) $ is a symmetry plane so
that one of the bulk modes, say with an index $\alpha =3$, is the uncoupled
SH mode (i.e. $\mathbf{a}_{3}\parallel \mathbf{t}$), then $\left( mn\right)
_{3\beta }=\left( mn\right) _{\beta 3}=0,$ whereupon Eq. (\ref{96.1})
substantially simplifies (it even allows for a direct tabulation of the sign
of $b_{n,\alpha }$ in isotropic plates).

Interestingly, there is a direct link between the leading-order dispersion
coefficient $b_{n,\alpha }\left( \theta \right) $ and the local shape of the
slowness surface of the bulk modes. Let $S_{\alpha }$ be the curve lying in
the cross-section of the $\alpha $th-mode slowness sheet by the sagittal
plane $\left( \mathbf{m}\left( \theta \right) \mathbf{,n}\right) ,$ and $%
\kappa _{\alpha }\left( \theta \right) $ be its curvature evaluated at the
point pinned by the tangent parallel to $\mathbf{n.}$ It turns out that 
\begin{equation}
W_{\alpha }^{\left( 1\right) }\left( \theta \right) =c_{\alpha }\kappa
_{\alpha }\left[ 1+\left( mn\right) _{\alpha \alpha }^{2}\right] ^{3/2},
\label{96.2}
\end{equation}%
i.e. the sign of $W_{\alpha }^{\left( 1\right) }\left( \theta \right) $ is
prescribed by the sign of the curvature $\kappa _{\alpha }\left( \theta
\right) $ \cite{ShP}. In turn, the term $W_{\alpha }^{\left( 1\right)
}\left( \theta \right) $ exceeds $W_{n,\alpha }^{\left( 2\right) }\left(
\theta \right) \sim n^{-1}$ for $n\gg 1.$ Hence, by (\ref{96.1}), the sign
of $W_{\alpha }^{\left( 1\right) }\left( \theta \right) $ decides the sign
of $b_{n,\alpha }\left( \theta \right) $ and therefore of $g_{n,\alpha
}^{\left( \mathbf{m}\right) }\left( \theta \right) $ near the cutoff.
Consequently, if the normal $\mathbf{n}$ corresponds to, specifically, a
saddle point on the slowness surface sheet of the $\alpha $th bulk mode,
then the in-plane group velocity $g_{n,\alpha }^{\left( \mathbf{m}\right)
}\left( \theta \right) $ near the $\left( n,\alpha \right) $th cutoffs of
sufficiently large orders $n$ varies from positive values in the
cross-sections with convex $S_{\alpha }$ to negative ones in the
cross-sections with concave $S_{\alpha }.$ A graphical interpretation and a
numerical example of the above feature are presented in \cite{ShP}.

In conclusion, let us mention two exceptional cases when Eq. (\ref{96.1})
must be modified. The first is when $\mathbf{n}$ is an acoustic axis ($%
c_{\alpha }=c_{\beta }$ for $\alpha ,\beta \in \left\{ 1,2,3\right\} $)
lying in the plane $\left( \mathbf{m}\left( \theta \right) \mathbf{,n}%
\right) $ that is not a symmetry plane. The second is when the value of $%
c_{\alpha }/c_{\beta }$ is close to a rational fraction of natural numbers,
and hence the tangent in (\ref{96.1})$_{3}$ approaches infinity, thus
signalling that the quadratic dispersion on $kH$ described by Eq. (\ref{96})$%
_{1}$ is replaced with a quasilinear dependence. A detailed treatment of
both cases is described in \cite{ShP}.

\subsubsection{High-frequency approximation\label{SSSec9.2.4}}

The high-frequency (shortwave) shape of the dispersion branches in a layered
plate is governed, first, by the transonic velocities associated with
partial bulk modes in each of the (homogeneous) layers and, secondly, by the
pair of Rayleigh velocities $v_{\mathrm{R}}$ and the (possibly existing)
Stoneley velocities $v_{\mathrm{St}}$, which in the high-frequency limit are
associated with the external plate faces and internal layer-layer
interfaces, respectively. In the case of a functionally graded plate, the
role of transonic velocities $\hat{v}$ is emulated by local minima $\min
v\left( y\right) \equiv v_{\min }$ of the modal velocity profiles. As the
frequency increases, the dispersion branches $v_{J}\left( \omega \right) $
lying above the absolute minimum $\mathrm{Min}\left( v\left( y\right) ,\hat{v%
}\right) \equiv v_{\mathrm{Min}}$ form flattened terracing patterns near
each of the above values ($\hat{v}~$or $v_{\min }$ and $v_{\mathrm{R}},$ $v_{%
\mathrm{St}}$ if the latter exceed $v_{\mathrm{Min}}$) and then collapse to
a similar pattern near a lower velocity value within this set. In turn,
those values $v_{\mathrm{R}}$ and least of $v_{\mathrm{St}},$ which are less
than $v_{\mathrm{Min}},$ provide the asymptotic limits for, usually, the
fundamental branches.

Because of abrupt drops from one plateau to another, the high-frequency
trajectory of an individual branch $v_{J}\left( \omega \right) $ defies
simple analytical description; at the same time, it is straightforward to
evaluate the trends $v_{p}\left( \omega \right) ,\ p=1,2,...,$ of each
collective plateau as a whole. The rate at which it asymptotically
approaches the given $\hat{v}$ or $v_{\min }$ in a layered or functionally
graded plate is proportional to an inverse power of $\omega $ and can be
estimated through the WKB approach as 
\begin{equation}
v_{p}\left( \omega \right) -\hat{v}\sim \omega ^{-2}\ \ \mathrm{or}\ \
v_{p}\left( \omega \right) -v_{\min }\sim {\large (}\frac{p+\frac{1}{2}}{%
\omega }{\large )}^{^{\frac{2m}{m+2}}}\ ,  \label{98}
\end{equation}%
respectively, where $\frac{1}{2}$ in the numerator of (\ref{98})$_{2}$ must
be replaced by $\frac{1}{4}$ if the local minimum $v_{\min }$ occurs at one
of the free plate faces, and $m$ is the order of the first non-zero
derivative of $v\left( y\right) $ at $v_{\min }$ (the limit $m\rightarrow
\infty $ reduces (\ref{98})$_{2}$ to (\ref{98})$_{1}$) \cite{ShPK}.

The high-frequency trend towards the values $v_{\mathrm{R}}$ and $v_{\mathrm{%
St}}$ is exponential. To spot it, let us note from (\ref{83})$_{1}$ that the 
$6\times 6$ two-point impedance $\mathbf{Z}(y_{2},y_{1})$ taken at growing
frequency tends exponentially to the limit $\mathrm{diag}(\mathbf{Z|}%
_{y_{1}},\mathbf{Z|}_{y_{2}}^{T})$, where $\mathbf{Z}|_{y_{1}}$ and $\mathbf{%
Z|}_{y_{2}}$ are the $3\times 3$ surface impedances of "fictitious"
homogeneous half-spaces (see \S \ref{Sec6}), each with the material
parameters of the actual plate taken at $y=y_{1}$ and $y=y_{2}$,
respectively (obviously, $\mathbf{Z}|_{y_{1}}=\mathbf{Z|}_{y_{2}}$ if the
layer $\left[ y_{1},y_{2}\right] $ is homogeneous). Hence, given a
traction-free plate $\left[ 0,H\right] $ consisting of $N$ homogeneous or
functionally graded layers $[y_{2}^{\left( j\right) },y_{1}^{\left( j\right)
}],\ j=1,...,N$ ($y_{1,2}^{\left( 1\right) }=0,y_{2}^{\left( N\right) }=H$),
the high-frequency limit of the dispersion equation expressed in the form (%
\ref{91})$_{2}$ or (\ref{99}) is approached exponentially and is equal to 
\begin{equation}
\det \mathbf{Z}|_{0}\det (\mathbf{Z}|_{y_{2}^{\left( 1\right) }}+\mathbf{Z}%
|_{y_{1}^{\left( 2\right) }}^{T})...\det (\mathbf{Z}|_{y_{2}^{\left(
N-1\right) }}+\mathbf{Z}|_{y_{1}^{\left( N\right) }}^{T})\det \mathbf{Z}%
|_{H}=0,  \label{99.2}
\end{equation}%
(or $\det \mathbf{Z}|_{0}\det \mathbf{Z}|_{H}=0$ if the entire plate is
functionally graded with no interfacial jumps). According to (\ref{45}) and (%
\ref{48*})$_{2}$, the leading-order equations (\ref{99.2}) are fulfilled by
the Rayleigh and (possibly existing) Stoneley velocities $v_{\mathrm{R}}$
and $v_{\mathrm{St}},$ which was to be demonstrated. The above concepts and
estimates are elaborated in \cite{AM}.

An eye-catching feature of the dispersion spectra of anisotropic homogeneous
plates is the possible phenomenon of the branch "weaving", which arises due
to the slowness curve concavity at the transonic state or due to the "nearly
fulfilled" conditions for the existence of the supersonic Rayleigh wave of
the symmetric type, see \cite{SA,LT} and \cite{AM,Sh3}. Remarkably, all
these geometrically intricate patterns may be shown to possess a common
invariant property \cite{Sh0}. It is that the branches $v_{J}\left(
k_{x}\right) $ and $v_{J}\left( \omega \right) $ cannot have extreme points
in the range of velocity values above the constant benchmark $V,$ which is
equal to the largest of the zero-frequency limits $v_{02}$ and $v_{03}$ ($=%
\sqrt{\lambda _{2,3}/\rho },$ see (\ref{93})) of, specifically, the
dispersive fundamental velocity branches (i.e. excluding the SH
non-dispersive branch, which exists if $\left( \mathbf{m,n}\right) $ is the
symmetry plane), i.e. 
\begin{equation}
v_{J}^{\prime }\left( k_{x}\right) ,\ v_{J}^{\prime }\left( \omega \right)
\neq 0\ \mathrm{at}\ v>V=\left\{ 
\begin{array}{l}
\max (v_{02},v_{03})\ \mathrm{if\ there\ is~no\ SH\ wave\ uncoupling,} \\ 
\mathrm{one\ of\ }(v_{02},v_{03})\ \mathrm{if\ the\ other\ is\ the\ SH\
wave\ velocity.}%
\end{array}%
\right.  \label{100}
\end{equation}%
The proof is provided in Appendix 2. Note that this property is in line with
the fact that the uppermost velocity branch, whose limit $V$ is involved in (%
\ref{100}), is guaranteed to have a downbent longwave onset (see e.g. \cite%
{Sh2}); at the same time, this assertion cannot be extended to transversely
inhomogeneous plates and nor can (\ref{100}).

\subsection{Related problems\label{SSec9.3}}

\subsubsection{Layer on a half-space\label{SSSec9.3.1}}

Many micro- and macroscale applications engage guided waves in a layer $%
\left[ 0,H\right] $ that is free on one side and bonded to a homogeneous
substrate on the other (the Love waves). Formally, this boundary-value
problem may be viewed either as one in a free half-space $y\geq 0,$ where
the system matrix $\mathbf{Q}\left( y\right) $ of (\ref{5}), possibly
varying within the layer $\left[ 0,H\right] ,$ has a jump at $y=H,$ and is
constant at $[H,\infty ),$ or as a similar problem in a free bilayered plate 
$\left[ 0,H_{\infty }\right] ,$ where the layer $\left[ 0,H\right] $ may be
discretely or functionally graded, while the layer $\left[ H,H_{\infty }%
\right] $ is homogeneous and its lower boundary $y=H_{\infty }$ extends
towards infinity. Seeking the localized guided waves, i.e., those that do
not leak energy into the substrate, restricts their velocity $v=\omega
/k_{x} $ by the transonic velocity (bulk-wave threshold) $\hat{v}^{\left(
s\right) } $ of the substrate material. Interpreting a substrate as an
infinitely thick layer within an overall plate helps to grasp why the three
fundamental branches of the Lamb wave spectrum come down to a single one in
the Love wave spectrum with a longwave origin at the Rayleigh wave $v_{%
\mathrm{R}}^{\left( s\right) }$ of the substrate (often referred to as the
Rayleigh-Love branch; let us denote it by $v_{\mathrm{R-L}}\left(
k_{x}\right) $).

Given that the layer surface $y=0$ is traction-free, the boundary value
equation can be written as 
\begin{equation}
\det \left( \mathbf{z}\left[ \omega ,k_{x}\right] +\mathbf{Z}\left[ v\right]
\right) =0,  \label{101}
\end{equation}%
where $\mathbf{z}\left[ \omega ,k_{x}\right] \equiv \mathbf{z}\left(
y_{1}\right) |_{\mathbf{z}\left( 0\right) =\mathbf{0}}=0$ is the $3\times 3$
conditional impedance for the layer and $\mathbf{Z}\left[ v\right] $ is the $%
3\times 3$ impedance for the substrate. The basic patterns of the Love wave
dispersion spectrum may be viewed as the hybridization of the spectra of the
layer and the substrate, which is governed by the relationship between their
Rayleigh and transonic velocities $v_{\mathrm{R}}^{\left( l,s\right) }$ and $%
\hat{v}^{\left( l,s\right) }$ (the superscript $l$ or $s$ indicates the
layer or substrate, respectively). Consider the typical situation where $v_{%
\mathrm{R}}^{\left( l\right) }$ is the lower bound of the high-frequency
velocity in the layer and $v_{\mathrm{R}}^{\left( l\right) }<\hat{v}^{\left(
l\right) },v_{\mathrm{R}}^{\left( s\right) }<\hat{v}^{\left( s\right) }$. If
the layer is relatively "fast" in the sense that $v_{\mathrm{R}}^{\left(
s\right) }<v_{\mathrm{R}}^{\left( l\right) }$, the spectrum consists solely
of the Rayleigh-Love branch $v_{\mathrm{R-L}}\left( k_{x}\right) $, which
goes up from $v_{\mathrm{R}}^{\left( s\right) }$ at $k_{x}=0$ and tends to $%
v_{\mathrm{R}}^{\left( l\right) }<\hat{v}^{\left( s\right) }$ or terminates
(as a real-valued branch) with $\hat{v}^{\left( s\right) }<v_{\mathrm{R}%
}^{\left( l\right) }$ at finite $k_{x}$. If the layer is relatively "slow",
i.e. $v_{\mathrm{R}}^{\left( l\right) }<v_{\mathrm{R}}^{\left( s\right) }$
and $\hat{v}^{\left( l\right) }<~\hat{v}^{\left( s\right) },$ then the
spectrum contains an infinite extent of the branch $v_{\mathrm{R-L}}\left(
k_{x}\right) ,$ going from $v_{\mathrm{R}}^{\left( s\right) }$ to $v_{%
\mathrm{R}}^{\left( l\right) }$, along with the continuum of descending
upper branches. A similar "common sense consideration" tells us that a
"light slow" or "dense" coating layer should cause the shortwave extent of
the branch $v_{\mathrm{R-L}}\left( k_{x}\right) $ to mimic the lowest branch
in the free/clamped layer or to approach the lowest (flexural) fundamental
branch in the free/free layer, respectively (see examples in \cite{ShE1}%
\footnote{{\footnotesize Note the misprints: the lowest short curve in Fig.
5a should be dashed, and a term is missing on the right-hand side of Eq.
(24).}}).

At the same time, the reasoning in general terms, such as relatively "slow"
or "fast" coating, is certainly not enough to capture some subtle spectral
features. For instance, knowing that the Rayleigh-Love branch $v_{\mathrm{R-L%
}}\left( k_{x}\right) $ starts from $v_{\mathrm{R}}^{\left( s\right) }$ and
tends to $v_{\mathrm{R}}^{\left( l\right) }$, it is natural to expect that
the longwave onset of this branch goes up or down when $v_{\mathrm{R}%
}^{\left( s\right) }$ is, respectively, less or greater than $v_{\mathrm{R}%
}^{\left( l\right) }$. \ Indeed, this is usually the case, but it turns out
that this is not always so. Provided the layer is homogeneous, the sought
slope is positive proportional to the following difference \cite{ShE2}: 
\begin{equation}
\frac{dv_{\mathrm{R-L}}\left( k_{x}\right) }{dk_{x}}|_{k_{x}=0}\sim \left( 
\frac{|\mathbf{u}_{\mathrm{R}}^{\left( s\right) T}\mathbf{m}|^{2}}{|\mathbf{u%
}_{\mathrm{R}}^{\left( l\right) T}\mathbf{m}|^{2}}v_{\mathrm{R}}^{\left(
l\right) 2}-v_{\mathrm{R}}^{\left( s\right) 2}\right) ,  \label{102}
\end{equation}%
where $\mathbf{u}_{\mathrm{R}}^{\left( s\right) },$ $\mathbf{u}_{\mathrm{R}%
}^{\left( l\right) }$ are the similarly normalized displacement vectors of
the Rayleigh wave propagating along the direction $\mathbf{m}$ in the free
half-spaces made of the substrate and layer materials, respectively. Thus,
the derivative's sign may not actually coincide with the sign of $\left( v_{%
\mathrm{R}}^{\left( l\right) }-v_{\mathrm{R}}^{\left( s\right) }\right) .$
If the sagittal plane $\left( \mathbf{m,n}\right) $ is the symmetry plane,
then the quantity on the right-hand side of (\ref{102}) may be simplified to
a form having the sign of $\left( \kappa ^{\left( l\right) }-\kappa ^{\left(
s\right) }\right) ,$ where $\kappa ^{\left( l,s\right) }=\sqrt{\mathbf{m}^{T}%
\mathbf{N}_{3}^{\left( l,s\right) }\mathbf{m/}12\rho ^{\left( l,s\right) }}$
are the Kirhhoff-like coefficients describing the origin of the flexural
branch $v_{1}^{\left( l,s\right) }\left( k_{x}\right) =\kappa ^{\left(
l,s\right) }k_{x}H+...$ in the free plates of a layer and substrate
materials (see \S \ref{SSSec9.2.2}).

\subsubsection{Reflection/transmission via impedance\label{SSSec9.3.2}}

Consider the reflection/transmission problem from an inhomogeneous layer $%
\left[ y_{1},y_{2}\right] $ bonded between two solid homogeneous substrates
1 ($y\leq y_{1}$) and 2 ($y\geq y_{2}$). It was already discussed in \S \ref%
{Sec5}, where the reflection and transmission matrices were derived in terms
of the matricant. Let us find their expression via the plate impedance,
which may be advantageous for computations in the large frequency-thickness
domain. Keeping the mode numbering as set in \S \ref{SSSec3.2.1}, assuming
the incidence from substrate 1,\ and appropriately rearranging the
continuity condition at the interfaces $y=y_{1}$ and $y=y_{2},$ we obtain%
\begin{equation}
\begin{array}{c}
\mathbf{R}^{\left( 11\right) }=\mathbf{\Xi }_{2}^{\left( 1\right) -1}\left( 
\mathbf{Z}^{\left( 1\right) T}+\mathbf{z}\left( y_{1}\right) \right)
^{-1}\left( \mathbf{Z}^{\left( 1\right) }-\mathbf{z}\left( y_{1}\right)
\right) \mathbf{\Xi }_{1}^{\left( 1\right) } \\ 
=-\mathbf{\Xi }_{2}^{\left( 1\right) -1}\mathbf{\Xi }_{1}^{\left( 1\right)
}+i\left[ \mathbf{\Xi }_{1}^{\left( 1\right) T}\left( \mathbf{G{\Large /}G}%
_{4}\right) \mathbf{\Xi }_{2}^{\left( 1\right) }\right] ^{-1}, \\ 
\mathbf{T}^{\left( 12\right) }=i\left[ \mathbf{\Xi }_{1}^{\left( 1\right)
T}\left( \mathbf{G{\Large /}G}_{2}\right) \mathbf{\Xi }_{2}^{\left( 1\right)
}\right] ^{-1},%
\end{array}
\label{89}
\end{equation}%
where $\mathbf{Z}^{\left( i\right) }=i\mathbf{\Xi }_{3}^{\left( i\right) }%
\mathbf{\Xi }_{1}^{\left( i\right) -1},~i=1,2,$ are the impedances of the
substrates\footnote{{\footnotesize Here, the impedances }$\mathbf{Z}^{\left(
1\right) }\left[ v\right] ${\footnotesize \ and possibly }$\mathbf{Z}%
^{\left( 2\right) }\left[ v\right] ${\footnotesize \ are defined in the
supersonic velocity interval and hence are not Hermitian impedances, unlike
the subsonic case discussed in \S \ref{Sec6}.}}, $\mathbf{z}\left(
y_{1}\right) =\mathbf{z}\left( y_{1}\right) |_{\mathbf{z}\left( y_{2}\right)
=\mathbf{Z}^{\left( 2\right) }}$ is the conditional impedance, and 
\begin{equation}
\begin{array}{l}
\mathbf{G{\Large /}G}_{4}=\mathbf{Z}_{1}+\mathbf{Z}^{\left( 1\right) T}-%
\mathbf{Z}_{2}\left( \mathbf{Z}_{4}+\mathbf{Z}^{\left( 2\right) }\right)
^{-1}\mathbf{Z}_{3}, \\ 
\mathbf{G{\Large /}G}_{2}=\mathbf{Z}_{2}-\left( \mathbf{Z}_{1}+\mathbf{Z}%
^{\left( 1\right) T}\right) \mathbf{Z}_{3}^{-1}\left( \mathbf{Z}_{4}+\mathbf{%
Z}^{\left( 2\right) }\right)%
\end{array}
\label{89.0}
\end{equation}%
are the Schur complements of the right off-diagonal and diagonal blocks $%
\mathbf{G}_{4}$ and $\mathbf{G}_{2}$ of the $6\times 6$ matrix $\mathbf{G}=%
\mathbf{Z}\left( y_{2},y_{1}\right) +\mathrm{diag}\left( \mathbf{Z}^{\left(
1\right) T},\mathbf{Z}^{\left( 2\right) }\right) .$ The equation $\det 
\mathbf{G}=0$ is an equivalent to the (\ref{R1}) form of the dispersion
equation for the guided wave in a free layer.

For completeness, we also present the impedance formulation of the
reflection and transmission matrices from the interface of two substrates
which are in direct contact (without an intermediate layer). Inserting the
interface scattering matrix $\mathbf{S}^{\left( 12\right) }=\mathbf{\Xi }%
^{\left( 1\right) T}\mathbf{\mathbb{T}\Xi }^{\left( 2\right) }$ into the
left-hand side of Eq. (\ref{48.S}) yields 
\begin{equation}
\begin{array}{c}
\mathbf{R}^{\left( 11\right) }=\mathbf{\Xi }_{2}^{\left( 1\right) T}\left( 
\mathbf{Z}^{\left( 2\right) }-\mathbf{Z}^{\left( 1\right) }\right) \left( 
\mathbf{Z}^{\left( 1\right) T}+\mathbf{Z}^{\left( 2\right) }\right) ^{-1}%
\mathbf{\Xi }_{1}^{\left( 1\right) -T},\  \\ 
\mathbf{T}^{\left( 12\right) }=-\left[ \mathbf{\Xi }_{2}^{\left( 1\right)
T}\left( \mathbf{Z}^{\left( 1\right) T}+\mathbf{Z}^{\left( 2\right) }\right) 
\mathbf{\Xi }_{1}^{\left( 2\right) }\right] ^{-1}.%
\end{array}
\label{89.2}
\end{equation}

\subsubsection{Immersed plate\label{SSSec9.3.3}}

Consider guided waves in a transversely inhomogeneous plate $\left[ 0,H%
\right] $ immersed in an ideal fluid with density $\rho _{f}$ and speed of
sound $c_{f}.$ In this context, it is suitable to utilize the trace velocity 
$v=\omega /k_{x}$ (or $s=v_{x}^{-1}$ at $k_{x}=0$) as one of the two
dispersion parameters, the other being $\omega $ or $k_{x}$. Guided waves
carrying energy along the plate and vanishing in the fluid depth propagate
with a real velocity in the range $v<c_{f}$ called subsonic (in the present
context). Supersonic waves propagate with a complex velocity ($\func{Re}%
v>c_{f}$) and hence are leaky waves transmitting energy flux into the fluid,
with $\func{Im}v$ and energy leakage remaining small as long as $\rho
_{f}/\rho _{\mathrm{plate}}$ is small. Note that complex $v$ permits either $%
\omega $ and $k_{x}$ or both to be complex; when addressing leaky waves
below, we will keep $\omega $ real as assumed throughout the following text.

\paragraph{Subsonic spectrum\label{SSSSec9.3.3.1}}

Extending the idea \cite{BGL} of using the admittance matrix for the
fluid-solid interfacial contact, the dispersion equation for an immersed
plate can be written as 
\begin{equation}
\begin{array}{c}
(Y_{1}^{\left( \mathbf{n}\right) }-Y_{f})(Y_{4}^{\left( \mathbf{n}\right)
}-Y_{f})=Y_{2}^{\left( \mathbf{n}\right) }Y_{3}^{\left( \mathbf{n}\right) }\
\ \Leftrightarrow \\ 
Y_{f}=\frac{1}{2}\left[ Y_{1}^{\left( \mathbf{n}\right) }+Y_{4}^{\left( 
\mathbf{n}\right) }\pm \sqrt{\left( Y_{1}^{\left( \mathbf{n}\right)
}-Y_{4}^{\left( \mathbf{n}\right) }\right) ^{2}+4Y_{2}^{\left( \mathbf{n}%
\right) }Y_{3}^{\left( \mathbf{n}\right) }}\right] {\Large ,}%
\end{array}
\label{103}
\end{equation}%
where 
\begin{equation}
Y_{1,...,4}^{\left( \mathbf{n}\right) }\left[ v,\cdot \right] =\mathbf{n}^{T}%
\mathbf{Y}_{1,...,4}\mathbf{n},\ \ Y_{f}\left( v\right) =\frac{\sqrt{%
1-v^{2}/c_{f}^{2}}}{\rho _{f}v^{2}},  \label{104}
\end{equation}%
$\mathbf{n}$ is a unit normal to the plate face, $\cdot $ here and below
stands for $\omega $ or $k_{x}$, the sign in front of $Y_{f}$ implies the
choice of decreasing modes in fluid (at $v>0$), and $\mathbf{Y}_{1,...,4}$
are the blocks of the plate admittance matrix $\mathbf{Y=Z}^{-1}\left(
H,0\right) $ (see (\ref{83})). Equation (\ref{103}) admits complex-valued $%
\omega $ and $k_{x}$ (thus remaining valid for leaky waves in the supersonic
domain, see below). It simplifies in the case of real at $\omega $ and $%
k_{x},$ when $\mathbf{Y}$ is Hermitian, so $Y_{1}^{\left( \mathbf{n}\right)
} $ and $Y_{4}^{\left( \mathbf{n}\right) }$ are real and $Y_{2}^{\left( 
\mathbf{n}\right) }=Y_{3}^{\left( \mathbf{n}\right) \ast }.$ If the plate is
homogeneous or if the variation of its material properties is symmetric
about the midplane (see \S \ref{SSSec3.2.4}), then $Y_{1}^{\left( \mathbf{n}%
\right) }=Y_{4}^{\left( \mathbf{n}\right) }$. According to \S \ref{SSec9.1},
for any fixed $k_{x}\neq 0$, the functions $Y_{1}^{\left( \mathbf{n}\right) }%
\left[ v\right] $ and $Y_{4}^{\left( \mathbf{n}\right) }\left[ v\right] $
begin with positive values at $v=0$ and increase monotonically between the
poles. By (\ref{83}), each of $Y_{i}^{\left( \mathbf{n}\right) }\left[
v,\cdot \right] $ has poles at $\det \mathbf{M}_{3}=0$, i.e. on the
free-plate dispersion branches $\check{v}_{J}\left( \cdot \right)
,~J=1,2,... $, which are thereby important markers in the sought immersed
plate spectrum. Henceforth in this subsection, we use the inverted hat
symbol for the benchmark parameters of the free-plate spectrum.

The impact of fluid loading is largely characterized by the leading-order
asymptotic of the plate admittance contractions (\ref{104})$_{1}$%
\begin{equation}
Y_{1,...,4}^{\left( \mathbf{n}\right) }\left[ v,\cdot \right] \varpropto 
\frac{a_{J\left( 1,...,4\right) }\left( \cdot \right) }{\check{v}%
_{J}^{2}\left( \cdot \right) -v^{2}}\   \label{106}
\end{equation}%
at $v$ close to the poles $\check{v}_{J}\left( k_{x}\right) $ or $\check{v}%
_{J}\left( \omega \right) $ (we are barring extraordinary events of
2nd-order poles arising due to branch crossing or group velocity vanishing).
Using the formulas given in Appendix 2, the residue $a_{J\left( 1\right)
}\left( \cdot \right) $ for $Y_{1}^{\left( \mathbf{n}\right) }\left[ v,\cdot %
\right] $ is found to be of the form 
\begin{equation}
a_{J\left( 1\right) }\left( k_{x}\right) =\dfrac{\check{v}_{J}}{4{\large %
\langle }\overline{\mathcal{\check{K}}}_{J}{\large \rangle }}\left\vert 
\check{u}_{J}^{\left( \mathbf{n}\right) }\left( 0\right) \right\vert
^{2}\left( >0\right) ,\ a_{J\left( 1\right) }\left( \omega \right) |_{1}=%
\dfrac{\check{v}_{J}}{\check{g}_{J}^{\left( \mathbf{m}\right) }}a_{J\left(
1\right) }\left( k_{x}\right) ,\   \label{107}
\end{equation}%
where ${\large \langle }\overline{\mathcal{\check{K}}}_{J}{\large \rangle }$
is the time- and thickness-averaged kinetic energy, $\check{u}_{J}^{\left( 
\mathbf{n}\right) }\equiv \mathbf{\check{u}}_{J}^{T}\left( 0\right) \mathbf{n%
}$ is the normal component of displacement on the free face $y=0,$ and $%
\check{g}_{J}^{\left( \mathbf{m}\right) }=\mathbf{\check{g}}_{J}^{T}\mathbf{m%
}$ is the in-plane component of the group velocity, all referred to a free
plate and taken on the branch $\check{v}_{J}$. The residues for $%
Y_{4}^{\left( \mathbf{n}\right) }$ and $\sqrt{Y_{2}^{\left( \mathbf{n}%
\right) }Y_{3}^{\left( \mathbf{n}\right) }}=\left\vert Y_{2}^{\left( \mathbf{%
n}\right) }\right\vert $ differ from (\ref{107})$_{1}$ by replacing $|\check{%
u}_{J}^{\left( \mathbf{n}\right) }\left( 0\right) |^{2}$ with $|\check{u}%
_{J}^{\left( \mathbf{n}\right) }\left( H\right) |^{2}$ and $|\check{u}%
_{J}^{\left( \mathbf{n}\right) }\left( 0\right) ||\check{u}_{J}^{\left( 
\mathbf{n}\right) }\left( H\right) |$, respectively, so that the leading
order of the square root on the right-hand side of (\ref{103})$_{2}$ equals $%
|Y_{1}^{\left( \mathbf{n}\right) }+Y_{4}^{\left( \mathbf{n}\right) }|$.
Hence, one of the two curves corresponding to different signs in (\ref{103})$%
_{2}$ tends to $+\infty $ as it approaches the pole $\check{v}_{J}\left(
\cdot \right) $ from the left, while the other is not affected by the pole
(if $k_{x}H\ll 1,$ then the former curve tends to the pole $\check{v}_{1}$
as $\sim \left[ \rho \left( \check{v}_{1}^{2}-v^{2}\right) kH\right] ^{-1},$
while the latter trails as $\sim kH$). According to (\ref{103}), it remains
to figure out where these curves cross the curve $Y_{f}\left( v\right) $ (%
\ref{104})$_{1}$ to identify the principal configuration of the subsonic
spectrum without any calculations. An elementary graphics readily shows
that, regardless of the materials involved, it always includes two
fundamental branches $v_{1}\left( \cdot \right) $ and $v_{2}\left( \cdot
\right) $: one starts at $v=0$ and extends below the free-plate flexural
branch $\check{v}_{1}\left( \cdot \right) $; the other starts at the lesser
of the values $c_{f}$ and $\check{v}_{2}\left( 0\right) \equiv \check{v}%
_{02} $ and extends below $\check{v}_{2}\left( \cdot \right) $ (here $\check{%
v}_{2}\left( \cdot \right) $ is the lower of the two free-plate branches
with the origin (\ref{93}) or the only one of them if the other is a
dispersionless branch of the uncoupled SH wave). The branches $v_{1,2}\left(
\cdot \right) $ asymptotically tend to the Scholte wave velocity(ies) at the
plate/fluid interfaces.

The above considerations, based on the plate admittance approach, are
applicable to both homogeneous and transversely inhomogeneous plates. As
regards the explicit estimates, they are available for a homogeneous plate 
\cite{Sh4} but defy compact form in the case of inhomogeneous plates, except
for the flexural branch, whose longwave asymptotic 
\begin{equation}
v_{1}\left( k_{x}\right) =\frac{\overline{\kappa }\left( k_{x}H\right) ^{3/2}%
}{\sqrt{k_{x}H+2\rho _{\mathrm{f}}/\left\langle \rho \right\rangle }}
\label{107.1}
\end{equation}%
takes over the well-known Osborne and Hart's approximation \cite{OH}, but
with $\left\langle \rho \right\rangle $ instead of $\rho $ and the
coefficient $\overline{\kappa }$ instead of its counterpart for a
homogeneous plate $\kappa $, see (\ref{95.0}) and (\ref{95}). Formula (\ref%
{107.1}) can be modified to fit the entire branch extent via the Pad\'{e}
approximation similarly to \cite{PShK}.

A similar methodology applies to the cases where a plate is in contact with
two different fluids on opposite sides and is fluid-loaded on one side while
being free of traction on the other (cf. Love waves). The dispersion
equation (\ref{103})$_{1}$ modifies by replacing $Y_{f}$ with $Y_{f1}\neq
Y_{f2}$ in the former case and takes the form $Y_{1}^{\left( \mathbf{n}%
\right) }=Y_{f}$ in the latter. For more details, see \cite{Sh4}.

Note in conclusion that while the waves decreasing into the depth of the
fluid must propagate with a real subsonic velocity $v<c_{f}$, the inverse is
not true, that is, a wave with a real subsonic velocity may not be
evanescent. For instance, the flexural-type velocity branch permits real
values if the increasing mode is chosen in one or both fluid half-spaces.
Interestingly, the latter setting may lead to the formation of a real-valued
loop on this branch, see \cite{ShPD12}$_{1}$.

\paragraph{Supersonic spectrum\label{SSSSec9.3.3.2}}

In a typical case of a relatively light fluid, $\rho _{f}/\rho \ll 1,$ the
supersonic dispersion spectrum of the leaky-wave velocity $v_{J}\left(
\omega \right) =\func{Re}v_{J}+i\func{Im}v_{J}$ can be regarded as a
perturbation of the free-plate branches $\check{v}_{J}\left( \omega \right) $
in the range above $c_{f}.$ Afar from the cutoffs, the imaginary part $\func{%
Im}v_{J}$, which is the measure of leakage, and the difference between $%
\func{Re}v_{J}$ and $\check{v}_{J}$ are small of the order $\rho _{f}/\rho $
and $\left( \rho _{f}/\rho \right) ^{2}$, respectively. The leaky wave
incorporates the fluid modes increasing away from the plate, and so the
value $\func{Im}v_{J}$ is negative whenever the in-plane group velocity $%
\check{g}_{J}^{\left( \mathbf{m}\right) }$ associated with the referential
free-plate branch $\check{v}_{J}$ is positive, which is a predominant case.
However, $\check{g}_{J}^{\left( \mathbf{m}\right) }$ may be negative, as
often occurs for the free-plate modes in the vicinity of the cutoff
(vertical resonance) frequencies, see \S \ref{SSSec9.2.3}. When the plate is
immersed in a fluid, such modes give rise to "unusual" leaky waves that
incorporate decreasing fluid modes and have positive $\func{Im}v_{J}.$

One more interesting feature is the existence of two drastically different
dispersion patterns occurring for a $J$th mode near its fluid-uncoupled and
fluid-coupled resonances. The former is when the polarization $\mathbf{%
\check{a}}_{J}$ of the resonant free-plate mode is orthogonal to the plate
normal $\mathbf{n;}$ in this case, the slowness $s_{J}\left( \equiv
v_{J}^{-1}\right) $ in the immersed plate reaches zero at the cutoff
frequency, as it does in the free plate. The latter is when $\mathbf{\check{a%
}}_{J}^{T}\mathbf{n}=0$, in this case, $s_{J}$ at the cutoff remains
non-zero and has commensurate real and imaginary parts.

Analytical estimates of the leaky-wave spectrum far and near the cutoffs
obtained in detail for homogeneous immersed plates are available in \cite%
{ShPD12}$_{2}$.

\paragraph{Reflection/transmission\label{SSSSec9.3.3.3}}

Assume that a bulk mode propagating in a fluid with real $\omega ,~k_{x}$
and (supersonic) velocity $v=c_{f}\sin ^{-1}\theta _{\mathrm{inc}}$ impinges
on an immersed transversely inhomogeneous plate $\left[ 0,H\right] .$ The
reflection and transmission coefficients are determined by the formulas 
\begin{equation}
\begin{array}{c}
R\left[ v,\omega \right] =\dfrac{(Y_{1}^{\left( \mathbf{n}\right)
}+Y_{f})(Y_{4}^{\left( \mathbf{n}\right) }-Y_{f})-|Y_{2}^{\left( \mathbf{n}%
\right) }|^{2}}{(Y_{1}^{\left( \mathbf{n}\right) }-Y_{f})(Y_{4}^{\left( 
\mathbf{n}\right) }-Y_{f})-|Y_{2}^{\left( \mathbf{n}\right) }|^{2}}, \\ 
T\left[ v,\omega \right] =\dfrac{-2Y_{3}^{\left( \mathbf{n}\right) }Y_{f}}{%
(Y_{1}^{\left( \mathbf{n}\right) }-Y_{f})(Y_{4}^{\left( \mathbf{n}\right)
}-Y_{f})-|Y_{2}^{\left( \mathbf{n}\right) }|^{2}}e^{i\varphi },%
\end{array}
\label{108}
\end{equation}%
where $\func{Im}Y_{1,4}^{\left( \mathbf{n}\right) }=0$ and $Y_{2}^{\left( 
\mathbf{n}\right) \ast }=Y_{3}^{\left( \mathbf{n}\right) }$ due to $\mathbf{%
Y=Y}^{+},$ $Y_{f}=-i\frac{1}{\rho _{f}v^{2}}\sqrt{v^{2}/c_{f}^{2}-1}$ is
purely imaginary (see (\ref{104})$_{2}$ at $v>c_{f}$), $\varphi =\frac{%
\omega }{v}H\cos \theta _{\mathrm{inc}}$, and the normalization of the modes
in fluid to unit energy-flux normal component is used. The latter ensures
the identity $\left\vert R\right\vert ^{2}+\left\vert T\right\vert ^{2}=1$.
The denominator of (\ref{108}) can be recognized as the left-hand side of
the dispersion equation (\ref{103}) with $Y_{2}^{\left( \mathbf{n}\right)
\ast }=Y_{3}^{\left( \mathbf{n}\right) }$ (due to real $\omega ,$ $k_{x}$)$.$
If the plate $\left[ 0,H\right] $ is in contact with fluid on one side (say,
at $y=0$) and free of traction on the other, the reflection coefficient
simplifies to the form $R=(Y_{1}^{\left( \mathbf{n}\right)
}+Y_{f})(Y_{1}^{\left( \mathbf{n}\right) }-Y_{f})^{-1}.$ Replacing the plate
with a (solid) half-space $y\geq 0$ retains the above formula for $R$ except
that the plate admittance block is replaced with the $3\times 3$ half-space
admittance (see \S \ref{Sec6}).

Consider the possible zeros of the reflection and transmission coefficients (%
\ref{108}). In a general situation, i.e. if the plate has an arbitrary
profile of inhomogeneity and unrestricted anisotropy, the values of $%
Y_{1}^{\left( \mathbf{n}\right) }$ and $Y_{4}^{\left( \mathbf{n}\right) }$
are not equal and $Y_{3}^{\left( \mathbf{n}\right) }$ is complex; hence, the
zeros of $R$ and $T$ are determined by complex-valued equations and may
therefore occur only at isolated points on the parameter plane $\left(
v,\omega \right) $. At the same time, if the variation of elastic properties
is symmetric about the midplane of the plate (in particular, if it is
homogeneous), then $Y_{1}^{\left( \mathbf{n}\right) }=Y_{4}^{\left( \mathbf{n%
}\right) }$ and hence the condition for zero reflection $R\left[ v,\omega %
\right] =0$ reduces to the (real) equation 
\begin{equation}
Y_{1}^{\left( \mathbf{n}\right) 2}-|Y_{2}^{\left( \mathbf{n}\right)
}|^{2}=Y_{f}^{2},  \label{109}
\end{equation}%
defining the curves of zero reflection $v\left( \omega \right) \left(
>c_{f}\right) .$ Provided that the ratio $\rho _{f}/\rho $ is small enough,
one of these curves forms a closed arch, which starts at $\omega =0$ and
extends in between $c_{f}$ and one of the fundamental branches of the
free-plate spectrum (see example in \cite{ShPD12}$_{1}$). Besides, there is
a set of zero-reflection curves interlacing with the higher-order branches $%
\check{v}_{J}\left( \omega \right) $ of the free plate.

According to (\ref{108})$_{2}$, the condition for zero transmission $T\left[
v,\omega \right] =0$ is 
\begin{equation}
Y_{3}^{\left( \mathbf{n}\right) }\left[ v,\omega \right] =0,  \label{110}
\end{equation}%
where, by (\ref{32}) and (\ref{83}), $Y_{3}^{\left( \mathbf{n}\right) }$ is
real (and equal to $Y_{2}^{\left( \mathbf{n}\right) }$) if a transversely
inhomogeneous plate possesses a symmetry plane orthogonal to $\mathbf{e}_{1}$
and/or $\mathbf{e}_{2}$. In this case, (\ref{110}) is a real equation that
defines continuous curves of zero transmission on the $\left( v,\omega
\right) $-plane $\left( v,\omega \right) $. They can be shown to be confined
to the area strictly above the free-plate fundamental branch $\check{v}%
_{J}\left( \omega \right) \left( >c_{f}\right) $ (see the intersection of
the curves $Y_{1}^{\left( \mathbf{n}\right) }\pm Y_{2}^{\left( \mathbf{n}%
\right) }$ plotted in Fig. 3 of \cite{Sh4}).

It is noteworthy that Eq. (\ref{110}) depends solely on the layer parameters
and hence is the same regardless of whether the given layer is immersed into
fluid as considered above, or embedded between two solid substrates with
sliding contact at the interfaces, in which case (\ref{110}) nullifies all
components of the transmission matrix (see \S \ref{SSSec4.5.2}).

\section{Conclusion\label{Sec10}}

This review was intended to outline the application of the Stroh formalism
to current mainstream problems of theoretical solid-state acoustics.
Relevant topics that can be addressed by the same methodology but are not
covered here are mentioned in the Introduction; their list includes both
well-established and innovative research axes. Further developments are
anticipated due to the emergence of a new generation of materials and, in
parallel, the enhancement of computing power. Such advances call for an
appropriately inclusive formulation of the physical model and an adequate
reinforcement of the analytical techniques. The Stroh formalism is the right
platform for that, and the present effort is hoped to help engage its
fruitful potential.

\section{Acknowledgements\label{Sec11}}

The author is grateful to all his coauthors of publications mentioned in
this review. Special thanks go to Michel Destrade for encouraging the idea
of this review and David Barnett for reading the manuscript and providing
valuable comments.

\pagebreak

\pagebreak

\section*{Appendix 1. Two related setups 
\addcontentsline{toc}{section}{Appendix 1. Two related setups}}

\subsection*{A1.1 PDF formulation 
\addcontentsline{toc}{subsection}{A1.1
PDF
formulation}}

Stroh's development \cite{Stroh} proceeded from assuming the steady wave
solution of the form $\mathbf{u=u}\left( x-vt,y\right) ,$ whose substitution
in Eq. (\ref{1}) leads to%
\begin{equation}
\frac{\partial }{\partial x}\left( \mathbf{t}_{1}-\rho v^{2}\frac{\partial 
\mathbf{u}}{\partial x}\right) +\frac{\partial }{\partial y}\mathbf{t}_{2}=%
\mathbf{0}  \label{21}
\end{equation}%
and hence to 
\begin{equation}
\mathbf{t}_{1}-\rho v^{2}\frac{\partial \mathbf{u}}{\partial x}=\frac{%
\partial \mathbf{\phi }}{\partial y},\ \mathbf{t}_{2}=-\frac{\partial 
\mathbf{\phi }}{\partial x},  \label{22}
\end{equation}%
where $\mathbf{\phi =\phi }\left( x-vt,y\right) $ is the stress-function
vector. Furthermore, Eqs. (\ref{22}) may be manipulated into the form \cite%
{ChadS} 
\begin{equation}
\left( \mathbf{N}\left[ v^{2}\right] \frac{\partial }{\partial x}-\frac{%
\partial }{\partial y}\right) \left( 
\begin{array}{c}
\mathbf{u}\left( x-vt,y\right) \\ 
\mathbf{\phi }\left( x-vt,y\right)%
\end{array}%
\right) =\mathbf{0},  \label{23}
\end{equation}%
where $\mathbf{N}\left[ v^{2}\right] $ is defined in (\ref{8})$_{1}$.

The above derivation of (\ref{23}) allows the spatial dependence $\rho =\rho
\left( y\right) $ and $c_{ijkl}=c_{ijkl}\left( x,y\right) $. Once the latter
is restricted to $c_{ijkl}\left( y\right) $, the solution can be sought in
the form $\mathbf{\phi }\left( x-vt,y\right) =\mathbf{\phi }\left( y\right)
e^{ik_{x}\left( x-vt\right) },$ which reduces PDS (\ref{23}) to the ODS
equivalent to (\ref{5}) with $\omega =kv$. In the case of a homogeneous
medium, Eq. (\ref{23}) admits partial solutions of d'Alembert's form $\left( 
\mathbf{u\ \phi }\right) ^{T}=\mathbf{\xi }f\left( x-vt+py\right) ,$ where $%
\mathbf{\xi }$ and $p$ are the eigenvector and eigenvalue of $\mathbf{N}%
\left[ v\right] $ and $f\left( \cdot \right) $ is an arbitrary function. The
d'Alembert-type solution of the same PDS but formulated with respect to the
state vector $\left( \mathbf{u}_{,x}\mathbf{\ \mathbf{t}}_{2}\right) ^{T}$
has underlaid an elegant derivation of \cite{Pa} for the surface waves of
general profile in the sense that they are not proportional to $%
e^{ik_{x}\left( x-vt\right) }.$

\subsection*{A1.2 ODS with temporally modulated coefficients 
\addcontentsline{toc}{subsection}{A1.2
ODS
with
temporally
modulated
coefficients}%
}

A hot topic, recently emerged in optics and acoustics, is the wave
propagation in time-modulated metamaterials \cite{NXNH}-\cite{T...C}.
Consider how one such model affects the Stroh formalism. Assume that the
material coefficients of the medium are functions of the composite
space-time variable%
\begin{equation}
\varsigma =y-ct  \label{t1}
\end{equation}%
and seek the solution in the form (\ref{2}) with $\varsigma $ instead of $y$%
. Replacing the right-hand side of Eq. (\ref{1})$_{1}$ with $\frac{\partial 
}{\partial t}\left( \rho \frac{\partial u_{i}}{\partial t}\right) $ retains
Stroh's form (\ref{5}) of the ODS in $\varsigma ,$ but the state vector $%
\mathbf{\eta }\left( \varsigma \right) $ and the system matrix $\mathbf{Q}%
\left( \varsigma \right) $ change their form due to 
\begin{equation}
\begin{array}{c}
\mathbf{t}_{2}=\left[ \left( e_{2}e_{2}\right) -\rho c^{2}\right] \mathbf{u}%
^{\prime }+i\left[ k_{x}\left( e_{2}e_{1}\right) -\omega c\rho \right] 
\mathbf{u,} \\ 
\mathbf{R}=\left( e_{1}e_{2}\right) -\frac{\omega }{k_{x}}\rho c\mathbf{I},\ 
\mathbf{T}=\left( e_{2}e_{2}\right) -\rho c^{2}\mathbf{I}%
\end{array}
\label{t2}
\end{equation}%
replacing, respectively, the traction $\mathbf{t}_{2}$ (\ref{4}) and
submatrices $\mathbf{R}$ and $\mathbf{T}$ (\ref{7}) of the Stroh matrix $%
\mathbf{N}$.

It is observed that the ODS with the modified matrix $\mathbf{Q}\left(
\varsigma \right) $ preserves the Hamiltonian structure, and hence its
solutions possess the same algebraic properties as those of the standard
Stroh's ODS, except that the matrix $\mathbf{T}$ (\ref{t2}) is no longer
unreservedly positive definite. This lifting of what looks like a minor
formal limitation actually unlocks new properties and scenarios for the ODS
solutions, which may extend beyond the conventional Floquet-Bloch theory and
are therefore highly inviting for further studies.

\section*{Appendix 2. Energy identities 
\addcontentsline{toc}{section}{Appendix 2. Energy identities}}

\subsection*{A2.1 General basics 
\addcontentsline{toc}{subsection}{A2.1
General
basics}}

Consider an inhomogeneous viscoelastic medium with the constitutive relation 
$\mathbf{\sigma =c\nabla u+\text{\ng }\nabla \dot{u},}$ where $\mathbf{\text{%
\ng }}$ is the 4th-rank tensor of (local) viscosity satisfying strong
ellipticity and the same symmetry in indices as the stiffness tensor $%
\mathbf{c}$. Multiplying the source-free wave equation $\mathbf{\nabla \cdot
\sigma }=\rho \mathbf{\ddot{u}}$ by $\mathbf{\dot{u}}$ yields the
instantaneous balance%
\begin{equation}
\mathcal{\dot{K}}+\mathcal{\dot{W}}+2\mathcal{D}+\func{div}\mathbf{P}=0%
\mathbf{,}  \label{e1}
\end{equation}%
where $\mathcal{K}=\frac{1}{2}\rho \mathbf{\dot{u}}^{2}\ $and $\mathcal{W}=%
\frac{1}{2}\mathbf{cu}^{\prime }\mathbf{u}^{\prime }\ $\ are the kinetic and
stored energy densities, $\mathcal{D}=\frac{1}{2}\mathbf{\eta \mathbf{\dot{u}%
}^{\prime }\dot{u}}^{\prime }\ $is the dissipation function density, and $%
\mathbf{P}=-\mathbf{\sigma \dot{u}}$ is the energy flux density (the
physical interpretations assume real $\mathbf{u}$ indeed).

Assume a time-harmonic wave train $\mathbf{u}\left( \mathbf{r,}t\right) =%
\mathbf{u}\left( \mathbf{r}\right) e^{-i\omega t}$, where $\mathbf{u}\left( 
\mathbf{r}\right) $ may be complex while $\omega $ is real, and let the
overbar indicate averaging over the time period, i.e.%
\begin{equation}
\overline{\left( \cdot \right) }\equiv \frac{1}{T}\int_{0}^{T=2\pi /\omega
}\left( \cdot \right) dt.  \label{e0}
\end{equation}%
Applying (\ref{e0}) to (\ref{e1}) yields%
\begin{equation}
2\overline{\mathcal{D}}+\func{div}\overline{\mathbf{P}}=0,  \label{e2}
\end{equation}%
where $\overline{\mathcal{D}}=\tfrac{1}{4}\omega ^{2}\mathbf{\text{\ng }%
u^{\prime }u^{\prime \ast }\ }$and $\overline{\mathbf{P}}=-\tfrac{1}{4}%
i\omega \left( \mathbf{\sigma u}^{\ast }-\mathbf{\sigma }^{\ast }\mathbf{u}%
\right) .$ In turn, multiplying the wave equation above by $\mathbf{\dot{u}}%
^{\ast }\left( =i\omega \mathbf{u}^{\ast }\right) $ leads to the identity%
\begin{equation}
i\omega \overline{\mathcal{L}}-\overline{\mathcal{D}}+\tfrac{1}{4}i\omega 
\func{div}\left( \mathbf{\sigma u^{\ast }}\right) =0,  \label{e3}
\end{equation}%
where $\overline{\mathcal{L}}=\overline{\mathcal{K}}-\overline{\mathcal{W}}$
with $\overline{\mathcal{K}}=\frac{1}{4}\rho \omega ^{2}\mathbf{uu^{\ast }\ }
$and $\overline{\mathcal{W}}=\frac{1}{4}\mathbf{cu}^{\prime }\mathbf{u}%
^{\ast \prime }$ is the averaged Lagrangian function density. The real part
of (\ref{e3}) coincides with (\ref{e2}), while the imaginary part gives%
\begin{equation}
\overline{\mathcal{L}}+\tfrac{1}{8}\func{div}\left( \mathbf{\sigma u^{\ast
}+\sigma }^{\ast }\mathbf{\mathbf{u}}\right) =0.  \label{e4}
\end{equation}%
Note that if there is no dissipation, then Eqs. (\ref{e2})-(\ref{e4}) yield $%
\func{div}\overline{\mathbf{P}}=0$ and 
\begin{equation}
\overline{\mathcal{L}}+\tfrac{1}{4}\func{div}\left( \mathbf{\sigma u^{\ast }}%
\right) =0.  \label{e4.1}
\end{equation}

Let us refer subsequent considerations to 1D inhomogeneous media and the
wave field $\mathbf{u}\left( x,y,t\right) $ of the form (\ref{2}) satisfying
Stroh's ODS (\ref{5}\textbf{)} with possibly viscoelastic stiffness $\mathbf{%
c}\left( y\right) -i\omega \mathbf{\text{\ng }}\left( y\right) $ in place of
a purely elastic one (see (\ref{4.1}) and (\ref{7})). Suppose that this wave
propagates in a transversely inhomogeneous plate $\left[ y_{1},y_{2}\right] $
or a half-space $\left[ y_{1},\infty \right) $ cut orthogonally to the axis $%
Y$ and it maintains a zero $y$-component of the flux $\overline{P}_{y}$ at
the plate boundaries $y=y_{1},y_{2}$ or else at the surface $y=y_{1}$ and
infinite depth of the half-space. Then integrating (\ref{e2}) yields 
\begin{equation}
\left\langle \overline{\mathcal{D}}\right\rangle -k_{x}^{\prime \prime
}\left\langle \overline{P}_{x}\right\rangle =0,\   \label{e6}
\end{equation}%
where $\overline{P}_{x}\left( y\right) $ is the $x$-component of the flux, $%
k_{x}^{\prime \prime }\equiv \func{Im}k_{x}$, and%
\begin{equation}
\left\langle \cdot \right\rangle \equiv \int_{y_{1}}^{y_{2}}\left( \cdot
\right) dy  \label{e5}
\end{equation}%
(no division by the length of the integration interval allows incorporating
the case of a half-space). Equality (\ref{e6}) confirms that non-zero
dissipation $\mathcal{D}\neq 0$ necessitates non-zero $k_{x}^{\prime \prime
},$ which signifies the spatial leakage of the time-harmonic wave. If $%
\mathcal{D}=0$, then $k_{x}^{\prime \prime }\neq 0$ entails $\left\langle 
\overline{P}_{x}\right\rangle =0$, i.e. the leaky waves in the absence of
dissipation are "non-propagating" (do not carry energy) along the $X$%
-direction. In turn, the non-leaky waves ($k_{x}^{\prime \prime }=0$) have
non-zero flux component $\left\langle \overline{P}_{x}\right\rangle \neq 0$
except at the cutoffs $k_{x}=0$ of the upper dispersion branches
(standing-wave resonances) and their folding points where the group velocity
vector and hence the energy flux $\left\langle \overline{\mathbf{P}}%
\right\rangle $ are normal to $X$ (see \S \ref{SSSec9.2.3}). The notion of
propagating and non-propagating waveguide modes underlies Auld's integral
reciprocity relations \cite{A}.

\subsection*{A2.2 Case of no dissipation and no leakage 
\addcontentsline{toc}{subsection}{A2.2 Case of no dissipation and no leakage}%
}

\subsubsection*{A2.2.1 Unbounded 1D inhomogeneous medium 
\addcontentsline{toc}{subsubsection}{A2.2.1
Unbounded
1D
inhomogeneous
medium}%
}

Further, we restrict attention to the case of pure elasticity and non-leaky
waves ($\mathcal{D}=0$, $k_{x}^{\prime \prime }=0$). For definiteness,
explicit formulas below will be referred to the definition (\ref{8})$_{3}$
of the entries of ODS (\ref{5}\textbf{)} so that $\mathbf{\eta }\left(
y\right) =\left( \mathbf{u\ }i\mathbf{t}_{2}\right) ^{T}$. By (\ref{e2}),
the $y$-component of the energy flux $\overline{\mathbf{P}}\left( y\right) $
is independent of $y$:%
\begin{equation}
\overline{P}_{y}=-\frac{\omega }{4}\mathbf{\eta }^{+}\left( y\right) \mathbf{%
\mathbb{T}\eta }\left( y\right) =const\ \ \ \forall y  \label{e7.0}
\end{equation}%
(see the comment below (\ref{25})). In turn, combining the equation of
motion (\ref{3}) with identities (\ref{24}) and (\ref{e4.1}) yields the
energy balance involving the $x$-component of $\overline{\mathbf{P}}\left(
y\right) $, namely, 
\begin{equation}
\overline{P}_{x}\left( y\right) +\tfrac{1}{4}v\Pi \left( y\right) =v\left[ 
\overline{\mathcal{K}}\left( y\right) +\overline{\mathcal{W}}\left( y\right) %
\right] ,  \label{e7}
\end{equation}%
where $v=\omega /k_{x}$ and 
\begin{equation}
\Pi \left( y\right) =i\mathbf{\eta }^{+}\left( y\right) \mathbf{\mathbb{T}Q}%
\left( y\right) \mathbf{\eta }\left( y\right) ,  \label{e7.1}
\end{equation}%
which is a real-valued quantity by (\ref{25}). If $y$ lies within the range
where the medium is homogeneous so that the wave solution $\mathbf{\eta }%
\left( y\right) $ is a superposition of six exponential modes (\ref{11.01}),
then 
\begin{equation}
\tfrac{1}{4}\omega \Pi \left( y\right) =\sum\nolimits_{\alpha }k_{y\alpha }%
\overline{p}_{y\alpha }+\sum\nolimits_{\beta }\left( k_{y\beta }^{\ast }%
\overline{p}_{\beta ^{\ast }\beta }+k_{y\beta }\overline{p}_{\beta \beta
^{\ast }}\right) ,  \label{e7'}
\end{equation}%
where $\overline{p}_{y\alpha }=-\frac{1}{4}\omega \mathbf{\xi }_{\alpha }^{+}%
\mathbf{\mathbb{T}\xi }_{\alpha }$ is the $y$-component of the energy flux
generated by the $\alpha $th bulk partial mode ($\func{Im}k_{y\alpha }=0$)
and $\overline{p}_{\beta ^{\ast }\beta }=-\frac{1}{4}\omega \mathbf{\xi }%
_{\beta ^{\ast }}^{+}\mathbf{\mathbb{T}\xi }_{\beta }\left( =\overline{p}%
_{\beta ^{\ast }\beta }^{\ast }\right) $ is the $y$-component of the
time-averaged energy flux generated by the interference of the $\beta $th
and $\beta ^{\ast }$th partial modes with complex conjugated $k_{y\beta }$
and $k_{y\beta ^{\ast }}\equiv k_{y\beta }^{\ast }$ ($\func{Im}k_{y\beta
}\neq 0$), hence $\mathbf{\xi }_{\beta }$ and $\mathbf{\xi }_{\beta ^{\ast
}}\equiv \mathbf{\xi }_{\beta }^{\ast }$. Note also the identity 
\begin{equation}
\overline{\mathcal{K}}=\tfrac{1}{8}i\omega \frac{d}{dy}{\LARGE [}\mathbf{%
\eta }^{+}\left( y\right) \mathbf{\mathbb{T}}\frac{\partial \mathbf{\eta }%
\left( y\right) }{\partial \omega }|_{k_{x}}{\LARGE ]},  \label{e7.2}
\end{equation}%
which is readily traceable from ODS (\ref{5}) and (\ref{8})$_{3}$ \cite{IT}.

The absence of dissipation and leakage brings forth the two-point
(Hermitian) impedance matrix $\mathbf{Z}\left( y_{2},y_{1}\right) \ $(\ref%
{82}). Integrating (\ref{e4.1}) in $y$ between arbitrary $y_{1}$ and $y_{2}$
and then either using (\ref{e7.2}) or taking a shortcut $2\overline{\mathcal{%
K}}=\omega \left( \partial \overline{\mathcal{L}}/\partial \omega \right) ,$
where $\partial /\partial \omega $ is evaluated at an arbitrary fixed $%
\mathbf{u}$ \cite{LB}, provide the equalities%
\begin{equation}
\begin{array}{c}
\left\langle \overline{\mathcal{L}}\right\rangle =-\tfrac{1}{4}\left( 
\begin{array}{c}
\mathbf{u}\left( y_{1}\right) \\ 
\mathbf{u}\left( y_{2}\right)%
\end{array}%
\right) ^{+}\mathbf{Z}\left[ \omega ,k_{x}\right] \left( 
\begin{array}{c}
\mathbf{u}\left( y_{1}\right) \\ 
\mathbf{u}\left( y_{2}\right)%
\end{array}%
\right) , \\ 
\left\langle \overline{\mathcal{K}}\right\rangle =-\tfrac{1}{8}\omega \left( 
\begin{array}{c}
\mathbf{u}\left( y_{1}\right) \\ 
\mathbf{u}\left( y_{2}\right)%
\end{array}%
\right) ^{+}\dfrac{\partial \mathbf{Z}\left[ \omega ,k_{x}\right] }{\partial
\omega }\left( 
\begin{array}{c}
\mathbf{u}\left( y_{1}\right) \\ 
\mathbf{u}\left( y_{2}\right)%
\end{array}%
\right) .%
\end{array}
\label{e8}
\end{equation}%
The same may be expressed via the conditional impedance $\mathbf{z}\left(
y\right) $ (\ref{85}), e.g., 
\begin{equation}
\left\langle \overline{\mathcal{L}}\right\rangle =-\tfrac{1}{4}\mathbf{u}%
^{+}\left( y_{1}\right) \mathbf{z}\left[ \omega ,k_{x}\right] \mathbf{u}%
\left( y_{1}\right) ,\ \left\langle \overline{\mathcal{K}}\right\rangle =-%
\tfrac{1}{8}\omega \mathbf{u}^{+}\left( y_{1}\right) \dfrac{\partial \mathbf{%
z}\left[ \omega ,k_{x}\right] }{\partial \omega }\mathbf{u}\left(
y_{1}\right) ,  \label{e9}
\end{equation}%
where $\mathbf{z}\left[ \omega ,k_{x}\right] \equiv \mathbf{z}\left(
y_{1}\right) |_{\mathbf{z}\left( y_{2}\right) =\mathbf{0}}$ (see (\ref{88})$%
_{2}$). According to (\ref{e8}), the matrices $\mathbf{Z}\left(
y_{2},y_{1}\right) $ and $\mathbf{z}\left( y_{1}\right) |_{\mathbf{z}\left(
y_{2}\right) =\mathbf{0}}$ are positive definite at $\omega =0$ and their
frequency derivatives $\left( \partial /\partial \omega \right) _{k_{x}}$
evaluated at a fixed $k_{x}$ are negative definite. The derivative $\left(
\partial /\partial \omega \right) _{k_{x}}$ may certainly be replaced with $%
k_{x}^{-1}\left( \partial /\partial v\right) _{k_{x}}$ taken of $\mathbf{Z}%
\left[ v,k_{x}\right] $ and $\mathbf{z}\left[ v,k_{x}\right] ;$ at the same
time, invoking $\left( \partial /\partial v\right) _{\omega }$ leads to to
the expressions 
\begin{equation}
\begin{array}{c}
\overline{\mathcal{K}}=\tfrac{1}{8}iv\dfrac{d}{dy}{\LARGE [}\mathbf{\eta }%
^{+}\left( y\right) \mathbf{\mathbb{T}}\dfrac{\partial \mathbf{\eta }\left(
y\right) }{\partial v}|_{\omega }{\LARGE ]}+\tfrac{1}{8}\Pi ,\  \\ 
\left\langle \overline{\mathcal{K}}\right\rangle =-\tfrac{1}{8}v\mathbf{u}%
^{+}\left( y_{1}\right) \dfrac{\partial \mathbf{z}\left[ v,\omega \right] }{%
\partial v}\mathbf{u}\left( y_{1}\right) +\tfrac{1}{8}\left\langle \Pi
\right\rangle ,%
\end{array}
\label{e10}
\end{equation}%
whose right-hand side acquire an additional term relative to Eqs. (\ref{e7.2}%
) and (\ref{e8})$_{2}$, (\ref{e9})$_{1}$. Similar relations can be written
in terms of admittance matrices and traction vectors. Passing from
definition (\ref{8})$_{3}$\ to (\ref{8})$_{1}$\ (as in \cite{LB} and \cite%
{DSh18,Sh3}) amounts to multiplying the right-hand side of (\ref{e8}) and (%
\ref{e9}) by $k_{x}$\ (and of (\ref{e10}) by $k_{x}^{-1}$) and thus retains
the above signs, whereas using (\ref{8})$_{2}$\ inverts them.

The above derivation of the impedance sign-definiteness generalizes the
pioneering approach of \cite{IT,LB}, which dealt with the surface impedance
of a homogeneous half-space, to the two-point impedance of 1D inhomogeneous
media; it also admits straightforward extension to the cases of transversely
and laterally periodic media, see \cite{DSh18,ShWM}.

\subsubsection*{A2.2.2 Dispersion spectrum 
\addcontentsline{toc}{subsubsection}{A2.2.2 Dispersion spectrum}}

We continue with a transversely inhomogeneous plate or a half-space, now
assumed to be subjected to a homogeneous (traction-free or clamped) boundary
condition. This condition nullifies the normal component of the energy flux $%
\overline{P}_{y}$ at the surface(s) and hence, by (\ref{e7.0}), at any $y$:%
\begin{equation}
\overline{P}_{y,J}=0\ \forall y.  \label{e11}
\end{equation}%
Here and below, the additional subscript $J$ refers to the $J$th dispersion
branch $\omega _{J}\left( k_{x}\right) =v_{J}\left( k_{x}\right) k_{x}$
defined by the boundary-value problem. By (\ref{e11}), either of the now
equivalent Eqs. (\ref{e4}) or (\ref{e4.1}) yields 
\begin{equation}
\left\langle \overline{\mathcal{L}}_{J}\right\rangle =0\ \Leftrightarrow \
\left\langle \overline{\mathcal{K}}_{J}\right\rangle =\left\langle \overline{%
\mathcal{W}}_{J}\right\rangle ,  \label{e12}
\end{equation}%
where the symbol $\left\langle \cdot \right\rangle $ defined in (\ref{e5})
indicates here \textit{arbitrary} integration limits $y_{1},~y_{2}$, not
necessarily attached to the plate faces. Note aside that the
surface-localized wave fields satisfying the radiation condition fulfil Eqs.
(\ref{e11}) and (\ref{e12}) at any $\omega $ and $k_{x}$, i.e. regardless of
any boundary conditions.

The through-plate average energy balance (\ref{e7}) taken on the dispersion
spectrum reads 
\begin{equation}
\left\langle \overline{P}_{x,J}\right\rangle +\tfrac{1}{4}v_{J}\left\langle
\Pi _{J}\right\rangle =2v_{J}\left\langle \overline{\mathcal{K}}%
_{J}\right\rangle ,  \label{e13}
\end{equation}%
where (\ref{e12}) is used and the integration limits $y=y_{1},y_{2}$ in $%
\left\langle \cdot \right\rangle $ are set at the plate faces, or one of
them is at infinity in the case of a half-space. This balance can be linked
to the $x$-component, i.e. the sagittal plane projection, of the group
velocity $g_{x,J}=d\omega _{J}/dk_{x}$ ($\equiv g_{J}^{\left( \mathbf{m}%
\right) }$ in \S \ref{SSSec9.2.3}). With this purpose, consider a
traction-free plate $\left[ y_{1},y_{2}\right] $ and the corresponding
dispersion equation in the form (\ref{91})$_{3}.$ Let $z_{J}$ be one of the
three (real) eigenvalues of the conditional impedance $\mathbf{z}\left(
y_{1}\right) |_{\mathbf{z}\left( y_{2}\right) =\mathbf{0}},$ which turns to
zero on the $J$th branch $\omega _{J}\left( k_{x}\right) $. Note the formal
identity 
\begin{equation}
\frac{\partial z_{J}\left( v_{J},\omega \right) }{\partial v}=\frac{g_{x,J}}{%
v_{J}}\frac{\partial z_{J}\left( v_{J},k_{x}\right) }{\partial v}.
\label{e14}
\end{equation}%
Taking Eqs. (\ref{e9})$_{2}$ and (\ref{e10})$_{2}$ on the $J$th dispersion
branch and eliminating the impedance derivatives yields the relation between
the in-plane group velocity and trace (phase) velocity 
\begin{equation}
g_{x,J}-v_{J}{\Large (}=k_{x}\frac{dv_{J}\left( k_{x}\right) }{dk_{x}}%
{\Large )}=-v_{J}\dfrac{\left\langle \Pi _{J}\right\rangle }{8\left\langle 
\overline{\mathcal{K}}_{J}\right\rangle }.  \label{e15}
\end{equation}%
Plugging (\ref{e15}) into (\ref{e13}), we arrive at the equality between the 
$x$-components of the mean energy and group velocities: 
\begin{equation}
\frac{\left\langle \overline{P}_{x,J}\right\rangle }{2\left\langle \overline{%
\mathcal{K}}_{J}\right\rangle }=g_{x,J},  \label{e16}
\end{equation}%
which is a standard feature of guided wave propagation (see, e.g. \cite{A}).
From (\ref{e15}) or (\ref{e13}) and (\ref{e16}), it follows that the
condition for the zero of the in-plane group velocity $g_{x,J}$ and the flux 
$\left\langle \overline{P}_{x,J}\right\rangle $ is the equality 
\begin{equation}
\left\langle \Pi _{J}\right\rangle =8\left\langle \overline{\mathcal{K}}%
_{J}\right\rangle .  \label{e16.1}
\end{equation}%
The same conclusions hold for the other types of homogeneous boundary
conditions, such as clamped or clamped/free plates, in which case the
difference amounts to rephrasing the interim derivations in terms of an
appropriate impedance or admittance matrix.

On the "technical side", note that expressions (\ref{e15}) and (\ref{e16})
may also be obtained from (\ref{e9})$_{2}$ by substituting the relations 
\begin{equation}
\frac{\partial z_{J}\left( v_{J},k_{x}\right) }{\partial v}\frac{d\omega _{J}%
}{dk_{x}}=-k_{x}\frac{\partial z_{J}\left( v_{J},k_{x}\right) }{\partial
k_{x}}=-\left\langle \Pi _{J}\right\rangle ,\ \   \label{e17}
\end{equation}%
where the second equality in (\ref{e9})$_{2}$ takes into account the
definition $\mathbf{z}\left( y_{1}\right) |_{\mathbf{z}\left( y_{2}\right) =%
\mathbf{0}}=-i\mathbf{M}_{4}^{-1}\mathbf{M}_{3}$ (\ref{88})$_{2},$ the
identities $\mathbf{M}_{3}\mathbf{u}\left( y_{1}\right) =\mathbf{0}$\textbf{,%
} $\mathbf{u}^{+}\left( y_{2}\right) \mathbf{M}_{3}=\mathbf{0}$, $\mathbf{u}%
\left( y_{1}\right) =\mathbf{M}_{4}^{+}\mathbf{u}\left( y_{2}\right) $ for a
free plate $\left[ y_{1},y_{2}\right] $ and the formula for the matricant
derivative (see it, e.g., in \cite{P}). Some care is needed at the branch
cutoffs where $k_{x}=0$ and $v_{J}\rightarrow \infty $, and at the folding
points where $g_{x,J}=0$ and $dv_{J}/d\omega \rightarrow \infty $. Note also
that the inequality $\Pi _{J}>0,$ which is the necessary condition for the
vanishing of the in-plane group velocity $g_{x,J},$ is unlikely, but
seemingly is not impossible to hold on the fundamental branches.

The energy balance (\ref{e7}) and the ensuing results outlined above were
discussed in the context of homogeneous plates in \cite{Sh0}; here, they are
extended to the transversely inhomogeneous plates and substrates. Note minor
dissimilarities with the notations of \cite{Sh0}, where the factor $\tfrac{1%
}{4}v$ was included into $\Pi ,$ the plate thickness was taken to be $2h$
and the definition (\ref{8})$_{1}$ of $\mathbf{Q}$ was used.

Let us end up with highlighting an interesting feature that pertains
specifically to homogeneous plates. First, Eq. (\ref{e15}) shows that the
zero and the sign of the integral $\left\langle \Pi _{J}\right\rangle $
taken on the $J$th dispersion branch dictate those of the derivatives $%
dv_{J}/dk_{x}=g_{x,J}dv_{J}/d\omega ,$ and Eq. (\ref{e16.1}) tells us that $%
\left\langle \Pi _{J}\right\rangle >0$ is a necessary condition for the zero
group and mean energy velocity. Second, if the material is homogeneous, then
the value of $\Pi $ (\ref{e7.1}) remains constant and hence is the same at
any $y$ as it is at the plate faces. Given that both faces are free of
traction (i) or (at least) one of them is clamped (ii), this value taken on
the branch $v_{J}\left( k_{x}\right) $ is%
\begin{equation}
\Pi _{J}=\left\{ 
\begin{array}{l}
-k_{x}^{2}\mathbf{u}_{J}^{+}\left( \mathbf{N}_{3}-\rho v_{J}^{2}\mathbf{I}%
\right) \mathbf{u}_{J}\ \ \ \ \left( \mathrm{i}\right) , \\ 
-k_{x}^{2}\mathbf{t}_{2,J}^{+}\mathbf{N}_{2}\mathbf{t}_{2,J}\ \ \ \ \ \ \ \
\ \ \ \ \ \ \ \ \left( \mathrm{ii}\right) ,%
\end{array}%
\right.  \label{e18}
\end{equation}%
where $\mathbf{u}_{J}$ is the displacement and $\mathbf{t}_{2,J}$ is the
traction on either the free or clamped face, respectively, and $\mathbf{N}%
_{2}$, $\mathbf{N}_{3}$ are the blocks of the Stroh matrix $\mathbf{N}$ (\ref%
{7}). Recall that $\mathbf{N}_{2}$ is negative definite and $\mathbf{N}_{3}$
is positive semi-definite with eigenvalues $0$ and $\lambda _{2},~\lambda
_{3}>0$, so that $\Pi _{J}$ is positive for $\rho v_{J}^{2}$ greater than
both $\lambda _{2}$ and $\lambda _{3}$ in the case (i) and for any $v_{J}$
in the case (ii). Hence, the in-plane group velocity $g_{x,J}$ is always
less than the trace (phase) velocity $v_{J},$ and the dispersion curves $%
v_{J}\left( k_{x}\right) $ or $v_{J}\left( \omega \right) $ cannot have
extreme points in the velocity interval $\rho v^{2}>\max \left( \lambda
_{2},\lambda _{3}\right) $ of the Lamb wave spectrum in any traction-free
homogeneous plate. The same holds true throughout the dispersion spectrum in
a homogeneous plate with clamped and clamped/free boundary conditions. A
precise formulation of the above property of Lamb waves is given in (\ref%
{100}), and its graphical illustration is provided in \cite{Sh3}.

\end{document}